\documentclass[prd,
a4paper,
nofootinbib,
reprint,
]{revtex4}
\pdfoutput=1 

\usepackage{graphicx}
\usepackage{amsfonts}
\usepackage{amssymb}
\usepackage{amsmath}
\usepackage{slashed} 
\usepackage{enumerate}
\usepackage{color}
\usepackage{comment}
\usepackage{tikz}
\usepackage{hyperref}
\usepackage[utf8]{inputenc}
\usepackage{erewhon}
\definecolor{VioletRed4}{rgb}{0, 0, 0}

\usepackage{hyperref}

\def\eq#1{{eq.~(\ref{#1})}}

\definecolor{oucrimsonred}{rgb}{0.6, 0.0, 0.0}
\definecolor{persianblue}{rgb}{0.11, 0.22, 0.73}
\definecolor{forestgreen}{rgb}{0.13,0.35,0.13}
 \hypersetup{colorlinks, citecolor=oucrimsonred, linkcolor=persianblue, urlcolor=oucrimsonred}
\def\hhref#1{\href{http://arxiv.org/abs/#1}{#1}} % in bibliography
\newcommand{\be}{\begin{equation}}
\newcommand{\ee}{\end{equation}}
\newcommand{\bea}{\begin{eqnarray}}
\newcommand{\eea}{\end{eqnarray}}
\newcommand{\nn}{\nonumber}

\definecolor{oucrimsonred}{rgb}{0.6, 0.0, 0.0}
\newcommand{\fd}[2]{\parbox{#1}{\includegraphics[width=#1]{figs/#2}}}

\definecolor{mtcolor}{rgb}{.8,.3,.1}

\definecolor{violachiaro}{rgb}{1,0.6,1}
 
\definecolor{gbcolor}{rgb}{1,0,0}
 
\definecolor{gbcolor2}{rgb}{.9,.2,.6}
\definecolor{gbcolor3}{rgb}{.3,.2,.6}

\begin{document}
%%%%%%%%%%%%%%%%%%%%%%%%%%%%%%%%%%%%%%%%%%%%%%%%%%%%%%%%%%%  FRONT PAGE

\begin{flushright}
\footnotesize
{IFT-UAM/CSIC-20-10}
\end{flushright}

\title[]{Primordial black holes as dark matter and gravitational waves \\ 
from single-field polynomial inflation
}
\date{\today}
\author{Guillermo Ballesteros$^{a,b}$}
\author{Juli\'an Rey$^{a,b}$}
\author{Marco Taoso$^{c}$}
\author{Alfredo Urbano$^{d,e}$}
\affiliation{$^a$ Instituto de F\'isica Te\'orica UAM/CSIC, Calle Nicolás Cabrera 13--15 Cantoblanco E-28049 Madrid, Spain}
\affiliation{$^b$ Departamento de F\'isica Te\'orica, Universidad Aut\'onoma de Madrid (UAM) Campus de Cantoblanco, E-28049 Madrid, Spain}
\affiliation{$^c$I.N.F.N. sezione di Torino, via P. Giuria 1, I-10125 Torino, Italy}
\affiliation{$^d$I.N.F.N. sezione di Trieste, SISSA, via Bonomea 265, I-34132 Trieste, Italy}
\affiliation{$^e$I.F.P.U., Institute  for  Fundamental Physics  of  the  Universe, via  Beirut  2, I-34014 Trieste, Italy.
}

%%%%%%%%%%%%%%%%%%%%%%%%%%%%%%%%%%%%%%%%%%%%%%%%%%%%%%%%%%%%%%%%%%%%%
\begin{abstract}
\noindent  
We consider the possibility that the majority of dark matter in our Universe consists of black holes  of primordial origin. 
We determine the conditions under which such black holes may have originated from a single-field model of inflation characterized by a quartic polynomial potential. We also explore the effect of higher-dimensional operators. The large power spectrum of curvature perturbations that is needed for a large black hole abundance sources sizable second order tensor perturbations. 
The resulting stochastic background of primordial gravitational waves could be detected by the future space-based observatories LISA and DECIGO or 
--as long as we give up on the dark matter connection--
by the ground-based Advanced LIGO-Virgo  detector  network.

 \end{abstract}
%%%%%%%%%%%%%%%%%%%%%%%%%%%%%%%%%%%%%%%%%%%%%%%%%%%%%%%%%%%%%%%%%%%
\maketitle
%%%%%%%%%%%%%%%%%%%%%%%%%%%%%%%%%%%%%%%%%%%%%%%%%%%%%%%%%%%%%%%%%%%
 %%%%%%%%%%%%%%%%%%%%%%%%%%%%%%%%%%%%%%%%%%%%%%%%%%%%%%%%%%%%%%%%%%%
 
 \section{Motivations and main results}\label{sec:Mot}

The possibility that all dark matter in our Universe consists  of black holes  of primordial origin (PBHs) is exciting.
This is a viable idea in the mass window $10^{17} \lesssim M_{\rm PBH}\,[\,{\rm g}\,] \lesssim 10^{21}$, as recently discussed in refs.\,\cite{Niikura:2017zjd,Katz:2018zrn,Montero-Camacho:2019jte}.
Assuming the majority of dark matter (or at least a sizable fraction)  to be comprised of PBHs, in  this work we will investigate the mechanism that is responsible for its generation.
 We consider the case in which the inflaton potential features an approximate stationary inflection point a few $e$-folds before the end of inflation.
\begin{figure}[t]
\begin{center}
$$\includegraphics[width=.4\textwidth]{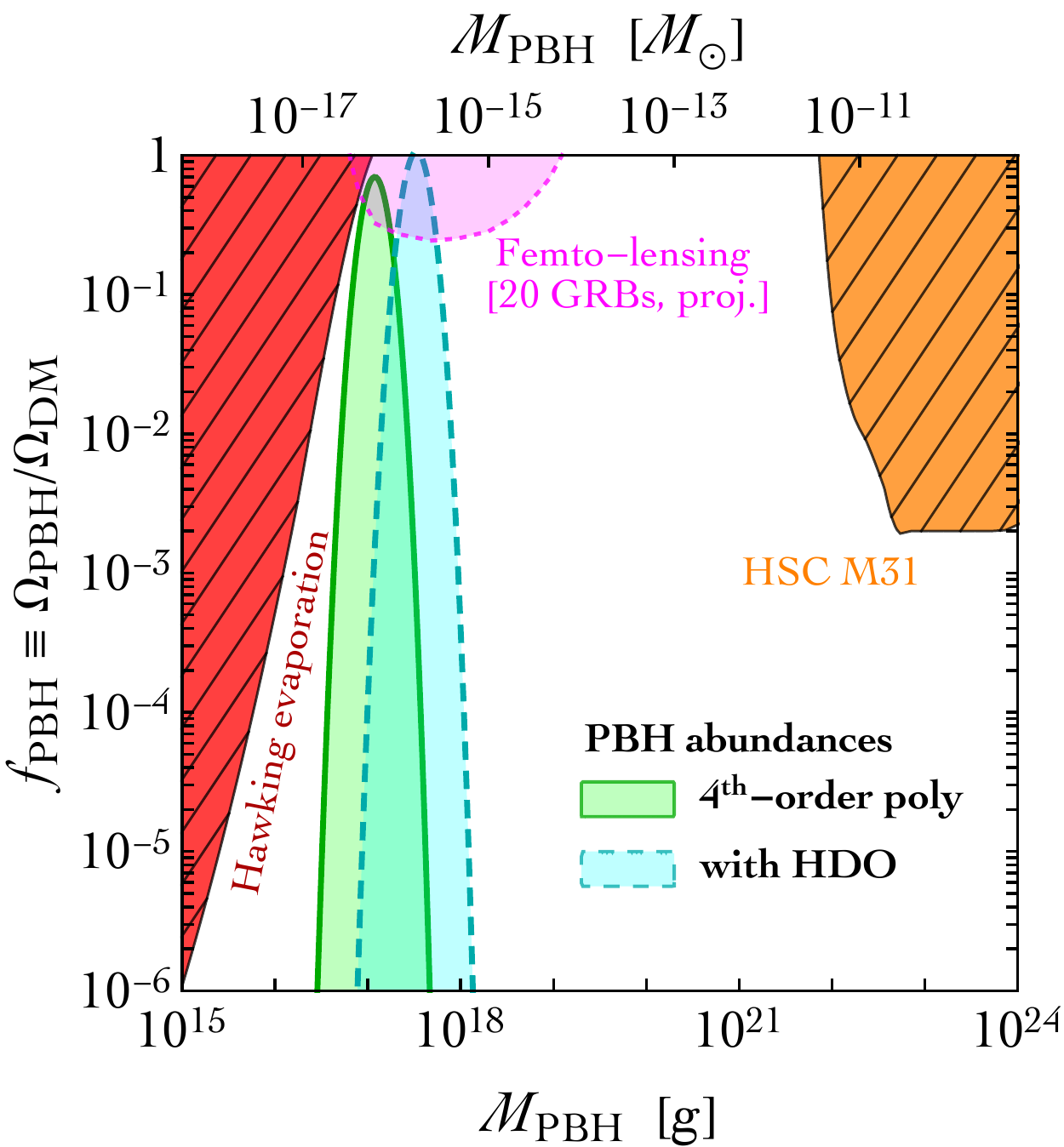}
\qquad\includegraphics[width=.41\textwidth]{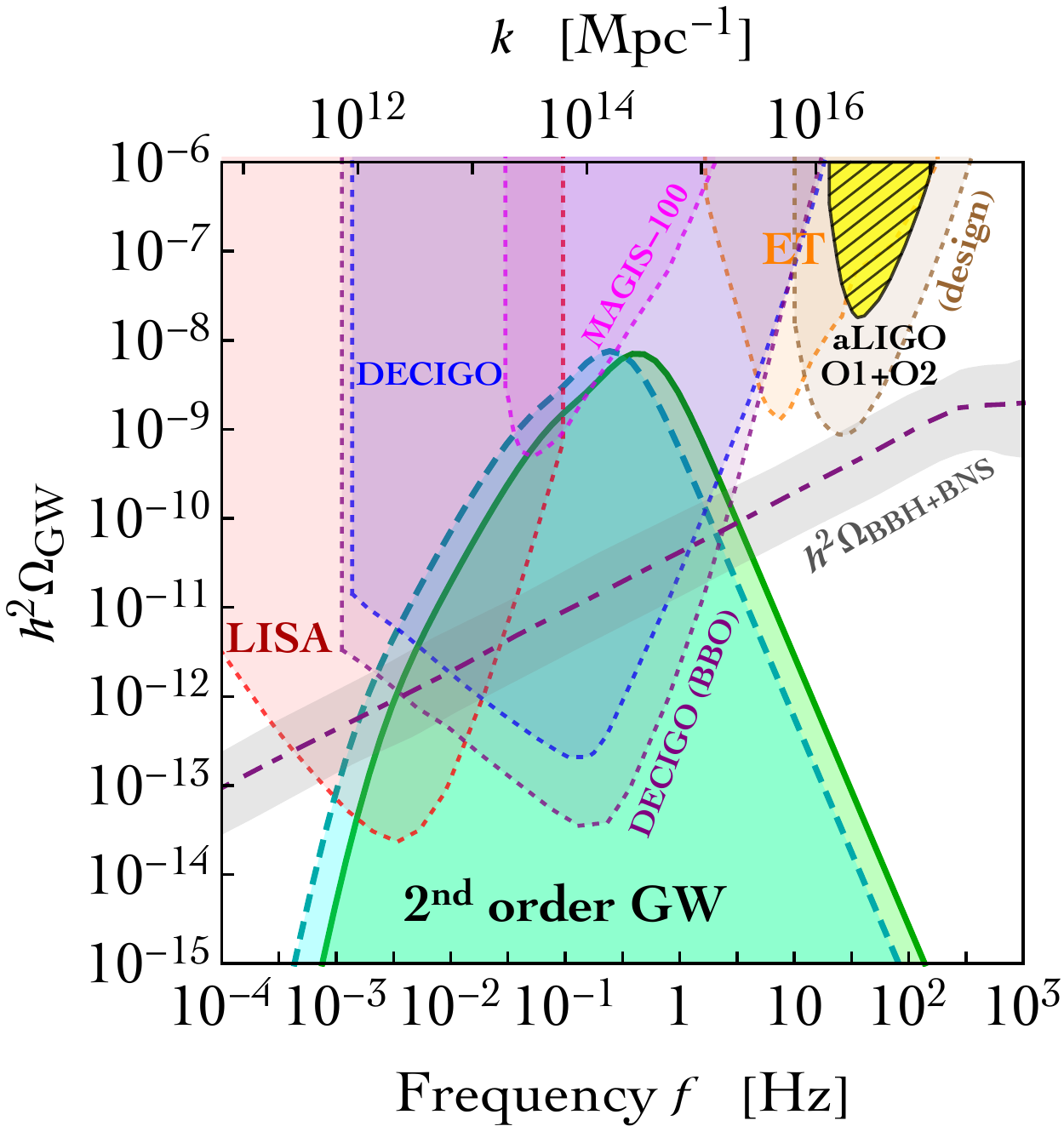}$$
\caption{\em \label{fig:Ka} 
{\color{VioletRed4}{Color code in this figure as in the rest of the paper. Colored regions with solid/dashed boundaries: signals. 
Gray region without boundary and central dot-dashed line: stochastic gravitational wave background from binary black holes and binary neutron stars with its uncertainty.
Colored regions with dotted boundaries: projected experimental/observational sensitivities. 
Colored regions with solid boundaries and diagonal meshes: existing bounds.
Left panel: Fractional abundance of PBHs with respect to the dark matter abundance as a function of the PBH mass for the parameter values  discussed in the text, with and without higher-dimensional operators (HDO)}. 
Right panel: Fraction 
of the energy density in gravitational waves relative to the critical energy density of the Universe as a function of the frequency (again, without and with HDO).
}}
\end{center}
\end{figure}   
As it is well-known, in this case it is possible to have a peak in the power spectrum of comoving curvature perturbations at scales much  smaller than 
those probed by cosmic microwave background (CMB) anisotropy measurements, $0.005 \lesssim k\,[\,{\rm Mpc}^{-1}\,] \lesssim 0.2$. 
If these fluctuations have a large enough amplitude, they may trigger the collapse of Hubble-sized regions into PBHs upon horizon re-entry during the radiation-dominated era; see ref.\,\cite{Sasaki:2018dmp} for a recent review.
What kind of inflaton potential possesses the above-mentioned property? 
Early attempts in the context of single-field inflation were already put forward for the production of PBH dark matter in refs.\,\cite{Ivanov:1994pa,Starobinsky:1992ts,Saito_2008}.
More recently, a model based on a ratio of polynomials was proposed in \cite{Garcia-Bellido:2017mdw}, initiating a search for single-field inflation models with an  approximate stationary inflection point for  PBH formation.\footnote{A phenomenological variant of this idea consists in adding an ad-hoc small bump to the inflaton potential \cite{mishra2019primordial}.}
Popular examples that can accommodate this feature are
motivated by string theory and supergravity, such as
axion-like potentials \cite{Ozsoy:2018flq,Ballesteros:2019hus}. Other examples are
potentials constructed within the framework of type IIB flux compactifications \cite{Cicoli:2018asa} or 
supersymmetric $\alpha$-attractor models \cite{Dalianis:2018frf}.
In the context of more standard (bottom-up) particle physics models, an interesting possibility is offered by 
the case  in which the approximate stationary inflection point has a radiative origin\,\cite{Ballesteros:2015noa}. 
In this  case, the inflaton  $\phi$ has a classical potential that is dominated by the quartic term while at the quantum level an
approximate stationary inflection point appears due to a precise balance between 
logarithmic and double-logarithmic corrections. Embedding this last idea in the context of PBH formation \cite{Ballesteros:2017fsr} requires the  addition of a non-minimal, but completely general, coupling to gravity \cite{Spokoiny:1984bd} that flattens the potential at large field values and allows us to fit the CMB observations.  
In this note, we are interested in the simplest possible scenario, in which the potential is the most general renormalizable one for a real scalar $\phi$:
\begin{align}\label{eq:RenPot}
V(\phi) = a_2\phi^2 + a_3\phi^3 + a_4\phi^4\,
\end{align}
and the flattening at large field values is also due to the non-minimal quadratic coupling of $\phi$ to the Ricci scalar \cite{Ballesteros:2017fsr}.

Let us anticipate our results straight away in fig.\,\ref{fig:Ka}. In the left panel, we show the fraction of the dark matter in the form of PBHs as a function of their mass $M_{\rm PBH}$. 
We find acceptable inflationary solutions in which the majority of dark matter --in the case of fig.\,\ref{fig:Ka} up to a fraction of order 70\%--
is comprised of PBHs (region shaded in green with solid boundary). The observationally excluded regions are depicted with diagonal vertical meshes and refer to limits based on extragalactic background photons from PBH Hawking evaporation\,\cite{Carr:2009jm} (left side, red) and 
micro-lensing searches\,\cite{Niikura:2017zjd} (right side, orange). 
As far as the bound based on Hawking evaporation is concerned, we assume Schwarzschild PBHs. See ref.\,\cite{Arbey:2019vqx} for an extension of this bound including a non-zero spin distribution, ref.\,\cite{Ballesteros:2019exr} for a recent re-analysis and future prospects, and refs.\,\cite{Clark:2016nst,Boudaud:2018hqb,DeRocco:2019fjq,Laha:2019ssq,Dasgupta:2019cae,Poulter:2019ooo} for additional constraints based on Hawking evaporation.
Furthermore, it is worth noticing that the Hawking evaporation bound reproduced in  fig.\,\ref{fig:Ka} is strictly applicable only for a monochromatic PBH mass function. However, our PBH mass distribution  
is well described by a log-normal function with width $\sigma \simeq 0.25$, and for such small value of $\sigma$ the monochromatic bound can be considered as approximately correct 
(see discussion in ref.\,\cite{Carr:2017jsz}).
The region shaded in pink with dashed boundary refers to a projected sensitivity of femto-lensing searches assuming 20 
 suitable gamma-ray burst events\,\cite{Katz:2018zrn}. 
 
As we shall discuss in section\,\ref{sec:Disc}, all the inflationary solutions that are capable of producing a sizable fraction of the dark matter in the form of PBHs obtained by means of the potential in eq.\,(\ref{eq:RenPot}) are characterized 
 by a spectral index at CMB scales $n_s \lesssim 0.95$. At face value, this number is slightly smaller compared to what is preferred by cosmological measurements, namely $n_s \simeq 0.96$, thus creating a 
 3\,$\sigma$ tension with the latest Planck constraints\,\cite{Aghanim:2018eyx}. 
As it was already pointed out in \cite{Ballesteros:2017fsr}, and as we shall argue in more detail in section\,\ref{sec:Disc}, this is not necessarily enough to rule out our solutions but it might be the indication (assuming the correctness of \eq{eq:RenPot}) of some non-standard cosmology beyond the base $\Lambda$CDM model.
 However, having a 3\,$\sigma$ tension may be unsettling, and invoking a non-standard cosmological setup may not be the most appealing solution.
 For this reason, we  will discuss in section\,\ref{sec:HDO} a simple --and arguably natural-- way of circumventing the aforementioned $n_s$ tension. 
The latter is based on just a slight deformation of the potential in eq.\,(\ref{eq:RenPot}), by including higher-dimensional operators (HDO) of the form
 \begin{align}\label{eq:HDOPot}
V(\phi) = a_2\phi^2 + a_3\phi^3 + a_4\phi^4 + \sum_{n\geqslant 5}a_n\phi^n\,.
\end{align}
We shall argue that a natural organization of the series of HDO leads to good inflationary solutions with a value of the spectral index in perfect agreement with Planck data. In fact, a single five-dimensional operator with a naturally small coefficient and negligible higher-order terms is sufficient.
We show the abundance of PBHs generated by one  of these solutions in the left panel of fig.\,\ref{fig:Ka} (region shaded in cyan with dashed boundary). 
In this case we find that having 100\% of dark matter in the form of PBHs is in excellent agreement with CMB observations.
It is also worth noticing that this population of PBHs satisfies the bound discussed in refs.\,\cite{Laha:2019ssq,Dasgupta:2019cae} based on the observation of the 511 keV gamma-ray line from positrons in the Galactic center, which is stronger than the Hawking evaporation bound obtained using the isotropic gamma-ray background.
  
 We also discuss gravitational wave
 signatures.
 In the right panel of fig.\,\ref{fig:Ka} --where we plot the 
 gravitational wave energy density  in  units  of  the critical energy density as a function of frequency--
 we show the gravitational wave signal that comoving curvature perturbations generate as a second-order effect\,\cite{Acquaviva:2002ud,Ananda:2006af,Baumann:2007zm,saito2008gravitational,Saito_2010,Kohri_2018}. 
 We superimpose the signal (region shaded in green with solid boundary for the quartic example in the left panel) on the expected sensitivity  curves of the future 
 gravitational wave detectors LISA (assuming the C1 configuration, see ref.\,\cite{Caprini:2015zlo}), DECIGO\,\cite{Yagi:2011wg}  
 and MAGIS-100\,\cite{Coleman:2018ozp} (shaded regions with dashed boundaries, see caption for details).  
 We find that the signal could be detected by LISA and DECIGO, and it stands out over the stochastic gravitational wave background from binary black holes (BBH) and binary neutron stars (BNS), which has been computed following ref.\,\cite{Chen:2018rzo}.
  
 The position of the peak amplitude of the power spectrum of curvature perturbations, the peak height in the PBH mass distribution and the frequency of the peak of the gravitational wave signal are related by:
 \begin{align}\label{eq:MasterScaling}
 \left(
 \frac{M_{\rm PBH}}{10^{17}\,{\rm g}}
 \right)^{-1/2}\simeq \frac{k}{2\cdot10^{14}\,{\rm Mpc}^{-1}}\simeq\frac{f}{{0.3\,\rm Hz}}\,.
 \end{align}
 A peak in the power spectrum of curvature perturbations at 
$k\simeq 2\cdot10^{16}$ Mpc$^{-1}$ generates a gravitational wave signal 
 with frequency
  $f\simeq 30$ Hz. 
 This is an interesting frequency for the ground-based Advanced LIGO-Virgo  detector  network, which 
already placed important limits
 on the energy density in gravitational waves by combining data from the first (O1) and second (O2) observing runs\,\cite{LIGOScientific:2019vic} (region shaded in yellow with diagonal meshes and solid boundary in fig.\,\ref{fig:Ka}; we also show the design sensitivity with dashed boundary). A gravitational wave signal at 
 frequency 
 $f \simeq 30$ Hz
 would correspond to PBHs with mass $M_{\rm PBH}\simeq 10^{13}$ g.
These PBHs cannot constitute the observed abundance of dark matter since they would have completely evaporated through the emission of Hawking radiation from their formation to the present day. 
Nevertheless, a population of PBHs with mass around  $M_{\rm PBH}\simeq 10^{13}$ g, although extinct today, is still subject to experimental constraints 
associated with the effects of their evaporation on big bang nucleosynthesis\,\cite{Carr:2009jm}.  
\begin{figure}[t]
\begin{center}
$$\includegraphics[width=.4\textwidth]{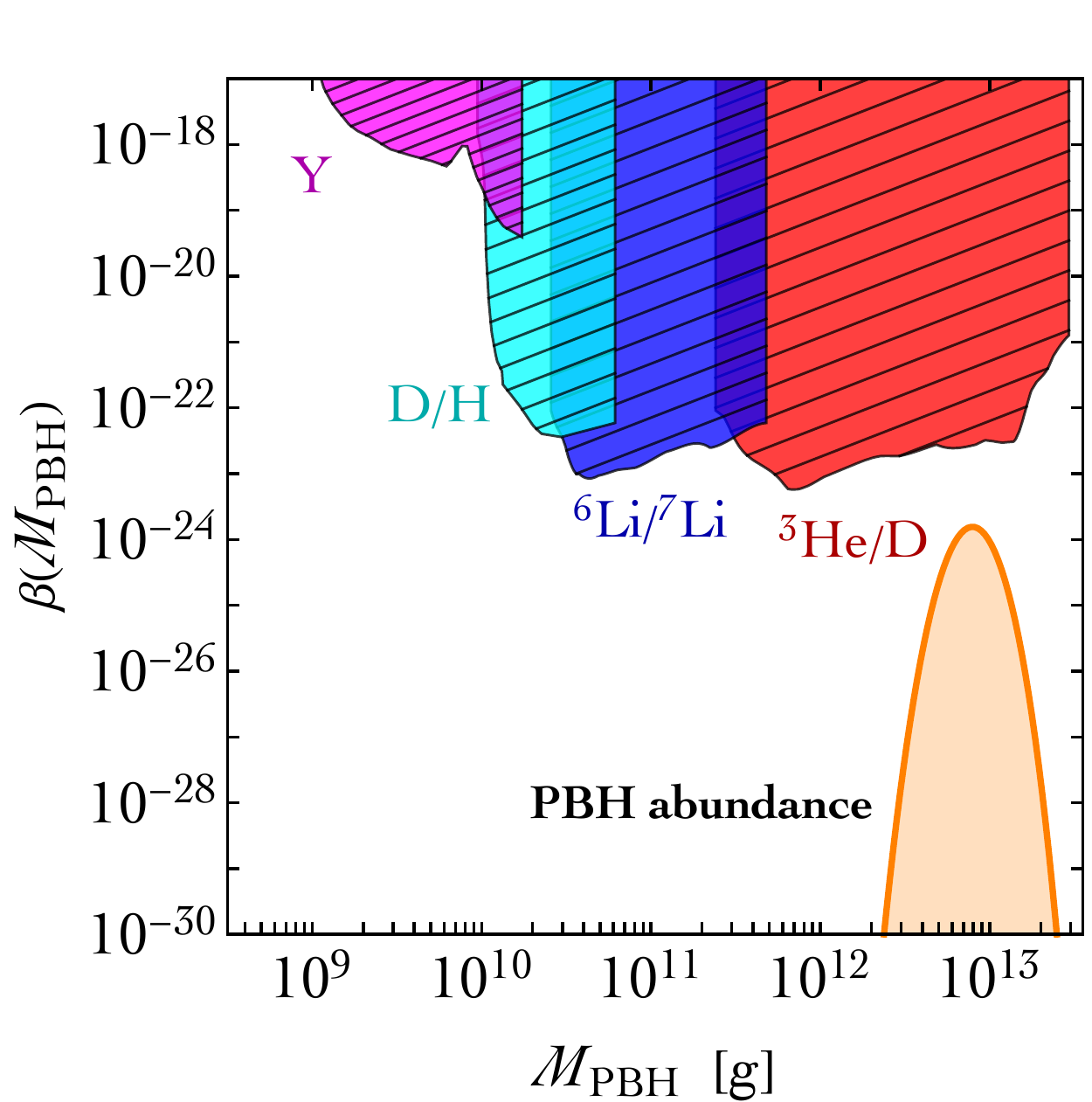}
\qquad\includegraphics[width=.41\textwidth]{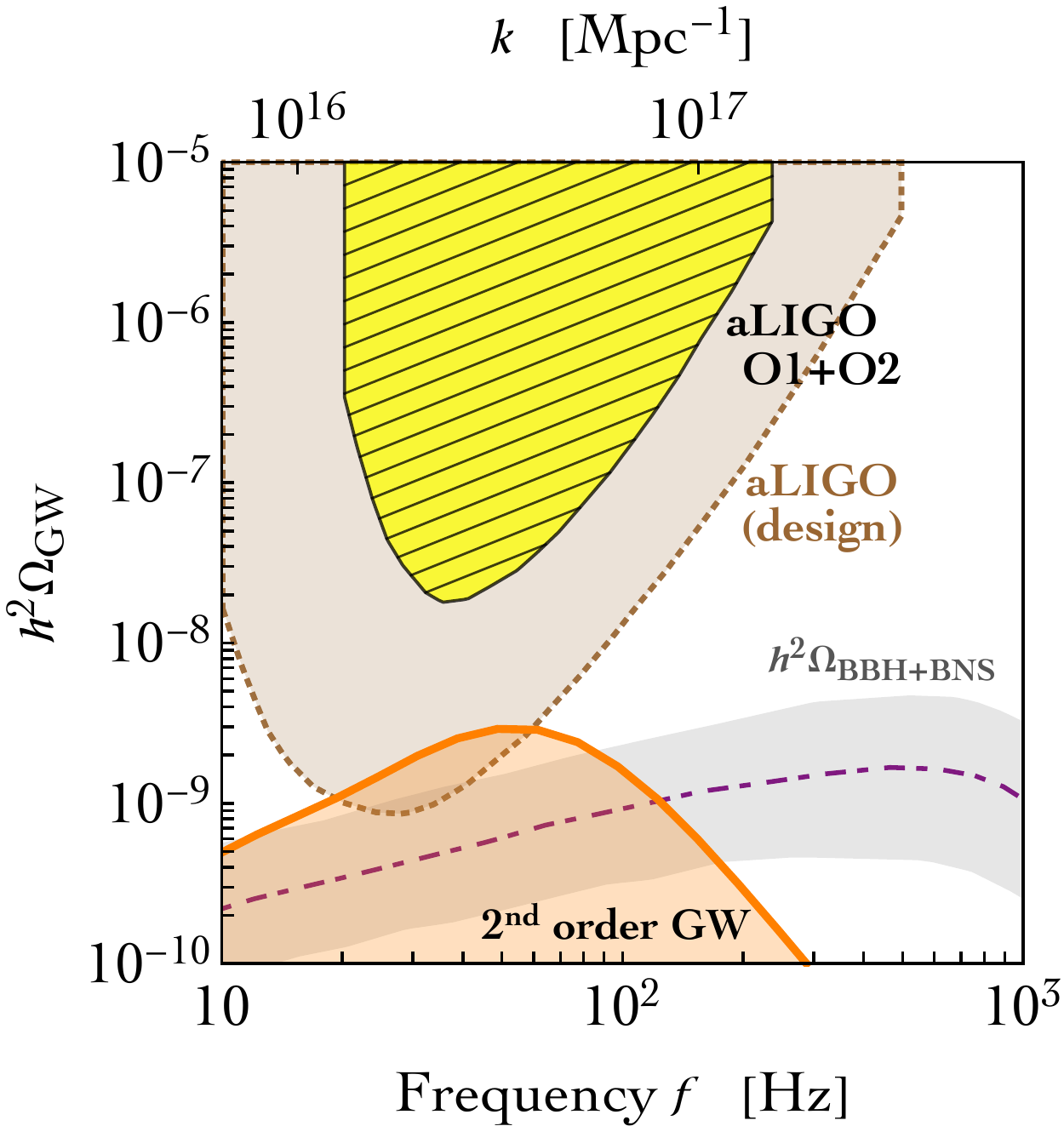}$$
\caption{\em \label{fig:Ka2} 
{\color{VioletRed4}{
Left panel:  Fraction of the Universe's mass in PBHs at their formation time as a function of the PBH mass. 
The bounds are taken from ref.\,\cite{Carr:2009jm}. 
Right panel. Fraction 
of the energy density in gravitational waves relative to the critical energy density of the Universe as a function of the frequency. 
The inflationary solution shown in this plot provides a good fit to all cosmological observables at CMB scales and can be obtained either with 
the quartic polynomial potential or with its generalization that includes HDO. 
In the first case, the solution is characterized by a spectral index $n_s \simeq 0.955$ while in the second case we find $n_s \simeq 0.965$, the latter in full agreement with the central value indicated by Planck data.
}}
 }
\end{center}
\end{figure} 
In the left panel of fig.\,\ref{fig:Ka2} we consider the case in which the PBH mass distribution peaks at $M_{\rm PBH}\simeq 10^{13}$ g. 
We show the corresponding bounds  in terms of the quantity $\beta(M_{\rm PBH})$ which is related to the fraction of the Universe's mass in PBHs at their formation time
(see ref.\,\cite{Carr:2009jm} for details).
In the right panel of fig.\,\ref{fig:Ka2} we show the corresponding second-order gravitational wave signal. In agreement with
the scaling of eq.\,(\ref{eq:MasterScaling}),  the frequency of the peak is around
$f\simeq 30$ Hz. 
This signal can be an appealing target for the updated Advanced LIGO sensitivity, as shown in fig.\,\ref{fig:Ka2}, where we superimpose on the signal 
the bound obtained by combining the first and second observing runs\,\cite{LIGOScientific:2019vic} (region shaded in yellow with diagonal meshes) and the design sensitivity curve (region shaded in brown with dotted boundary)\,\cite{aLIGOdesign}. To make contact with the analysis in ref.\,\cite{LIGOScientific:2019vic}, we have used the explicit value $H_0 = 67.9$ km\,s$^{-1}$\,Mpc$^{-1}$ for the present day Hubble expansion rate.
To fully establish the relevance of the proposed signal, it is important --if not crucial-- to understand to what extent it can be distinguished from the expected astrophysical 
stochastic gravitational wave background from coalescing (astrophysical) binary black holes (BBH) and binary neutron stars (BNS).
A comprehensive analysis of this issue is left for future work. 
In the present note, we just compare the signal with the prediction of the astrophysical stochastic BBH+BNS background. 
The latter is shown in the right panel of fig.\,\ref{fig:Ka2} with a purple dashed line together with a gray band that represents the statistical Poisson uncertainty in the local binary merger rate\,\cite{LIGOScientific:2019vic,Abbott:2017xzg}.
As this graphical comparison suggests, we may expect to be able to detect these signals, because the stochastic background from light PBHs (if present) lies well above the astrophysical one.

%Let us anticipate our results in fig.\,\ref{fig:Ka}.
 
%%%%%%%%%%%%%%%%%%%%%%%%%%%%%%%%%%%%%%%%%%%%%%%%%%%%%%%%%%%%%%%%%%%%
\section{Analysis}\label{sec:Ana}
%%%%%%%%%%%%%%%%%%%%%%%%%%%%%%%%%%%%%%%%%%%%%%%%%%%%%%%%%%%%%%%%%%%%

Our analysis follows the standard literature. We shall, therefore, write down explicitly only those equations that are necessary to understand our line of reasoning. 
We set the reduced Planck mass to 1 ($M_{\rm Pl}=1$).
Consider the inflaton $\phi$ with a quartic polynomial potential $V(\phi) = a_2\phi^2 + a_3\phi^3 + a_4\phi^4$. 
In full generality, we cannot avoid the presence of a non-minimal coupling to gravity of the form $ \xi  \sqrt{-g}\phi^2R$, where $R$ is the Ricci scalar.\footnote{A non-minimal coupling to gravity of the form $\phi R$ 
which is linear in the field $\phi$ can be eliminated by a field redefinition at the prize of 
redefining the Newton's constant in the Jordan frame.}
This term is in fact required to exist for an interacting scalar field in curved space.
 In the Einstein frame, the potential is
\begin{align}\label{eq:MasterPot}
\tilde{V}(\phi) = \frac{1}{(1+\xi \phi^2)^2}\left(a_2\phi^2 + a_3\phi^3 + a_4\phi^4\right)\,,
\end{align}
where  the field $\phi$ is not canonically normalized.
The potential $\tilde{V}(\phi)$ has the right properties to provide a working inflationary model
 since it has a minimum at $\phi=0$ where inflation ends, and it flattens, thanks to the presence of the non-minimal coupling, at large field values.
 We look for viable inflationary solutions in the special case in which the values of the parameters $a_i$ and $\xi$ are such that an approximate  
 stationary inflection point is present a few $e$-folds before the end of inflation at the field value $\phi = \phi_0$ (a stationary inflection point is defined by the two conditions 
 $\tilde{V}^{\prime}(\phi_0) = \tilde{V}^{\prime\prime}(\phi_0)= 0$). We refer to appendix\,\ref{app:A} for the exact definition of the parametrization 
 used in our numerical analysis.

\begin{figure}[t]
\begin{center}
$$\includegraphics[width=.44\textwidth]{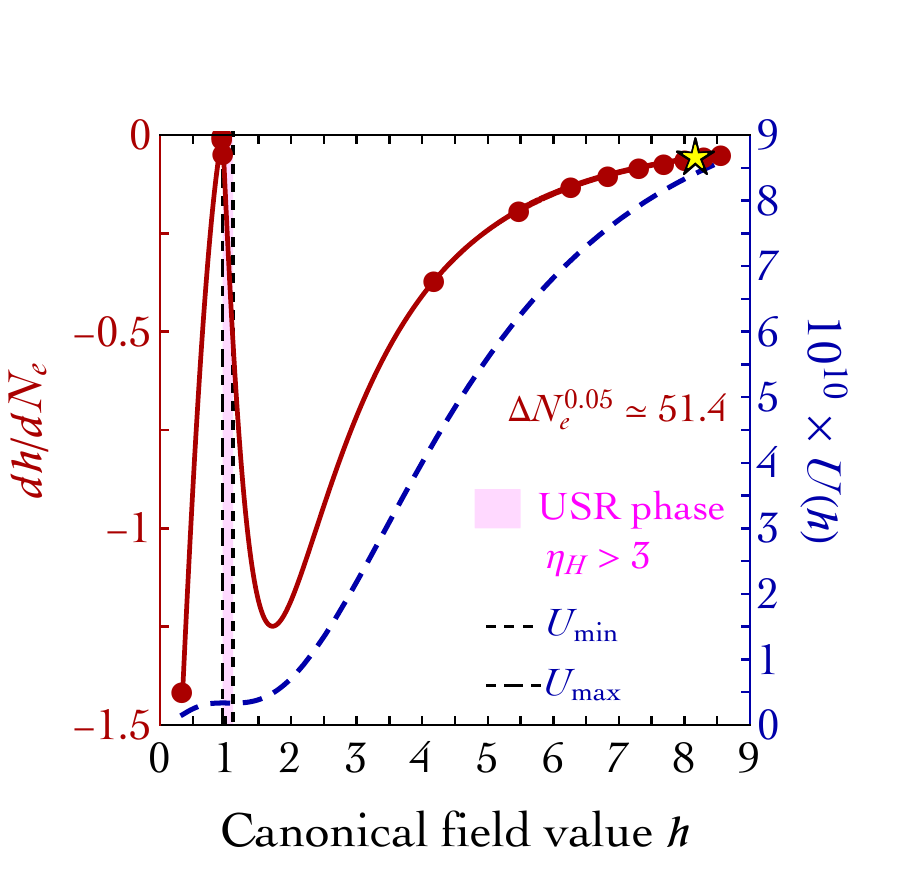}
\qquad\includegraphics[width=.4475\textwidth]{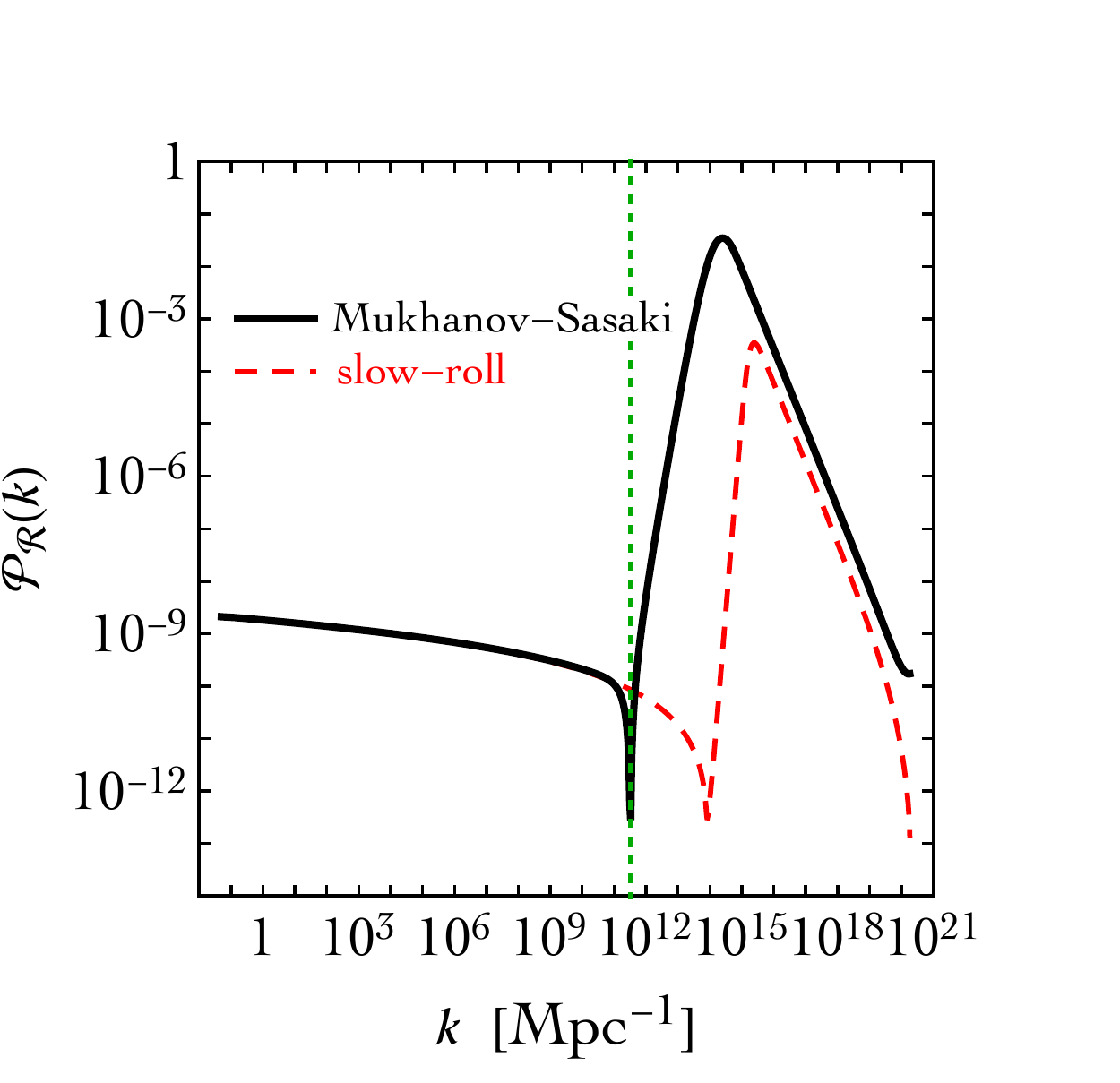}$$
\caption{\em \label{fig:StarPlot} 
{\color{VioletRed4}{
Left panel: Inflaton velocity (solid red, units given on the left-side $y$-axis) as a function of the (canonically normalized) 
inflaton field value (bottom $x$-axis). Physical potential (dashed blue, units given on the right-side $y$-axis) 
as a function of the (canonically normalized) 
inflaton field value. Right panel: Power spectrum of comoving curvature perturbations as a function of the comoving scale $k$. We show the result obtained by solving numerically the Mukhanov-Sasaki equation (solid black) and the approximation given by eq.\,(\ref{eq:PSgauge}) (dashed red).
}}
 }
\end{center}
\end{figure}

As it  is customary,  the kinetic term of $\phi$ can be canonically normalized by means of the field redefinition 
(using the boundary condition $h(\phi = 0) = 0$) 
\begin{align}\label{eq:fieldred}
\frac{dh}{d\phi} = \frac{\sqrt{1+\xi\phi^2(1 + 6\xi)}}{1+\xi\phi^2}\,\Longrightarrow\,
 h = \sqrt{\frac{1+6\xi}{\xi}}\sinh^{-1}\left[
 \phi\sqrt{\xi(1+6\xi)}
 \right] - \sqrt{6}\tanh^{-1}\left[
 \frac{\sqrt{6}\xi \phi}{\sqrt{1 + \xi(1+6\xi)\phi^2}}
 \right]\,,
\end{align}
and we indicate with $U(h) \equiv \tilde{V}[\phi(h)]$ the physical potential function of the canonically normalized field $h$.
The dynamics of $h$ can be obtained solving the equation of motion \cite{Ballesteros:2014yva}
\begin{align}\label{eq:EoM}
\frac{d^2h}{dN_e^2} + 3\frac{dh}{dN_e} - \frac{1}{2}\left(\frac{dh}{dN_e}\right)^3 
+\left[
3-\frac{1}{2}\left(\frac{dh}{dN_e}\right)^2
\right]\frac{d\log U}{dh}=0\,,
\end{align}
with slow-roll initial conditions. $N_e$ indicates the number of $e$-folds defined by $dN_e= Hdt$, where 
$H \equiv \dot{a}/a$ is  the Hubble rate, $a$ the scale factor of the Friedmann-Lema\^{\i}tre-Robertson-Walker metric, 
and $t$ the cosmic time (with $\dot{}\equiv~d/dt$).
In the left panel of fig.\,\ref{fig:StarPlot} we show the inflationary dynamics corresponding to 
the solution in fig.\,\ref{fig:Ka}, computed for the following parameters (see eq. (\ref{eq:MasterPot2}) for the parameterization of the potential)
\begin{equation}
\label{eq:Example1}
c_2=0.011,\quad c_3=0.0089,\quad \phi_0=1,\quad\mathrm{and}\quad \tilde{a}_{n\geq5}=0\,,
\end{equation}
while the values of $\lambda$ and $\xi$ are tuned in order to obtain, respectively,  the correct normalization of the power spectrum $A_s$ at CMB scales and the maximum amount of PBH abundance compatible with observations. The results of a scan over different values of $c_{2,3}$, with $\lambda$ and $\xi$ tuned accordingly, as discussed before, will be presented in section\,\ref{sec:Disc} but the results discussed in the rest of this section are of general validity. Typical values of $\lambda$ and $\xi$ for the solutions that we find
 are of order $\lambda\sim O(10^{-9})$ and $\xi \sim O(0.1)$.
The dashed blue line in  fig.\,\ref{fig:Ka} (which refers to the right-side $y$-axis) represents the physical potential $U(h)$ as a function of the 
canonically normalized field $h$. 
Inflation starts at large field values and ends at the absolute minimum 
of the potential (at $h=0$).
Before the end of inflation, the potential features the presence of an approximate stationary 
inflection point with a local minimum, marked by the vertical black dotted line close to $h=1$, and a subsequent local maximum, marked by the black vertical dot-dashed line (we refer to appendix\,\ref{app:A} for a better visualization of the potential close to the approximate stationary inflection point). 
The solid red line  (which refers to the left-side $y$-axis) represents the inflaton velocity $dh/dN_e$ as a function 
of the 
canonically normalized field $h$. Along the inflationary trajectory, the red dots count the number of $e$-folds (in steps of $5$).
It is instructive to follow the amount of inflation by 
tracking the comoving wavelengths $\lambda$ that --one after each other starting from the largest one (that is from small to large comoving wavenumber $k = 2\pi/\lambda$)-- exit the 
conformal Hubble horizon $1/(aH)$ as inflation proceeds.\footnote{It is common to use loosely the term "horizon" to refer to the boundary at a Hubble length $1/(aH)$. We will follow this practice here.}
The yellow star marks the number of $e$-folds
corresponding to the Hubble exit of the pivot scale $k_* = 0.05$ Mpc$^{-1}$ where we fit the CMB observables.
 From this time to the end of inflation, we count $\Delta N_e^{0.05} \simeq 51.4$ $e$-folds, where $\Delta N_e^{k}$ is the number of $e$-folds 
 between the time at which the scale $k$ (in Mpc$^{-1}$) exits the Hubble horizon and the end of inflation. 
 This is enough to solve the horizon and flatness problems of our observable Universe. 
We recall that at least from 50 to 60 $e$-folds of inflation are needed
 between the time that the largest observable scale, $k \sim 0.001$ Mpc$^{-1}$ exited the Hubble horizon and the time at which inflation ended. 
In the numerical solution explored in this section, we count $\Delta N_e^{0.001} \simeq 55.4$ $e$-folds.
When the inflaton draws near the approximate stationary inflection point, 
 its velocity suddenly decreases almost to zero. 
 The inflaton almost stops close to
  the approximate stationary inflection point of the potential but it has just enough inertia to overcome the barrier.
 This part of the dynamics is called ultra-slow roll (USR) phase  \cite{Kinney:2005vj}. It corresponds to the vertical region shaded in pink, which lasts for approximately $\Delta N_{\rm USR}\simeq 2.45$ $e$-folds.
 The USR phase is formally defined by the condition $\eta_H > 3$ on the Hubble parameter $\eta_H$ defined by
 \begin{align}\label{eq:HubbleEps}
\epsilon_H \equiv - \frac{\dot{H}}{H^2}  = \frac{1}{2}\left(\frac{dh}{dN_e}\right)^2\,,~~~~\eta_H \equiv -\frac{\ddot{H}}{2H\dot{H}} = \epsilon_H - \frac{1}{2}\frac{d\log \epsilon_H}{dN_e}\,.
\end{align}
It is also possible to introduce the so-called Hubble-flow parameters defined (for $i \geqslant 1$) by $\epsilon_{i} \equiv \dot{\epsilon}_{i-1}/(H\epsilon_{i-1})$, with the 
first parameter of the series given by $\epsilon_0 \equiv 1/H$. In this case we have
 $\epsilon_1 = \epsilon_H$ and $\epsilon_2 = 2\epsilon_H - 2\eta_H$, and, alternatively,  the USR phase can be identified with the region where $\epsilon_2<-6$.
\begin{figure}[t]
\begin{center}
$$\includegraphics[width=.65\textwidth]{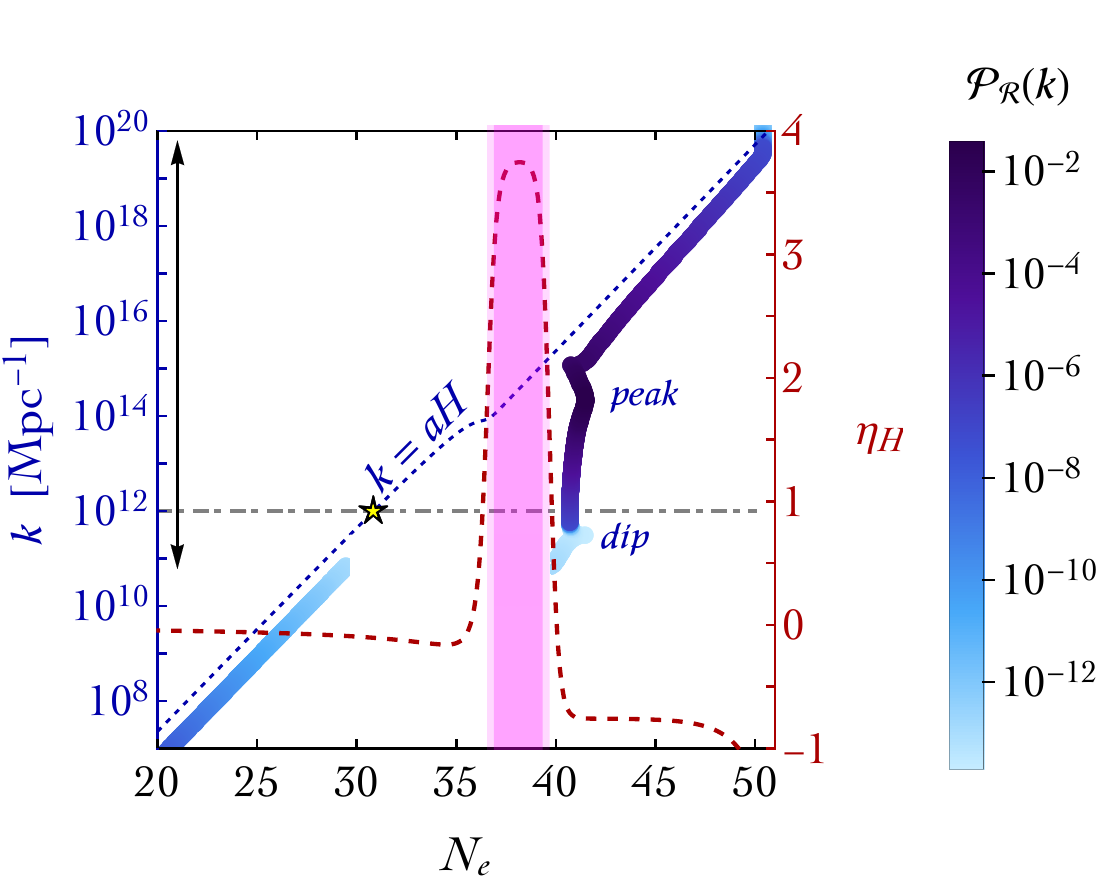}$$
\caption{\em \label{fig:keFold} 
{\color{VioletRed4}{
For each comoving wavenumber $k$ we plot the  value of $N_e$ (with $N_e = 0$ the $e$-fold time where we fit CMB observables) after which the corresponding mode 
 freezes to the final constant value that it maintains up to its horizon re-entry after the end of inflation (left-side $y$-axis, band shaded in blue). 
 The width of the blue band depends on the exact definition of the $e$-fold time at which the transition takes place 
 (for each $k$, the two extrema --from the left- to the right-end of the band-- correspond to the $e$-fold time after which the mode stays constant at the $0.01\%$ or $0.001\%$ level). 
 The different tonalities of blue follow the value of the power spectrum at each comoving wavenumber on the $y$-axis, according to the legend on the right. 
The $y$-coordinate of the label ``peak'' (``dip'') corresponds to the peak (dip) position of\, $\mathcal{P_R}$ at $k_{\rm peak} \simeq 2.5\times 10^{14}$ Mpc$^{-1}$ 
 ($k_{\rm dip} \simeq 3.2\times 10^{11}$ Mpc$^{-1}$).
 For comparison, we also show (dotted blue line) the value of $N_e$ obtained by means of the conventional horizon-crossing relation $k = aH$. 
 We superimpose  (right-side $y$-axis, dashed red line) the value of the Hubble parameter $\eta_H$ (see eq.\,(\ref{eq:HubbleEps})) and we highlight with a vertical pink band the region where the USR phase (defined by $\eta_H > 3$) takes place. The band shaded with a lighter tone of pink (slightly wider than the USR phase) is the region where the friction term in eq.\,(\ref{eq:Revo}) is negative: $2\eta_H - \epsilon_H > 3$ (roughly equivalent to $\eta_H > 3/2$).
 The vertical double-headed arrow indicates the interval of comoving wavenumbers over which we restrict the computation of the power spectrum in the left panel of fig.\,\ref{fig:GrowingModes}. 
 The yellow star marks the  horizon-crossing time given by the condition $k=aH$ for the mode with $k= 10^{12}$ Mpc$^{-1}$ (horizontal dot-dashed gray line) whose evolution is shown 
 in the left panel of fig.\,\ref{fig:FullModes}.
}}
 }
\end{center}
\end{figure} 

The presence of the USR phase boosts significantly the power spectrum of comoving curvature perturbations, $\mathcal{P}_{\mathcal{R}}(k)$. 
At the linear order in the Hubble parameters (in  the slow-roll approximation), we have 
  \begin{align}\label{eq:PSgauge}
  \mathcal{P}_{\mathcal{R}}(k) =  
  \frac{H^2}{8\pi^2 \epsilon_H}\left(\frac{k}{aH}\right)^{-4\epsilon_H + 2\eta_H} = A_s\left(\frac{k}{aH}\right)^{n_s - 1}\,,
  \end{align}
  with spectral index $n_s = 1-4\epsilon_H +2\eta_H$ and amplitude $A_s = H^2/8\pi^2 \epsilon_H$. 
 Although these expressions are  only approximate, they indicate that when the inflaton almost stops at the approximate stationary inflection point (where it accumulates almost 5 $e$-folds and we have $\epsilon_H \propto (dh/dN_e)^2 \simeq 0$) a sudden increase of $\mathcal{P}_{\mathcal{R}}(k) \propto 1/\epsilon_H$ occurs.
  This happens for comoving scales of order $k\sim 10^{14}$ Mpc$^{-1}$.
 As  it is well known, eq.\,(\ref{eq:PSgauge}) is not enough to fully describe the power spectrum at small scales in the presence of an USR phase\,\cite{Chongchitnan:2006wx,Motohashi:2017kbs,Germani:2017bcs,Ballesteros:2017fsr}. 
 To account  accurately for the dynamics, we solve numerically, for each mode $k$, the Mukhanov-Sasaki equation in Fourier space 
 \begin{align}\label{eq:Fulluk}
 \frac{d^2u_{\mathbf{k}}}{dN_e^2} + (1-\epsilon_H)\frac{du_{\mathbf{k}}}{dN_e} + 
 \left[
 \frac{k^2}{a^2 H^2} + (1+\epsilon_H - \eta_H)(\eta_H -2) - \frac{d}{dN_e}(\epsilon_H - \eta_H)
 \right]u_{\mathbf{k}} = 0\,,
  \end{align} 
  where $u_{\mathbf{k}}$ is related to the gauge-invariant comoving curvature perturbation $\mathcal{R}$ by means of the relation
   (in position space) $\mathcal{R} = -u/z$, with $z=a(dh/dN_e)$. We impose standard Bunch-Davies initial conditions. 
   The power spectrum of $\mathcal{R}$ is: 
    \begin{align}\label{eq:FullPS}
  \mathcal{P}_{\mathcal{R}}(k) = \frac{k^3}{2\pi^2}\left|\frac{u_{\mathbf{k}}}{z}\right|^2_{k\ll aH}\,,
  \end{align} 
  where for each mode $k$ the contribution to $\mathcal{P}_{\mathcal{R}}(k)$ is evaluated on super-Hubble scales $k\ll aH$ 
  where it remains frozen. We show the power spectrum corresponding to the numerical solution explored in this section 
  in the right panel of fig.\,\ref{fig:StarPlot}. The black line is the solution obtained by means of eq.\,(\ref{eq:FullPS}).
  At small $k$, the power spectrum is well-described by the slow-roll approximation in eq.\,(\ref{eq:PSgauge}) (dashed line). 
 At larger $k$, the  power  spectrum  has  a  pronounced dip (vertical dotted green line in fig.\,\ref{fig:StarPlot}) at scales that exit the Hubble horizon few $e$-folds before the onset of the USR phase. This sudden dip
 is followed by a rapid growth that can be well-described by a power law with scaling approximatively given by $\sim k^{3.49}$ (where the exponent is specific for the values of \eq{eq:Example1}).
The power spectrum peaks at scales $k\sim 10^{14}$ Mpc$^{-1}$ with a peak amplitude of order $A_s \sim 10^{-1}$.    As discussed in \cite{Ballesteros:2017fsr}, the difference in $\mathcal{P_R}(k)$ between the approximate and the Mukhanov-Sasaki solutions in this region is of a couple of orders of magnitude, which is very relevant for the PBH  abundance, as this is exponentially sensitive to $\mathcal{P_R}(k)$.
 
The relation between the USR phase and the growth of the power spectrum deserves an in-depth discussion, and a simple --although just approximated-- analytical understanding is possible. The equation of motion of the inflaton, eq.\,(\ref{eq:EoM}), can be recast in the form
\begin{align}\label{eq:USReq}
  \eta_H = 3 + (3-\epsilon_H) \frac{d\log U}{dh} \left(\frac{dh}{dN_e}\right)^{-1}\,.
\end{align}
In what follows, we will assume that $\epsilon_H<3$. From the previous equation we see that $\eta_H \leqslant 3$ are the possible values attainable  for a monotonically increasing potential, with $\eta_H = 3$ corresponding, for instance, to the presence of a stationary inflection point. Therefore, in the presence of an approximate stationary inflection point, the USR phase $\eta_H > 3$ is confined  to the small region in between the local minimum and the subsequent maximum of the potential (see the pink strip in the left panel of fig.\,\ref{fig:StarPlot}), where the second term on the right side of eq.\,(\ref{eq:USReq}) becomes positive and enhances $\eta_H.$ As far as the behavior of perturbations in eq.\,(\ref{eq:Fulluk}) is concerned, however, the condition $\eta_H > 3$ is not immediately illuminating. It is more instructive to look at the equation for the comoving curvature perturbation in Fourier space $\mathcal{R}_{\mathbf{k}}$ for the mode with comoving wavenumber $k$, which we write in the exact form
\begin{align}\label{eq:Revo}
  \frac{d^2\mathcal{R}_{\mathbf{k}}}{dN_e^2} + \left(
  3+\epsilon_H - 2\eta_H
  \right)\frac{d\mathcal{R}_{\mathbf{k}}}{dN_e} + \frac{k^2}{a^2 H^2} \mathcal{R}_{\mathbf{k}}=0\,.
\end{align} 
This is the differential equation of a damped harmonic oscillator. After horizon crossing ($k=aH$) during inflation, and for a standard positive friction term, the solution to eq.\,(\ref{eq:Revo}) freezes exponentially fast to a constant value. This is because in the limit $k^2/a^2 H^2 \ll 1$, in which we neglect the last term in eq.\,(\ref{eq:Revo}), the solution has first derivative given by $d\mathcal{R}_{\mathbf{k}}/dN_e\propto e^{-(3-2\eta_H)N_e}$ (assuming $\epsilon_H \ll |\eta_H|$ and the latter constant) with $N_e$ the number of $e$-folds  after horizon crossing. However, if the condition $2\eta_H - \epsilon_H > 3$ is met during the subsequent evolution of the inflaton field, a phase during which the friction term is negative (driving force) takes place. In such a situation, an exponential growth (or suppression) of the mode is possible. To better illustrate this point, let us  define $\Theta \equiv  3+\epsilon_H - 2\eta_H$ and  $\epsilon_k^2\equiv k^2/a^2H^2$. We indicate with $N_{\rm in}$ ($N_{\rm end}$) the $e$-fold time at which the negative friction phase starts (ends). For simplicity, we take $\Theta$ to be a negative constant in this range, and assume a constant $\epsilon_k^2$ equal, for instance, to the value it takes at $N_{\rm in}$. We also impose the boundary conditions $\mathcal{R}_{\mathbf{k}}(N_{\rm in}) = \mathcal{R}_{0}$ and $d\mathcal{R}_{\mathbf{k}}/dN_e(N_{\rm in}) = 0$, which simulate a mode freezing at some early time before the negative friction phase. These approximations are enough for a qualitative understanding of \eq{eq:Revo}.\footnote{A slightly better approximation can be obtained if we rewrite eq.\,(\ref{eq:Revo}) in the form
   \begin{align}\label{eq:BetterAna}
   \frac{d^2\mathcal{R}_{\mathbf{k}}}{dN_e^2} + \Theta\frac{d\mathcal{R}_{\mathbf{k}}}{dN_e} + 
   \bar{\epsilon}_k^2 e^{-2N_e}
    \mathcal{R}_{\mathbf{k}}=0\,,~~~~~~{\rm with}~~\bar{\epsilon}_k^2 \equiv \left(\frac{k}{k_*}\frac{H_*}{H_{\rm in}}\right)^2\,
       \end{align}
 with $k_* = 0.05$ Mpc$^{-1}$ the pivot scale at which $N_{\rm CMB} = 0$ where we fit CMB observables and $
 H_*\equiv H(N_{\rm CMB})$.   
 In this approximation, $H=H_{\rm in}$ is constant during the negative friction phase (equal to the value it takes at $N_{\rm in}$). 
 Eq.\,(\ref{eq:BetterAna}) admits the general solution
 \begin{align}\label{eq:BetterAna2}
 \tilde{\mathcal{R}}_{\mathbf{k}}(N_e) = \left(\frac{\bar{\epsilon}_k}{2}\right)^{\Theta/2}
 e^{-N_e\Theta/2}\left[
 c_1\Gamma(1-\Theta/2)J_{-\Theta/2}(\bar{\epsilon}_k e^{-N_e})
 +c_2\Gamma(1+\Theta/2)J_{\Theta/2}(\bar{\epsilon}_k e^{-N_e})
 \right]\,,
  \end{align}
  with $c_{1,2}$ integration constants, $J_{\alpha}(x)$ Bessel functions of the first kind and $\Gamma$ the Euler's gamma function. 
  However, considering this solution does not really add much to the qualitative results based on the simpler approximation in eq.\,(\ref{eq:CrucialCondition}), which we have thus preferred for the main discussion. 
  Furthermore, Bessel functions cannot be represented in general through elementary functions, and considering the solution 
  in eq.\,(\ref{eq:BetterAna2}) as analytical would be a bit like cheating. 
      }  
      Accordingly, we find the solution
  \begin{align}\label{eq:CrucialCondition}
\tilde{\mathcal{R}}_{\mathbf{k}}(N_e) = 
 \underbrace{\frac{\mathcal{R}_{0}e^{-\Theta(N_e-N_{\rm in})/2}}{\sqrt{\Theta^2 - 4\epsilon_k^2}}
 }_{e^{-\Theta N_e/2}{\rm\,growth\,if}\,\Theta<0}
 \bigg\{\underbrace{
 \sqrt{\Theta^2 - 4\epsilon_k^2}\cosh\left[
 \frac{(N_e-N_{\rm in})}{2}\sqrt{\Theta^2 - 4\epsilon_k^2}
 \right] + \Theta \sinh\left[
 \frac{(N_e-N_{\rm in})}{2}\sqrt{\Theta^2 - 4\epsilon_k^2}
 \right]}_{{\rm Non-oscillating\, superposition\,of\,}e^{\pm \sqrt{\Theta^2 - 4\epsilon_k^2}N_e/2}{\rm\,functions\,if}\,\Theta^2-4\epsilon_k^2 >0}
\bigg\}\,.
 \end{align} 
 To avoid confusion, let us remark that the real function $\tilde{\mathcal{R}}_{\mathbf{k}}$ defined by the solution above 
 must be considered as a proxy for either the real or the imaginary part of the actual $\mathcal{R}_{\mathbf{k}}$  (which is a complex variable). Even though the solution in eq.\,(\ref{eq:CrucialCondition}) is not fully representative of the actual numerical results, it is  possible to extract some interesting insights. We will now examine the qualitative behavior of this solution in four different regimes, $\epsilon_k^2\ll\Theta^2/4$, $\epsilon_k^2\lesssim\Theta^2/4$, $\epsilon_k^2\gtrsim\Theta^2/4$, and $\epsilon_k^2 \gg \Theta^2/4$. Note that in the last two cases we must consider a non-zero initial velocity. \footnote{More precisely, in these cases we solve eq.\,(\ref{eq:CrucialCondition}) with a non-zero initial condition for the velocity $d\mathcal{R}_{\mathbf{k}}/dN_e(N_{\rm in}) = \delta\mathcal{R} \neq 0$ so that we have
     \begin{align}
\tilde{\mathcal{R}}_{\mathbf{k}}(N_e) = 
 \frac{e^{-\Theta(N_e-N_{\rm in})/2}}{\sqrt{\Theta^2 - 4\epsilon_k^2}}
 \bigg\{\mathcal{R}_{0}
 \sqrt{\Theta^2 - 4\epsilon_k^2}\cosh\left[
 \frac{(N_e-N_{\rm in})}{2}\sqrt{\Theta^2 - 4\epsilon_k^2}
 \right] + (\mathcal{R}_{0}\Theta +2\delta\mathcal{R}) \sinh\left[
 \frac{(N_e-N_{\rm in})}{2}\sqrt{\Theta^2 - 4\epsilon_k^2}
 \right]
\bigg\}\,.
 \end{align}
   The precise value of $\delta\mathcal{R}$ is not relevant for the validity of our discussion.}
 
 \begin{figure}[t]
\begin{center}
$$\includegraphics[width=.44\textwidth]{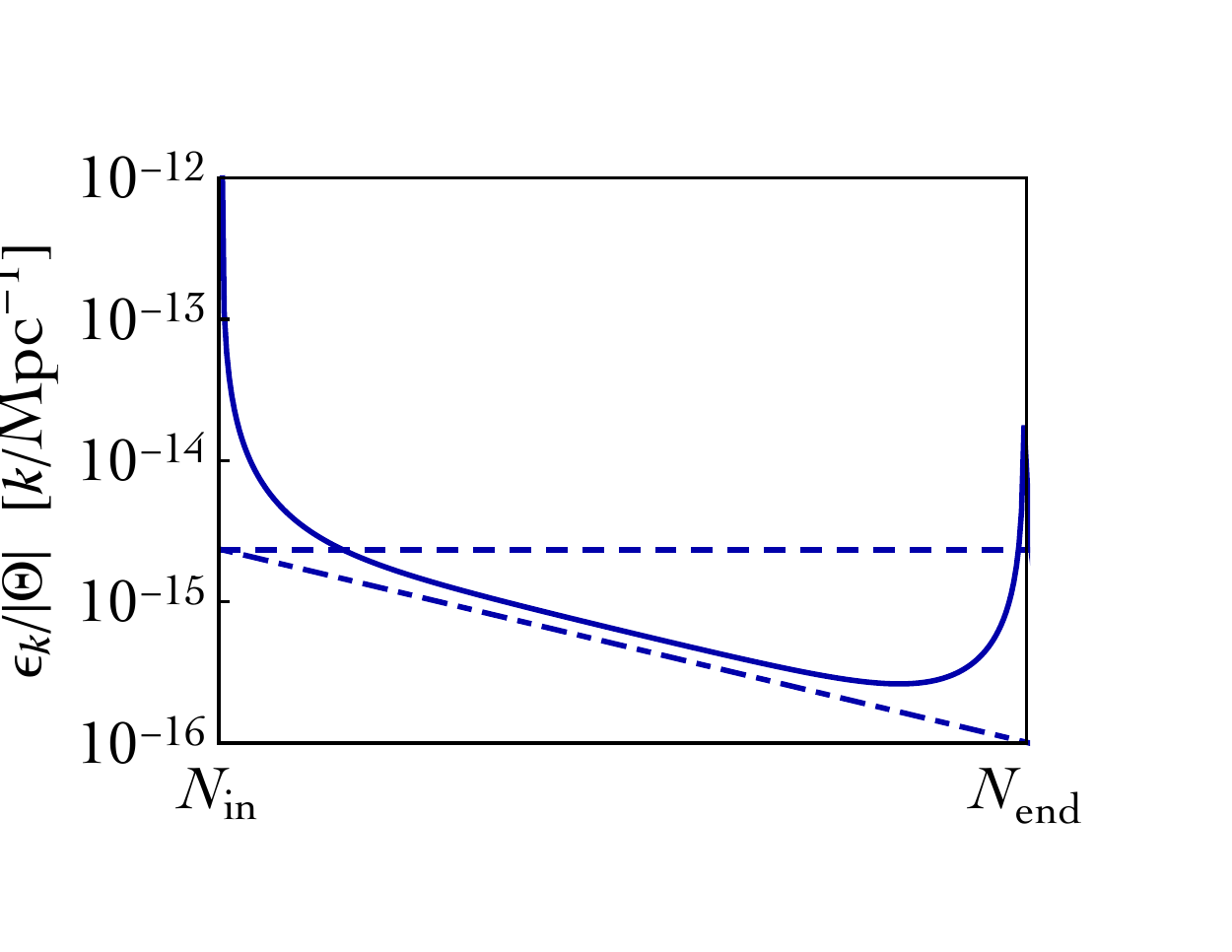}
\qquad\includegraphics[width=.4475\textwidth]{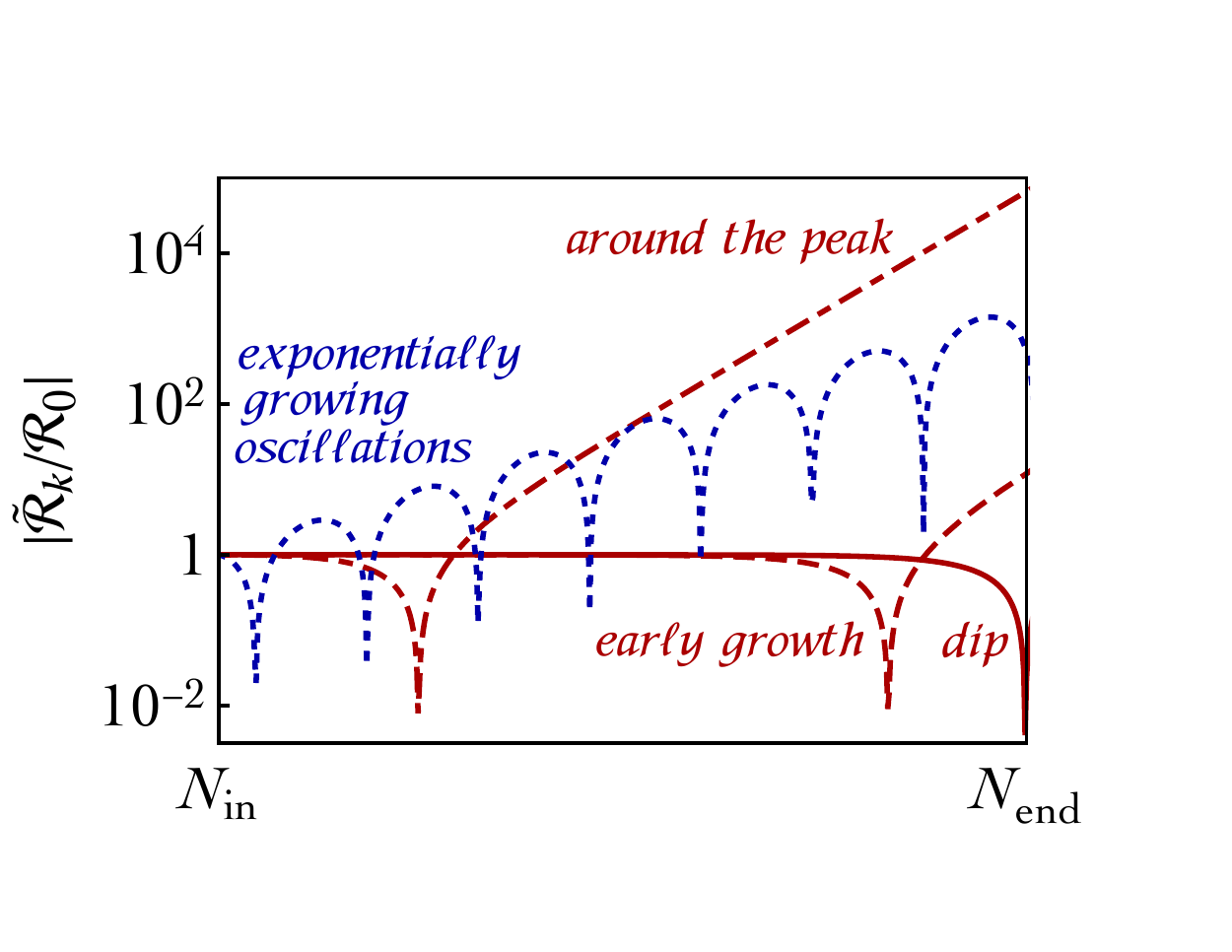}$$
\caption{\em \label{fig:AnalModes} 
{\color{VioletRed4}{
Left panel: We plot the quantity $\epsilon_k/|\Theta|$ (with $\Theta \equiv  3+\epsilon_H - 2\eta_H$ and  $\epsilon_k\equiv k/aH$) during the negative friction phase. In the explicit example discussed in section\,\ref{sec:Ana} we have $N_{\rm in}\simeq 36.5$ and $N_{\rm end}\simeq 39.7$.
 The solid line is the exact numerical result while the dashed line is the approximation obtained by taking 
 constant
$\Theta = -5$ and $\epsilon_k$ fixed at the value it takes at the beginning of the phase with negative friction. 
The dot-dashed line takes into account the redshift of $k/aH$ during the negative friction phase 
(assuming $H$ constant during the latter, see eq.\,(\ref{eq:BetterAna})) which explains the slope of the exact numerical result.
Right panel: We plot the absolute value of the solution in eq.\,(\ref{eq:CrucialCondition}) for three representative values of $\epsilon_k < 1$ (solid, dashed and dot-dashed red lines for, respectively, increasing values of $\epsilon_k < 1$) and for one representative value of $\epsilon_k > |\Theta|/2$ (dotted blue line).
}}
 }
\end{center}
\end{figure}
\begin{figure}[t]
\begin{center}
$$\includegraphics[width=.44\textwidth]{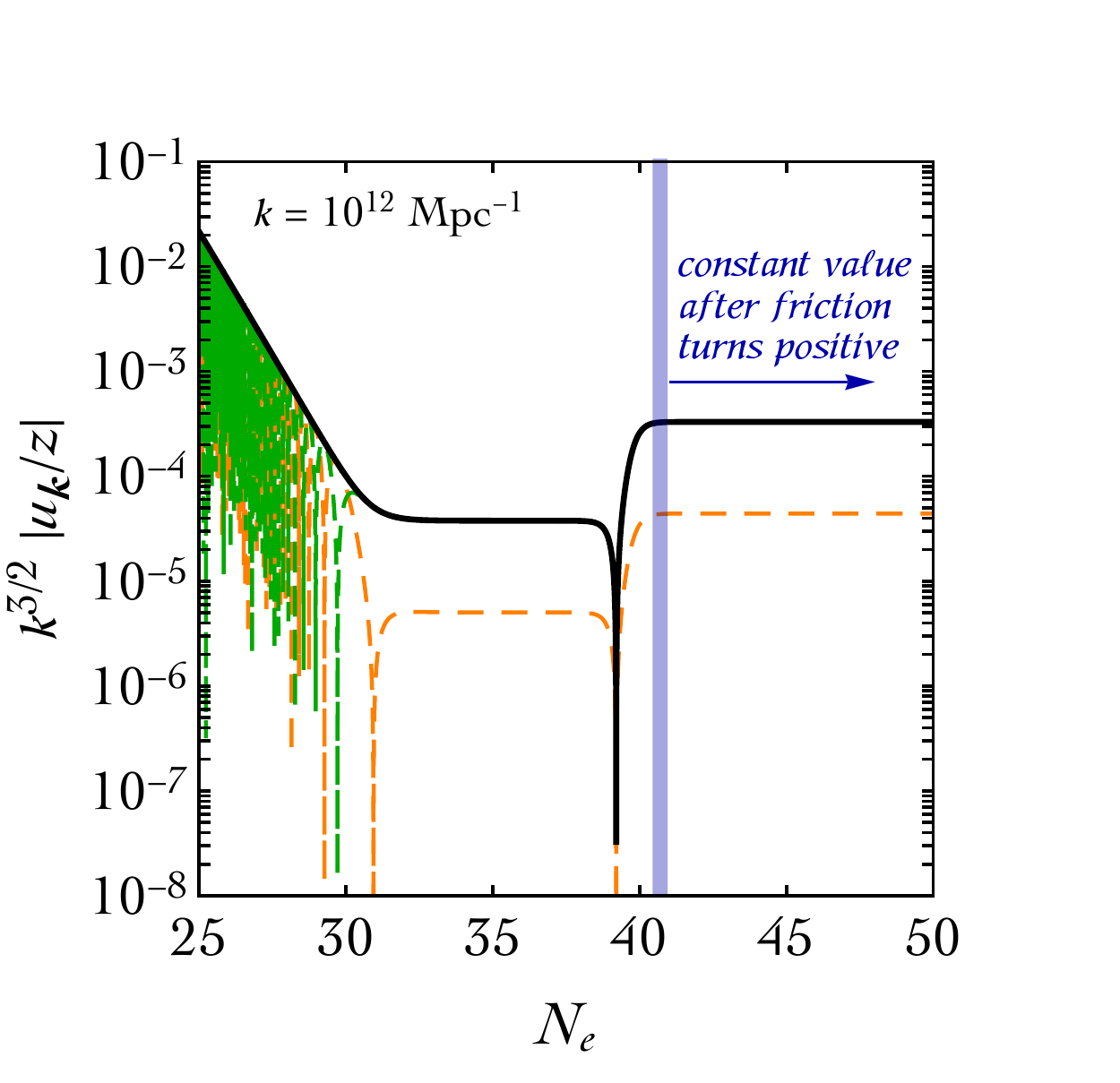}
\qquad\includegraphics[width=.4475\textwidth]{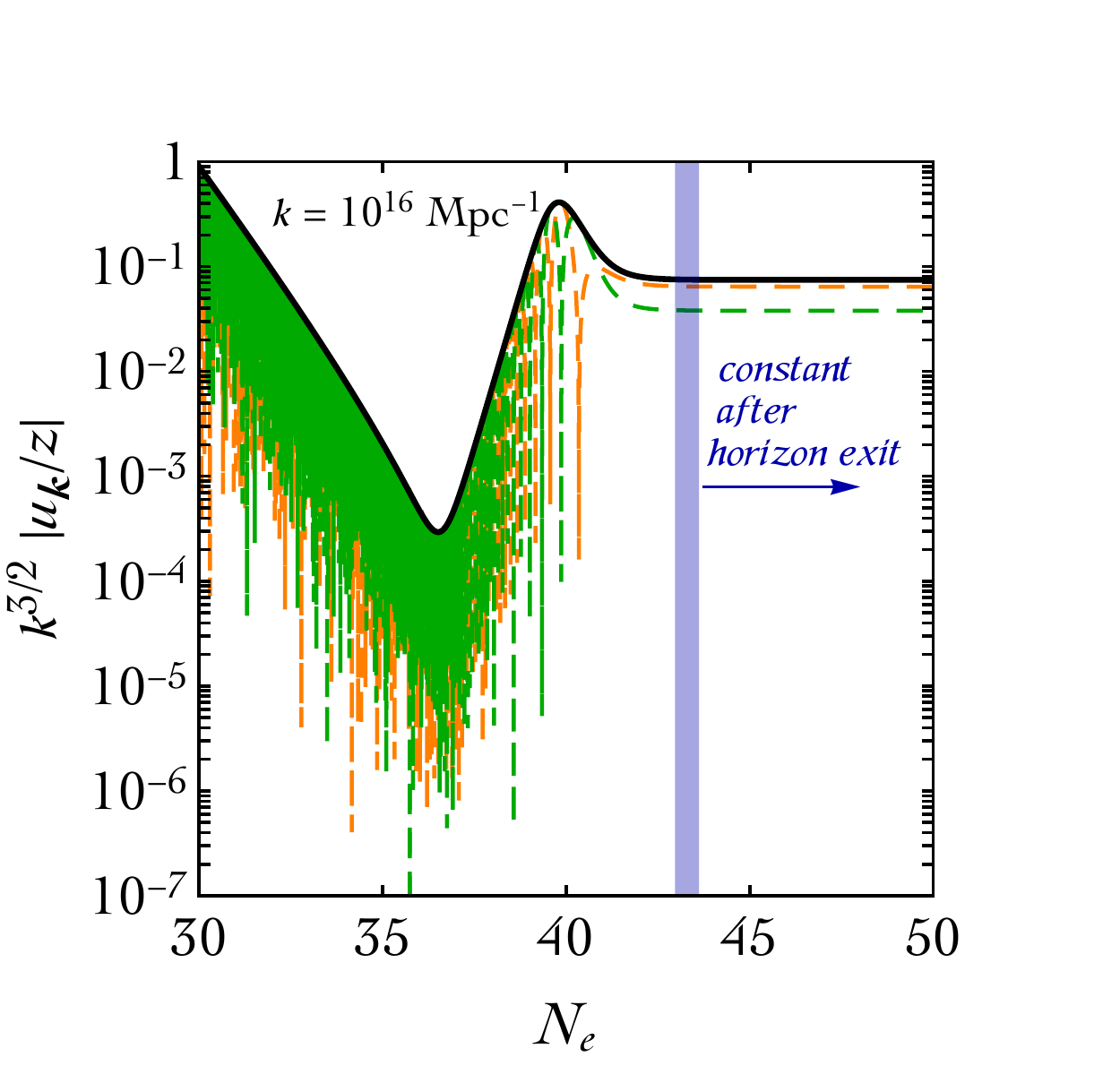}$$
\caption{\em \label{fig:FullModes} 
{\color{VioletRed4}{  
Left panel: Time evolution (with the number of $e$-folds as time variable), of the mode with benchmark value of comoving wavenumber $k = 10^{12}$ Mpc$^{-1}$. The dashed orange and green lines represent, respectively, 
the real and imaginary parts of the function plotted on the $y$-axis, while the solid black is its modulus.
The $e$-fold time after which the mode (modulus) freezes to its final constant value is marked by a vertical blue line. The width of the blue band depends on the exact definition of the $e$-fold time at which the final freezing of the mode takes place, and it corresponds to the width of the blue band plotted in figure \ref{fig:keFold}. Right panel: Same analysis, for the mode with $k = 10^{16}$ Mpc$^{-1}$.
}}}
\end{center}
\end{figure}
\begin{figure}[t]
\begin{center}
$$\includegraphics[width=.44\textwidth]{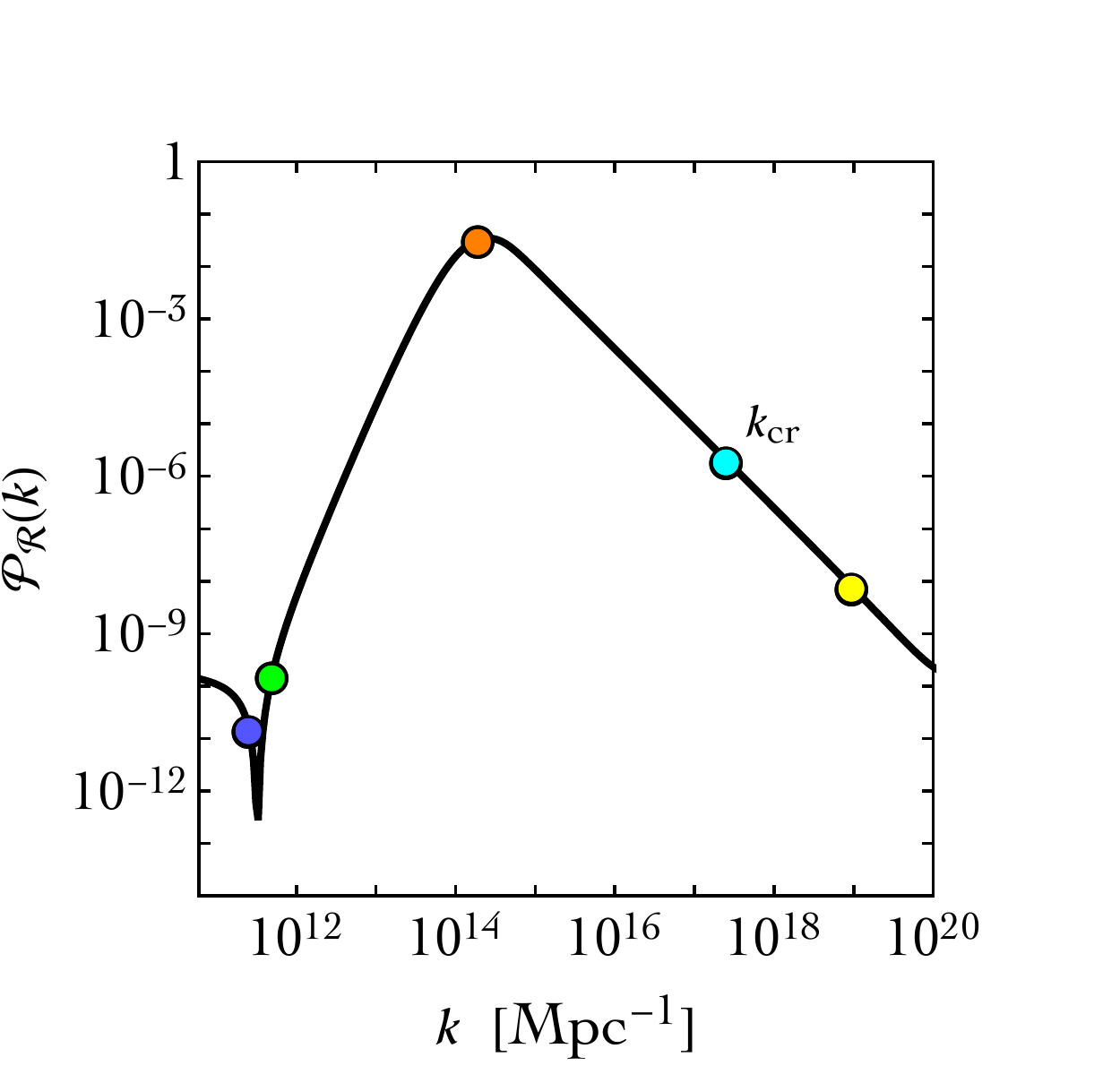}
\qquad\includegraphics[width=.4475\textwidth]{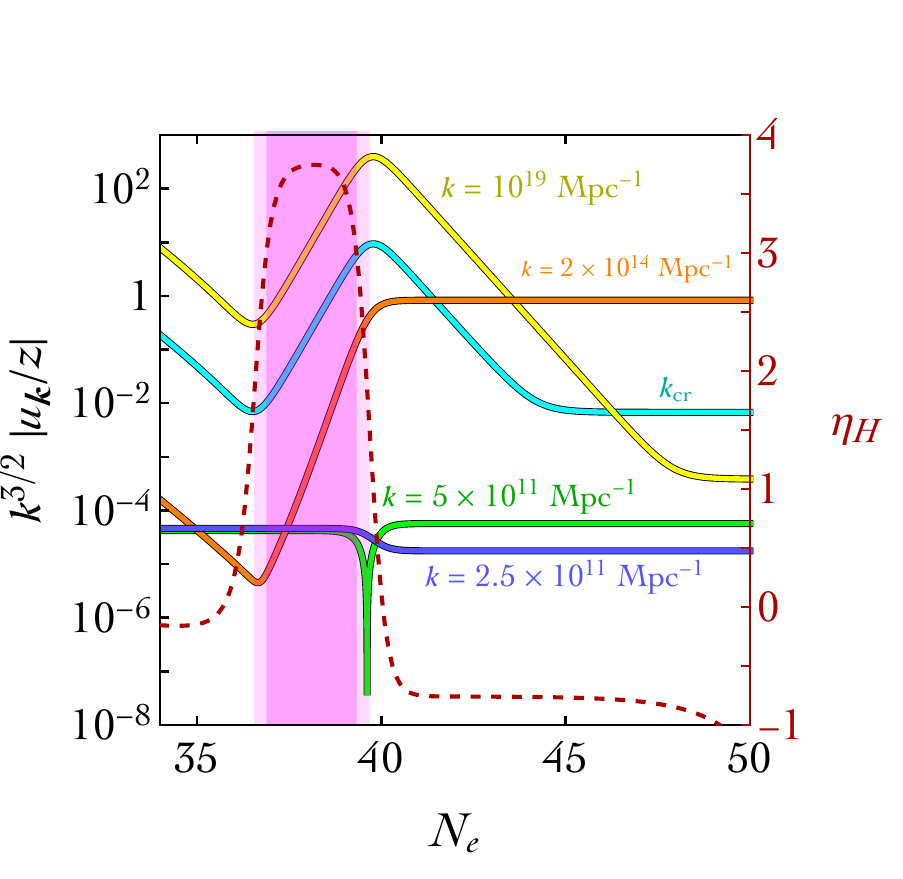}$$
\caption{\em \label{fig:GrowingModes} 
{\color{VioletRed4}{
Left panel (zoomed-in version of the power spectrum in the right panel in fig.\,\ref{fig:StarPlot}): Power spectrum for the modes that exit the horizon after the phase with 
negative friction (see eq.\,(\ref{eq:Revo}) and related discussion),
deviating from the conventional horizon-crossing relation $k=aH$ (valid in slow-roll approximation, see fig.\,\ref{fig:keFold}).
The modes marked by (for increasing values of $k$) the blue, green, orange, cyan and yellow dots are represented in the right panel (with the same color-code).
These modes are enhanced or suppressed in the proximity of the USR phase ($\eta_H > 3$, vertical band with a darker tonality of pink). 
The wider vertical band shaded with a lighter tonality of pink indicates the $e$-fold interval during which $2\eta_H -\epsilon_H > 3$ 
(see eq.\,(\ref{eq:Revo}) and related discussion). After this phase of negative friction,  the modes freeze exponentially fast to their final constant values. 
The mode in cyan has $k_{\rm cr} = 2.42\times 10^{17}$ Mpc$^{-1}$. 
In this case the exponential decay during the interval $N_{k=aH} - N_{\rm end} > N_{\rm end} - N_{\rm in}$ is such that it compensates the exponential growth 
accumulated during the negative friction phase. 
Notice that the exponential growth for the orange mode with $k = 2\times 10^{14}$ Mpc$^{-1}$ is faster than that of the cyan and yellow modes, with 
$k_{\rm cr}$ and $k = 10^{19}$ Mpc$^{-1}$ (see right panel of fig.\,\ref{fig:AnalModes} and the related analytical discussion).
}}
 }
\end{center}
\end{figure}

\begin{itemize}
\item [$\circ$] For modes with $\epsilon_k^2 \ll \Theta^2/4$, we have the approximation    
\begin{align}\label{eq:CrucialConditionApp}
\tilde{\mathcal{R}}_{\mathbf{k}} \approx \mathcal{R}_{0}\left[
   1 - \frac{\epsilon_k^2}{\Theta^2}e^{-\Theta(N_e-N_{\rm in})}
   \right]\,,
\end{align}
meaning that modes that exit the horizon early enough are not affected by the negative friction phase and remain frozen to their original values
 --this happens for modes with $k \lesssim 6\times 10^{10}$ Mpc$^{-1}$ in fig.\,\ref{fig:keFold}--   unless the duration of the phase with negative $\Theta$ is enough to compensate the suppression  given by $\epsilon_k^2/\Theta^2 \ll 1/4$. 
For this to occur, one needs $(N_e-N_{\rm in}) \gtrsim (-1/\Theta)\log\Theta^2/\epsilon_k^2$. Typical values of $\epsilon_k/|\Theta|$ are shown in the left panel of fig.\,\ref{fig:AnalModes} for the explicit solution studied in this section. If we consider, for illustrative purposes,  $\Theta = -5$ (meaning that $\eta_H = 4 \gg \epsilon_H\simeq 0$) and $k = 10^{11}$ Mpc$^{-1}$ (that is $\epsilon_k/|\Theta|\simeq 10^{-4}$) we find that we need at least $(N_e-N_{\rm in}) \gtrsim 3$ to sizably affect this mode. In the particular example that we are using for illustration in this section, the duration of the negative friction phase is approximately $N_{\rm end} - N_{\rm in}\simeq 3$, and indeed only the scales $k \gtrsim 10^{11}$ Mpc$^{-1}$ are altered by the negative friction phase (see fig.\,\ref{fig:keFold}), in agreement with the analytical argument.

\item [$\circ$] For modes with larger values of $k$, which satisfy $\epsilon_k^2 \lesssim \Theta^2/4$, the solution of eq.\,(\ref{eq:CrucialCondition}) with $\Theta$ negative depends on $N_e$ through a superposition of exponential functions, and $|\tilde{\mathcal{R}}_{\mathbf{k}}|$ can be either enhanced or suppressed, as we will now discuss. In the right panel of fig.\,\ref{fig:AnalModes} we show the behavior of $|\tilde{\mathcal{R}}_{\mathbf{k}}|$ in eq.\,(\ref{eq:CrucialCondition}) during the negative friction phase for three different values of $\epsilon_k$. The solution with smaller $\epsilon_k$ (solid line) is supressed towards $N_{\rm end}$ and it is, therefore, a proxy for the typical mode contributing to the dip of the power spectrum. On the other hand, for larger values of $\epsilon_k$ (dashed and dot-dashed lines) the corresponding modes are exponentially enhanced, and contribute to the growth  of the power spectrum. 
Notice that the solution in eq.\,(\ref{eq:CrucialCondition}) vanishes at 
$N_0 = N_ {\rm in} + 2\,{\rm arccosh}(-\Theta/2\epsilon_k)/\sqrt{\Theta^2-4\epsilon_k^2}$ and becomes negative for $N_e > N_0$.
This is because under the assumption $\Theta^2 - 4\epsilon_k^2 > 0$ the arguments of the exponential functions in 
eq.\,(\ref{eq:CrucialCondition}) are always positive, but $\tilde{\mathcal{R}}_{\mathbf{k}}$ is driven towards negative values because of the sinh term, which is proportional to $\Theta<0$.
However, since it is the modulus $|\tilde{\mathcal{R}}_{\mathbf{k}}|$ that matters for the primordial power spectrum, the exponential suppression towards negative values of $\tilde{\mathcal{R}}_{\mathbf{k}}$ results in an enhancement of $|\tilde{\mathcal{R}}_{\mathbf{k}}|$ for $N_e > N_0$.
  As can be seen in the right panel of 
fig.\,\ref{fig:AnalModes}, $N_0 < N_{\rm end}$ for the modes contributing to the growth of the power spectrum, whereas the mode contributing to the dip is attracted towards zero --which explains the suppression-- but has $N_0 > N_{\rm end}$. 
Despite the simplicity of our approximation, this feature is present also in the full numerical solutions (left panel of fig.\,\ref{fig:FullModes} and fig.\,\ref{fig:GrowingModes} that we shall discuss later). 
In particular, we remark that also in the numerical solutions the enhancement in $|\tilde{\mathcal{R}}_{\mathbf{k}}|$ actually comes from 
an exponential suppression towards negative values of its real or imaginary part.
  At the end of the phase with negative friction, $\Theta$ becomes positive again and the modes,
  after being enhanced or suppressed, freeze a second time to reach the final constant value that is maintained until horizon re-entry after the end of inflation (see the left panel of fig.\,\ref{fig:FullModes} for a representative example).

\item [$\circ$]  Let us now consider the case with $\epsilon_k^2 \gtrsim \Theta^2/4$. This can be thought of as a proxy for the case in which the horizon crossing happens after the negative friction phase. The term in curly brackets in eq.\,(\ref{eq:CrucialCondition}) now consists of a combination of oscillating functions with exponentially growing amplitude controlled by the overall coefficient in front. We show one representative solution of this kind in the right panel of fig.\,\ref{fig:FullModes}. The negative friction phase triggers an exponential growth also in this case, but the enhancement occurs more slowly than for $\epsilon_k^2 \lesssim \Theta^2/4$. This can be seen from eq.\,(\ref{eq:CrucialCondition}), since for $\epsilon_k^2 \gtrsim \Theta^2/4$  the quantity $\tilde{\mathcal{R}}_{\mathbf{k}}$ grows as $\propto e^{-\Theta(N_e-N_{\rm in})/2}$, whereas for $\epsilon_k^2 \lesssim \Theta^2/4$ the additional positive contributions coming from the $\cosh$ and $\sinh$ functions make the solution grow much faster. Indeed, in the limit $\epsilon_k^2\ll\Theta^2/4$, $\tilde{\mathcal{R}}_{\mathbf{k}}$ grows as $\propto e^{-\Theta(N_e-N_{\rm in})},$ see eq.\,(\ref{eq:CrucialConditionApp}). Interestingly, this behavior can be noticed also in the full numerical solutions, as we shall see later in the right panel of fig.\,\ref{fig:GrowingModes}. The modes for which $\epsilon_k^2 \gtrsim \Theta^2/4$ are the ones with $k \gtrsim 2\times 10^{15}$ Mpc$^{-1}$ in fig.\,\ref{fig:keFold}.
  In this case the modes are still exponentially enhanced during the negative friction phase but after its end --being still sub-Hubble-- they keep oscillating until 
  their horizon exit with an amplitude that is now exponentially suppressed because $\Theta$ turns positive. This is well illustrated by the mode with $k = 10^{16}$ Mpc$^{-1}$ shown in the right panel of fig.\,\ref{fig:FullModes} where $|\tilde{\mathcal{R}}_{\mathbf{k}}|$ grows exponentially during the negative friction phase and then decays until it freezes to a constant value after horizon exit takes place. 
If the modes freeze right after the end of the negative friction phase  (i.e.\ if $\epsilon_k$ is not too large)
we have a final net  enhancement, since there is not enough time before horizon crossing to completely erase the growth gained during this phase.  
This is precisely what happens in the case shown in the right panel of fig.\,\ref{fig:FullModes}, and  modes like this one contribute to the right-side of the peak in the power spectrum.

\item [$\circ$]  For $\epsilon_k^2 \gg \Theta^2/4$, the transition between the end of the negative friction phase and the horizon crossing will be, eventually, long enough to completely erase the effect of the exponential enhancement.
We estimate that this occurs if  $N_{k = aH} - N_{\rm end}  \gtrsim (N^{\rm cr}_{k = aH} - N_{\rm end}) \equiv (N_{\rm end} - N_{\rm in})(-\Theta)/\Theta_{>}$, where 
$N_{k = aH}$ denotes the $e$-fold time at which the horizon crossing for the mode $k$ takes place, and 
$\Theta_{>}$ is the positive value of $\Theta$ after the negative friction phase.
Since we have in general $\Theta_> < |\Theta|$, as can be deduced if one looks at the values of $\eta_H$, the exponential decay for $N_e \in [N_{\rm end},N_{k = aH}]$ will be slower than the preceding growth. 
In terms of their comoving wavenumber, the modes for which the effect of the negative friction phase is erased by the subsequent decay have a comoving scale
$k \gtrsim k_{\rm cr} \simeq k_*(H^{\rm cr}_{k=aH}/H_*)e^{N^{\rm cr}_{k=aH}}$ with $H^{\rm cr}_{k=aH}$ being the value of the Hubble rate at $N^{\rm cr}_{k = aH}$ and $k_{\rm cr}=a_{\rm cr} H^{\rm cr}_{k=aH}$. Of course, it would be incorrect to conclude that modes with $k \gtrsim k_{\rm cr}$ are not affected by the negative friction phase. The reason is that these modes need  to spend $N_{k=aH} - N_{\rm in}$ $e$-folds to come back to the value they had at $N_{\rm in}$, an interval of time during which --if the negative friction phase were absent-- they would have continued decreasing down to a value at horizon crossing much smaller than  the one they actually reach in the presence of the negative friction phase. 
This means that also for these modes the power spectrum is enhanced if compared to that of standard slow-roll dynamics.
We give an illustrative example of this kind of modes in the right panel of fig.\,\ref{fig:GrowingModes}.

\end{itemize}

  We will discuss a different analytical approximation that captures the relevant features of the power spectrum in appendix\,\ref{app:C}.
  With the previous discussion, the most relevant point that we aim to stress is that the ``negative friction condition'' $2\eta_H - \epsilon_H > 3$ (which is approximately $\eta_H > 3/2$) determines, more precisely than $\eta_H > 3$, the phase during which  
  perturbations can be enhanced or suppressed. Let us now further corroborate this fact.
 In the left panel of fig.\,\ref{fig:GrowingModes} we show the part of the power spectrum composed by modes with comoving wavenumbers that are affected by the negative friction phase. 
Indeed, the latter is responsible for the dip, the peak and the subsequent decrease of the power spectrum. 
 In the right panel of fig.\,\ref{fig:GrowingModes}, we consider explicitly the evolution of five modes. The first of these modes is responsible for the dip of the power spectrum (with $k = 2.5\times 10^{11}$ Mpc$^{-1}$),
 the second for its early growth (with $k = 5\times 10^{11}$ Mpc$^{-1}$), the third for the peak (with  $k = 2\times 10^{14}$ Mpc$^{-1}$), 
 the fourth corresponds to the critical value of $k$ (with $k_{\rm cr} = 2.42\times 10^{17}$~Mpc$^{-1}$)  and the last one being
 responsible for the right-side end of the peak (with $k = 10^{19}$~Mpc$^{-1}$). The vertical bands mark the intervals of $e$-folds during which the USR (darker pink) and the negative friction (lighter pink, wider) phases occur. We see that the USR phase does not fully capture the sudden changes of the modes with $k = 5\times 10^{11}$ Mpc$^{-1}$ and
 $k = 2.5\times 10^{11}$ Mpc$^{-1}$,
   which, as explained by means of the analytical model, take place
 within the negative friction phase. 
A final comment about the orange mode with $k = 2\times 10^{14}$ Mpc$^{-1}\approx k_{\rm peak}$ that contributes to the peak of the power spectrum is in order. 
As it is clear from our analytical discussion --and confirmed by the exact numerical computation in fig.\,\ref{fig:GrowingModes}-- the location of the peak of the power spectrum comes from modes for which horizon crossing takes place close  to the beginning of the negative friction phase (see also fig.\,\ref{fig:keFold}).
These modes have indeed at their disposal the whole negative friction phase to grow exponentially fast.
  
After this discussion about the properties of the primordial power spectrum, we now return to its relation to the PBH abundance. In order to connect the two, we assume Gaussian fluctuations and
apply the Press-Schechter formalism of gravitational collapse to compute the probability that a given horizon-sized volume forms a PBH when a large curvature fluctuation 
reenters the horizon during the radiation era.\footnote{If $\mathcal{P}_\mathcal{R}$ peaks at scales which re-enter the horizon during a period of matter domination, the formation of PBHs is enhanced  with respect to the case of radiation domination. Our potential has an approximately quadratic absolute minimum, which allows for an early period of matter domination  if the inflaton is weakly coupled to other species, preventing  violent preheating from happening. Following the analysis of \cite{Ballesteros:2019hus}, we find examples compatible with the CMB and capable of producing $f_{\rm PBH}\simeq 1$ in an adequate range of mass in this scenario.} The fraction of PBHs in the form of dark matter is\footnote{We have used eq.\ (2.7) of ref.\ \cite{Ballesteros:2017fsr} with  the standard value $\gamma = 0.2$ for the efficiency factor and $g=106.75$. See eq. (2.2) of ref.\ \cite{Ballesteros:2019hus} for a more precise expression, which reduces to the previous one assuming $g_s\simeq g$ (equality between the energy density and entropy degrees of freedom). Notice than uncertainties in $\delta_c$ and $\gamma$ can be compensated with slight variations of the parameters of the potential to maintain the same $f_{\rm PBH}$.}
\begin{equation}\label{eq:MasterAbundance}
f_{\mathrm{PBH}}=\frac{\Omega_{\rm PBH}}{\Omega_{\rm DM}}\simeq\left(\frac{M_{\mathrm{PBH}}}{10^{18}\,{\rm g}}\right)^{-1/2}\frac{1.25\times 10^{15}}{\sqrt{2\pi\sigma^2}}\int_{\delta_c}^\infty\exp\left[-\frac{\delta^2}{2\sigma^2}\right]d\delta
\end{equation}
In this expression the integral is over
is the radiation density contrast in the total matter gauge (see \cite{Ballesteros:2018wlw}), $\delta_c = 0.45$ is
the standard value for the PBH formation threshold in a radiation-dominated era (see e.g. \cite{Green:2004wb,Musco:2004ak})\footnote{It has been argued that the threshold for collapse is non-universal and depends on the  (radial) mass profile of the collapsing overdensity \cite{Musco:2018rwt}. See however \cite{Escriva:2019phb} for a formulation (in terms of the so-called compaction function) which may allow to define an approximately universal $\delta_c\simeq 0.4$.}
\begin{equation}
\sigma^2=\frac{16}{81}\int\frac{dq}{q}\left(\frac{q}{k}\right)^4\mathcal{P}_\mathcal{R}(q)W^2(q/k)\,,
\end{equation}
where $W$ is a window function that we choose as $W(q/k)=\exp(-(q/k)^2/2)$. Our final result for $f_{\mathrm{PBH}}$ is shown in the left panel of fig.\,\ref{fig:Ka}.

Notice that eq.\,(\ref{eq:MasterAbundance}) is based on the assumption of a Gaussian statistic. 
Non-Gaussian effects are in general expected\,\cite{Saito_2008,Atal:2019erb}, and they alter the Gaussian estimate of the abundance by 
introducing corrections controlled by higher-order 
cumulants of the probability density functions (of order equal of higher than three, which in the Gaussian
case vanish)\,\cite{Franciolini:2018vbk}. 
If the higher-order cumulants are positive, 
non-Gaussian corrections enhance the abundance computed by means of the Gaussian approximation. 
As already anticipated in section\,\ref{sec:Mot}, we are mostly interested in the position of the peak (since it is directly related to the observed value of $n_s$) rather than its amplitude. For this reason, we stick to the Gaussian approximation introduced before, and leave
to a future work a proper quantification of non-Gaussianities in this model.

\begin{figure}[t]
\begin{center}
$$\includegraphics[width=.41\textwidth]{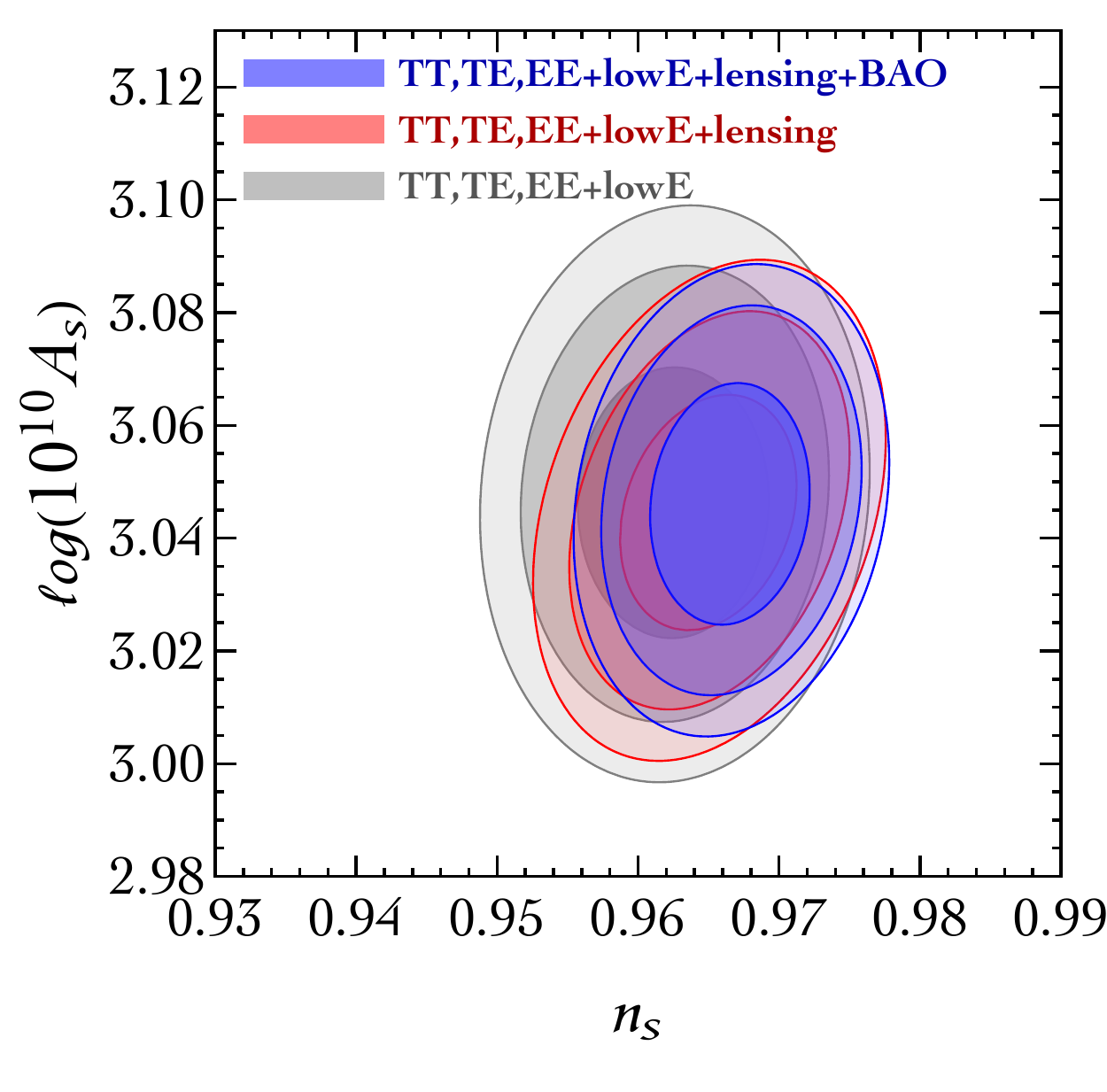}
\qquad\includegraphics[width=.41\textwidth]{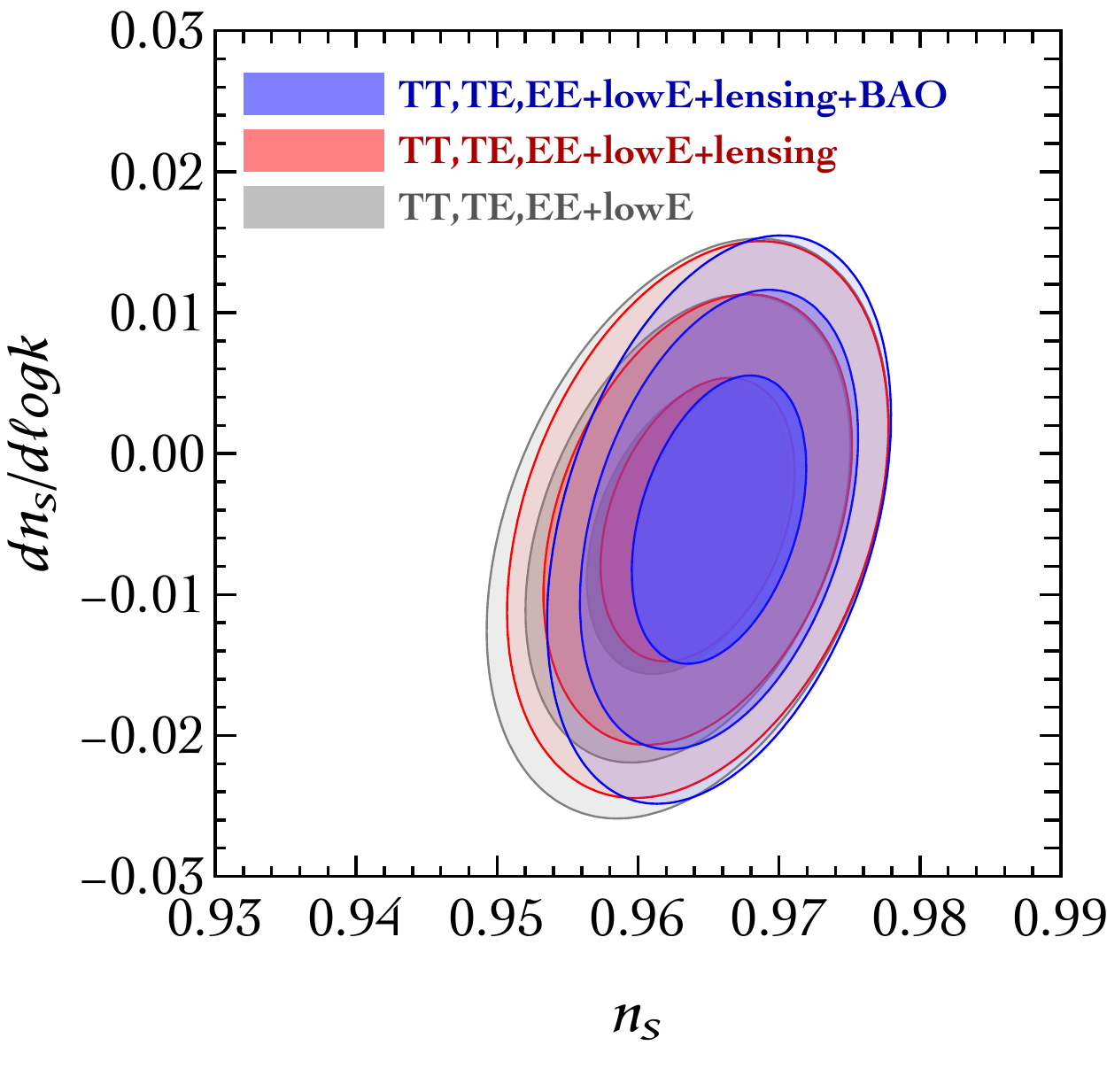}$$ 
\caption{\em \label{fig:Fit} 
{\color{VioletRed4}{
Constraints  on the  Base $\Lambda$CDM  model and some of its extensions.
Contours  show the 68\%, 95\% and 99\% confidence regions for Planck\,TT,TE,EE+lowE (grey), Planck\,TT,TE,EE+lowE+lensing  (red),  and Planck\,TT,TE,EE+lowE+lensing+BAO  (blue).
Left panel: We use the 6-parameters Base $\Lambda$CDM  model, and we show the constraint in the plane 
$(n_s, A_s)$. Right panel: The running of the spectral index is added to the 6-parameters Base $\Lambda$CDM  model, and we show the constraint in the plane 
$(n_s,dn_s/dlog k)$.
}}
 }
\end{center}
\end{figure}

In order to make contact with CMB observables,  at scales $10^{-4} \lesssim k\,[\,{\rm Mpc}^{-1}\,] \lesssim 0.5$, we fit our power spectrum against the parametric function\,\cite{Akrami:2018odb}
\begin{equation}\label{eq:ParametricPS}
\log \mathcal{P}_{\mathcal{R}}(k) =\log A_s+\left({n_s - 1 +\frac{\alpha}{2}\log\frac{k}{k_*}+
\frac{\vartheta}{6}\log^2\frac{k}{k_*}+\dots}\right)\log \frac{k}{k_*}\,,
\end{equation}      
with $\alpha = dn_s/d\log k$, $\vartheta= d^2n_s/d\log k^2$.
At the pivot scale $k_* = 0.05$ Mpc$^{-1}$, we find\footnote{We always refer to CMB parameters (such as $n_s$) at the scale $k_* =0.05$ Mpc$^{-1}$, even if the latter is not mentioned explicitly.}
    \begin{align}\label{eq:CMBfit}
 \log(10^{10}A_s) \simeq 3.06\,,~~~~n_s \simeq 0.9491\,,~~~~\alpha \simeq -10^{-3}\,,~~~~\vartheta \simeq 2\times 10^{-4}\,.
    \end{align} 
    For the tensor-to-scalar ratio, we find (by means of the slow-roll approximation) $r\simeq 0.03$.
     All these values but the spectral index, $n_s$, are in good agreement with observations.
  The fit of the spectral index $n_s$ results in a $\sim 3\,\sigma$ tension with the latest Planck constraints 
  if one takes the analysis obtained assuming the 6-parameters Base $\Lambda$CDM model or the extension in which 
 the running of the spectral index is added as an additional free parameter\,\cite{Akrami:2018odb}:\footnote{The addition of BAO data increases the best fit value of $n_s$ in  both cases just at the  level of 0.2\%, see \cite{Akrami:2018odb}.}
 \begin{align}
 {\rm Base}~\Lambda{\rm CDM}:~~~~&n_s = 0.9649\pm 0.0042\,,~~~~[68\%\,{\rm CL,\,Planck\,TT,TE,EE+lowE+lensing}]\,,\label{eq:nsPlanck}\\
  {\rm Base}~\Lambda{\rm CDM}+ \frac{dn_s}{d\log k}:~~~~&n_s = 0.9641\pm 0.0044\,,~~~~[68\%\,{\rm CL,\,Planck\,TT,TE,EE+lowE+lensing}]\,.\label{eq:nsPlanck2}
   \end{align}
   To be even more concrete, we show in fig.\,\ref{fig:Fit} the two-dimensional 65\%, 95\% and 99\% confidence contours 
    for the parameters $n_s$ and $A_s$ in the Base $\Lambda$CDM model (left panel)
 and    for the parameters $n_s$ and $dn_s/d \log k$ for the case in which the running of the spectral index is added to the Base $\Lambda$CDM.\footnote{We smooth the 99\% confidence contours by means of a Gaussian approximation which works extremely well for our illustrative purposes.}
A scan over the parameters of the potential\footnote{A less intensive scan was already performed for \cite{Ballesteros:2017fsr}, with the same qualitative result, which we now confirm.} seems to suggest that values of $n_s$ slightly smaller than the one expected on the basis of eqs.\,(\ref{eq:nsPlanck}, \ref{eq:nsPlanck2}) is
 a general result, and not just a vice of the specific numerical solution analyzed in this section.
 More in detail, we find that increasing the value of $n_s$ in order to reduce the tension with eqs.\,(\ref{eq:nsPlanck}, \ref{eq:nsPlanck2}) results in a shift of the peak of the PBH mass
 distribution towards smaller values of $M_{\rm PBH}$. 
Even though we find that in these cases it is still possible to obtain a peak in the power spectrum of order $\sim 10^{-1}$, 
the constraint from Hawking evaporation, as evident from fig.\,\ref{fig:Ka}, kicks in and very rapidly forbids 
sizable abundances of PBHs.
On the contrary, moving the peak of the PBH mass distribution towards larger values of $M_{\rm PBH}$ 
 implies a decrease of the spectral index below $n_s \simeq 0.95$, thus exacerbating the tension with CMB observables.
 Having a better understanding of the implications of this tension is an important point that we shall discuss in 
 more detail in section\,\ref{sec:Disc}.
For the sake of the present discussion,
 let us consider the results of the fit in eq.\,(\ref{eq:CMBfit}) as acceptable, and continue with the description of our analysis.

Given the power spectrum of curvature perturbations computed by means of eq.\,(\ref{eq:FullPS}) and shown in fig.\,\ref{fig:StarPlot}, 
we compute the induced second-order gravitational wave spectrum. 
Our result is shown on the right panel of fig.\,\ref{fig:Ka}, where we plot the fraction 
of the energy density of gravitational waves relative to the critical energy density, $\Omega_{\rm GW}$. 
We follow the computation recently revisited in ref.\,\cite{Espinosa:2018eve} 
(see refs.\,\cite{Ananda:2006af,Baumann:2007zm} for earlier analyses),
and we validate our 
numerical results against the analysis in ref.\,\cite{Inomata:2018epa}. 
 For illustrative purposes, we superimpose to the signal the expected sensitivity curves computed for the  gravitational wave detectors LISA, DECIGO and 
 MAGIS-100. 
 The sensitivity curves are obtained by converting the noise spectra $S_h(f)$, a function of the frequency $f$,
  into the corresponding fractional energy density by means of $\Omega_{\rm GW} = (4\pi^2/3H_0^2)f^3S_h$.
 We find that the signal could be in principle detected.  In particular, we point out that the gravitational wave signal peaks in the region covered by the DECIGO  sensitivity curve but it features a ``shoulder''
 within the reach of LISA sensitivity.
 
Before moving on to the next section, we should remark that our calculation of the power spectrum of curvature perturbations is based on the classical roll of the inflaton along its potential. Quantum diffusion may modify the result for large deviations from slow-roll, playing a role in the generation of curvature perturbations.
The importance of quantum diffusion for the formation of PBHs  has been investigated in refs.\,\cite{Biagetti:2018pjj,Cruces:2018cvq,Ezquiaga:2018gbw,Ezquiaga:2019ftu}. 
Ref.\,\cite{Biagetti:2018pjj} focuses on scenarios with an USR phase. 
This effect can have a large impact on the PBHs abundance, which is very sensitive to the amplitude of the power spectrum of curvature perturbations. 
On the other hand, the latter can also be easily modified by a small tuning of the parameters of the model, as stressed in appendix \ref{app:A}.
Ref.\,\cite{Ezquiaga:2019ftu} develops a framework to compute the probability distribution of curvature perturbations including quantum diffusion. This analysis is valid for slow-roll dynamics and therefore it can not be directly applied to our case. We leave for future work a thorough investigation of quantum diffusion during the USR phase.

%%%%%%%%%%%%%%%%%%%%%%%%%%%%%%%%%%%%%%%%%%%%%%%%%%%%%%%%%%%%%%%%%%%%
\section{Discussion}\label{sec:Disc}
%%%%%%%%%%%%%%%%%%%%%%%%%%%%%%%%%%%%%%%%%%%%%%%%%%%%%%%%%%%%%%%%%%%%
We now come back to the case illustrated in fig.\,\ref{fig:Ka} in which the PBHs are responsible for the majority of dark matter. 
As already discussed below eqs.\,(\ref{eq:nsPlanck},\ref{eq:nsPlanck2}) the inflationary solutions corresponding to this situation are characterized by $n_s \simeq 0.95$, a value that is smaller than the one suggested by the most recent Planck analysis. As anticipated in the previous section, if we try to alleviate the tension by increasing the value of $n_s$, the peak of the PBH mass distribution shifts towards smaller values of $M_{\rm PBH}$ where the Hawking evaporation bound excludes sizable fractions of dark matter in the form of PBHs.
\begin{figure}[t]
\begin{center}
$$\includegraphics[width=.39\textwidth]{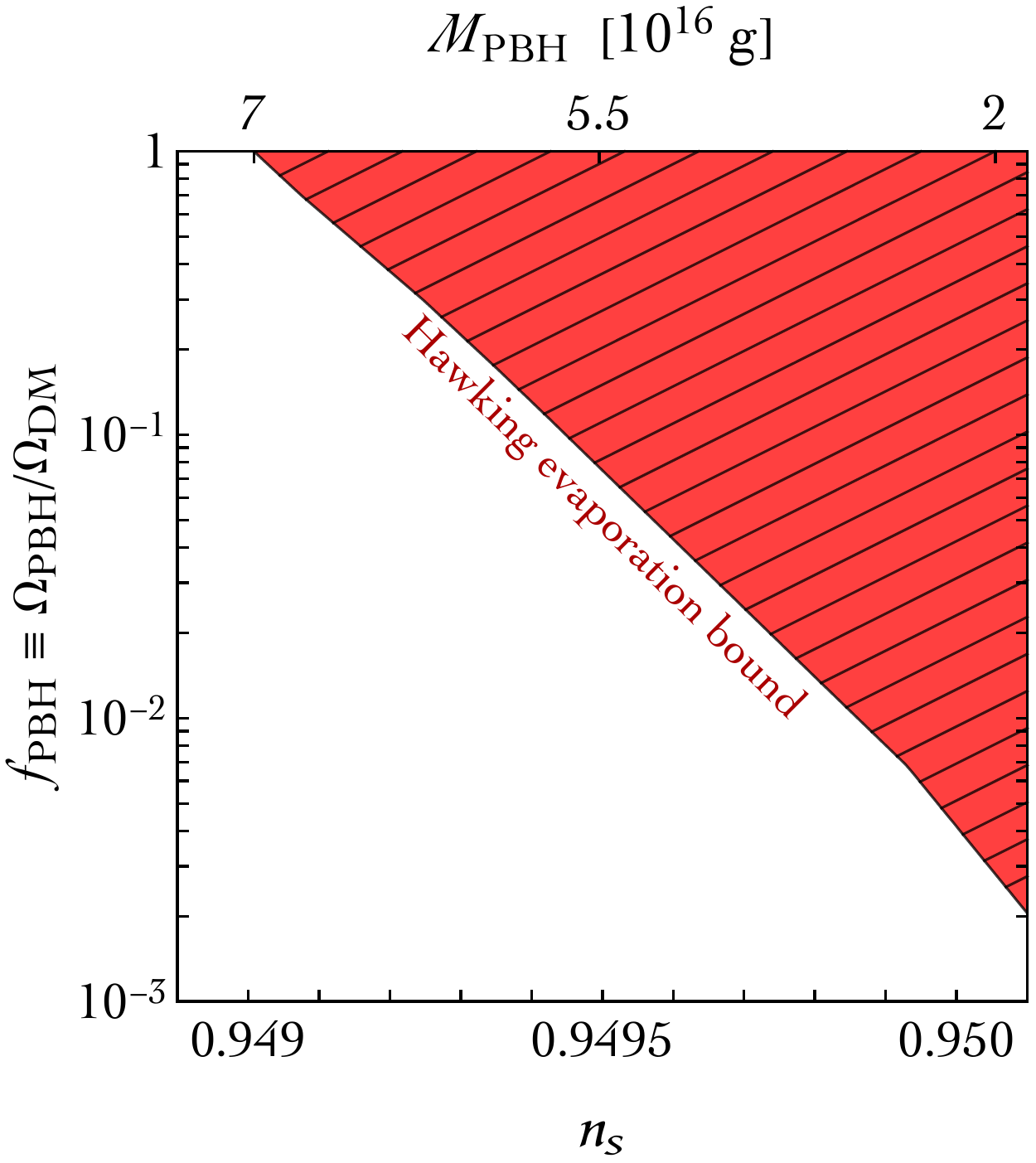}$$
\caption{\em \label{fig:CMBTension} 
{\color{VioletRed4}{
Fraction of the dark matter in the form of PBHs as a function of the spectral index $n_s$ (bottom $x$-axis). 
Solutions with acceptable values for the cosmological parameters $A_s,$ $r,$ $\alpha,$ $\vartheta,$ $\Delta N_e^{0.05}$ and satisfying the Hawking evaporation constraint populate the white region.
On the top $x$-axis we show the value of the PBH mass corresponding to the peak of the mass distribution that saturates the Hawking evaporation bound for a given $n_s$.
}}
 }
\end{center}
\end{figure} 
To be more quantitative, in fig.\,\ref{fig:CMBTension} we show the result of a scan over the parameter space of the model. 
All solutions found are characterized by acceptable values for the CMB observables $A_s$, $r$, $\alpha$, $\vartheta$ in eq.\,(\ref{eq:ParametricPS}) and by the condition $\Delta N_e^{0.05}>50$ $e$-folds.
The plot confirms that values of the spectral index $n_s > 0.95$ are compatible with a fraction of the dark matter in the form of PBHs smaller than  $10^{-3}$ while the totality 
of dark matter in the form of PBHs can be obtained only if $n_s \simeq 0.949$ or smaller.
The tendency for a value of $n_s$ smaller than suggested by Planck seems to be a rather common property of single-field inflationary models with an approximate stationary inflection point. 
See e.g.\ the discussions in ref.\,\cite{Ballesteros:2017fsr} relative to the model we study here and  the model with a radiatively induced inflection point.
In the model discussed in ref.\,\cite{Ozsoy:2018flq}, some inflationary solutions with $n_s \gtrsim 0.96$ are presented. 
However, these solutions -- although consistent as far as the value of the spectral index is concerned -- are all characterized by a large running ($\alpha \simeq -0.03$) that is in conflict with the Planck bound (see the right panel of fig.\,\ref{fig:Fit}). Examples consistent with  the CMB can however be found with the model recently put forward in \cite{Ballesteros:2019hus}.

\begin{figure}[t]
\begin{center}
$$\includegraphics[width=.41\textwidth]{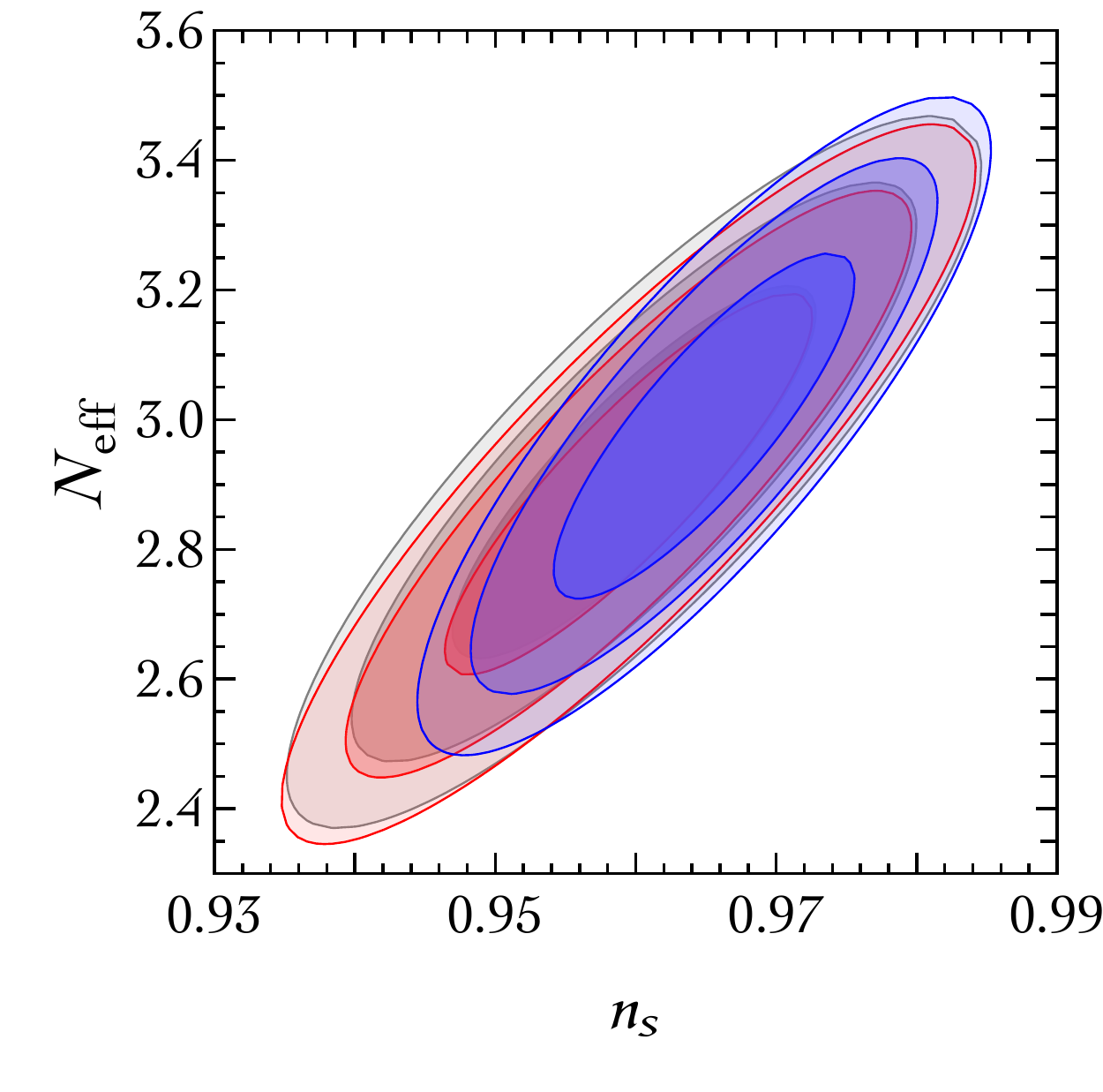}
\qquad\includegraphics[width=.41\textwidth]{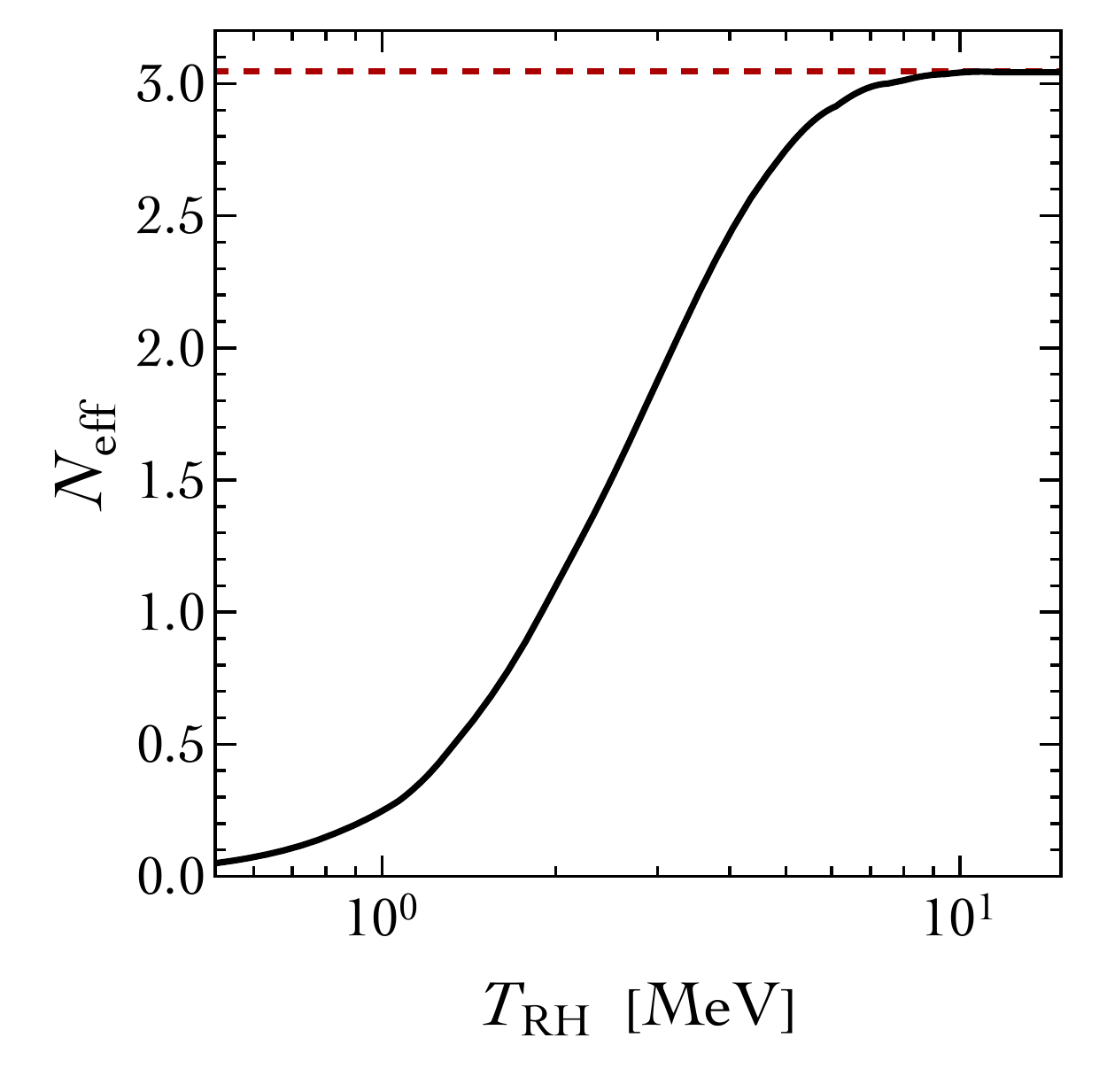}$$ 
\caption{\em \label{fig:Fit2} 
{\color{VioletRed4}{
Left panel: The effective number of relativistic degrees of freedom is added to the 6-parameters Base $\Lambda$CDM  model, and we show the constraint in the plane 
$(n_s,N_{\rm eff})$ (same color code as in fig.\,\ref{fig:Fit}). Right panel: 
Evolution of the effective number of relativistic degrees of freedom as function of the reheating temperature of the Universe, from \cite{deSalas:2015glj}.
}}
 }
\end{center}
\end{figure}

In light of the above discussion, it is important to interpret the actual relevance of this tension.
The value of $n_s$ quoted in eq.\,(\ref{eq:nsPlanck}) refers to the 6-parameter Base $\Lambda$CDM model while the one in eq.\,(\ref{eq:nsPlanck2}) to the extension in which also the running of the spectral index is added as a free parameter. Let us consider other popular extension of the 6-parameters Base $\Lambda$CDM model\,\cite{Aghanim:2018eyx}. 
There are indeed valid motivations to believe that the Base $\Lambda$CDM model does not capture all the relevant physics throughout the evolution of the Universe, from inflation to the present day.
For instance, in the  Base $\Lambda$CDM model there are three massless neutrinos. 
This is a reasonable first-order approximation.
However, neutrino oscillations  indicate that neutrinos have a small but non-zero mass.
It is, therefore, more than legitimate to extend the Base $\Lambda$CDM model by including the sum over the active neutrino masses, $\sum m_{\nu}$. 
This parameter is negatively correlated with $n_s$, and smaller values of $n_s$ correspond to increasing values of $\sum m_{\nu}$.
Another plausible extension includes the effective number of relativistic degrees  of  freedom, $N_{\rm eff}$. 
Quoting from the official Planck 2018 release \cite{Aghanim:2018eyx}:
\begin{align}
  {\rm Base}~\Lambda{\rm CDM}+ N_{\rm eff}:~~~~&n_s = 0.9589 \pm 0.0084\,,&[68\%\,{\rm CL,\,Planck\,TT,TE,EE+lowE+lensing}]\,,\label{eq:nsPlanckNeff}\\
   {\rm Base}~\Lambda{\rm CDM}+ N_{\rm eff}+ \sum m_{\nu}:~~~~&n_s = 0.9587\pm 0.0086 \,,&[68\%\,{\rm CL,\,Planck\,TT,TE,EE+lowE+lensing}]\,,\label{eq:nsPlanckNeffmnu}\\
    {\rm Base}~\Lambda{\rm CDM}+ N_{\rm eff}+ \frac{dn_s}{d\log k}:~~~~& n_s = 0.950 \pm 0.011\,,&[68\%\,{\rm CL,\,Planck\,TT,TE,EE+lowE+lensing}]\,,\label{eq:nsPlanckNeffRun}
\end{align}
and in all these motivated extensions the tension with respect to the value $n_s \simeq 0.95$ is significantly reduced. 
It is indeed known that including a 
marginalization over the  total  neutrino  mass or the number of relativistic degrees of freedom could induce a shift towards lower values in the determination of $n_s$\,\cite{Gerbino:2016sgw}.
Of course, one should take the one-parameter confidence intervals quoted in eqs.\,(\ref{eq:nsPlanckNeff},\ref{eq:nsPlanckNeffmnu},\ref{eq:nsPlanckNeffRun}) with a grain of salt.
For instance, $N_{\rm eff}$ is positively correlated with $n_s$, and smaller values of $n_s$ can be reached only at the prize of lowering $N_{\rm eff}$ thus obtaining $N_{\rm eff}<3$. 
This correlation --not explicit in eq.\,(\ref{eq:nsPlanckNeff})-- can be better appreciated by looking at the two-dimensional marginalized constraint contours for $n_s$ and $N_{\rm eff}$ that we show in the left panel of fig.\,\ref{fig:Fit2}.
A very similar comment also applies to the primordial Helium abundance $Y_p$. This is because both the Helium abundance and 
relativistic degrees of freedom affect the CMB damping tail and they are, therefore, partially degenerate.
The prediction of the standard neutrino decoupling model is $N_{\rm eff} = 3.046$\,\cite{Mangano:2005cc}. 
It is worth mentioning that deviations
 from the standard value  of  $N_{\rm eff}$  due only to neutrino physics are possible. 
For instance,  in very low-reheating scenarios in which the reheating temperature
is as low as  few MeV, neutrinos do not have time to thermalize. 
This affects their contribution to the radiation content of the Universe, and results in a value of $N_{\rm eff}$ that can be lower than otherwise expected, see \cite{Kawasaki_1999,deSalas:2015glj,Hasegawa_2019}.  
This is shown in the right panel of fig.\,\ref{fig:Fit2}.
We admit that this is a very coerced possibility but --since physics is based on facts and not only personal taste-- it is worth mentioning this as a valid scenario. 
On a more speculative ground, the increasing  statistical significance (that is now at the level of 5.3\,$\sigma$\,\cite{Wong:2019kwg}) of the so-called ``Hubble tension'' --the discrepancy between the values of the present day Hubble expansion rate $H_0$ derived from the local distance ladder and the CMB, see, e.g.,\,\cite{Riess:2019cxk,Verde:2019ivm} and references therein-- further motivates the need of some new physics beyond the Base $\Lambda$CDM model.
Following this line of reasoning, ref.\,\cite{DiValentino:2019dzu} considered a global analysis of current cosmological data in a cosmological scenario
that is significantly more extended 
 than the one provided by the Base $\Lambda$CDM model --They included as free parameters $\alpha$, $N_{\rm eff}$, $\Sigma m_\nu$ and  the equation of state of dark energy-- finding a preferred value of the spectral index of order $n_s \simeq 0.95$ 
 (but, importantly, again in combination with a slight preference for
  $N_{\rm eff}< 3$, due to the degeneracy discussed before).\footnote{This preference for small $n_s$ is lost if the lensing amplitude $A_L$ is also left free, see ref.\,\cite{DiValentino:2019dzu}.}

In light of these results, we argue that values of the spectral index of order $n_s \simeq 0.95$ --favored by our analysis 
assuming the majority of dark matter to be comprised of PBHs-- could consistently fit in the context of a cosmological model that extends the standard Base $\Lambda$CDM one.
A dedicated analysis of what  are the possible and best motivated extensions is underway.

Finally, let us mention that also the theoretical uncertainty on the calculation of the PBH mass could be relevant to determine whether a solution is viable or not.
As shown in fig.\,\ref{fig:CMBTension}, shifting the mass by a factor $\gtrsim 3$ would push solutions with $n_s=0.95$ (at the limit of the 99\% C.L. contours in fig.\,\ref{fig:Fit}) away from
 the Hawking evaporation constraints. In this way it would be possible to explain all the DM in terms of PBHs with
these solutions. In our calculation, we have assumed that the mass of the PBHs  is proportional to the mass contained in one Hubble volume at horizon crossing.
Numerical simulations in combination with the application of peaks theory suggest that the mass of the PBHs depends on the shape of the perturbation from which it is formed \cite{Germani:2018jgr}. Moreover it is related to the mass in an Hubble volume at a time slightly different from the horizon crossing time. Unfortunately, these calculations carry
uncertainties of $\mathcal{O}(1),$ which (at the moment) prevent a reliable determination of the fate of the models with low $n_s$.

Independently on this uncertainty, in the next section we will show that the effect of the higher dimensional operators of the potential in eq. (2), which we have neglected so far, can easily solve the $n_s$ tension. Before that we will briefly comment on earlier work.

Ref.\,\cite{Garcia-Bellido:2017mdw}, proposed a model characterized by a potential with the functional form of eq.\,(\ref{eq:MasterPot}). In ref.\,\cite{Garcia-Bellido:2017mdw} the field $\phi$ of eq.\,(\ref{eq:MasterPot}) is canonically normalized, whereas in our case it has a non-canonical kinetic term arising from the transformation from the Jordan to the Einstein frame. In simple words: the inflaton field in the Einstein frame in the two models is different, see eq.\,(\ref{eq:fieldred}). The denominator of eq.\,(\ref{eq:MasterPot}) appears in our model as a consequence of the metric redefinition required to go from one  frame to the other, while it is instead postulated from the start in ref.\ \cite{Garcia-Bellido:2017mdw}. This difference has a relevant phenomenological consequence: the examples provided in \cite{Garcia-Bellido:2017mdw} lead to PBHs several orders of magnitude heavier, which are allowed only at the level of $f_{\rm PBH}\lesssim 10\%$ due to the microlensing bounds from the EROS project \cite{Tisserand:2006zx}.
Another related work is ref.\,\cite{Kannike:2017bxn}. This differs from ours in two respects: a running of the quartic coupling of the inflation above a mass threshold is introduced, and the power spectrum is computed only in the slow-roll approximation.

%%%%%%%%%%%%%%%%%%%%%%%%%%%%%%%%%%%%%%%%%%%%%%%%%%%%%%%%%%%%%%%%%%%%
\section{On the role of higher-dimensional operators}\label{sec:HDO}
%%%%%%%%%%%%%%%%%%%%%%%%%%%%%%%%%%%%%%%%%%%%%%%%%%%%%%%%%%%%%%%%%%%% 

In models of large-field inflation one should generically take into consideration HDOs in the inflaton potential. It is then natural to question whether these corrections might spoil solutions that lead to a considerable abundance of PBHs at the renormalizable level, either by lowering the abundance, or by changing the power spectrum parameters at CMB scales.
Consider the scalar potential
\begin{align}\label{eq:TempHDO1}
\tilde{V}(\phi) = 
\frac{1}{(1+\xi \phi^2)^2}\left(a_2\phi^2 + a_3\phi^3 + a_4\phi^4 + \sum_{n=5}^{\mathcal{N}}a_n\phi^n\right)\,,
\end{align} 
where we remind that all dimensionful quantities are expressed in units of $M_{\rm Pl}$ which is set to $1$. 
A precise characterization of the HDOs is of course possible only if one knows the ultraviolet completion of gravity.
Nevertheless, it is still possible to gain an interesting insight from a minimal set of assumptions.
To validate our construction, we shall follow two simple rules:
\begin{itemize}
\item [{\it i)}] The HDOs must be subdominant compared to the  leading renormalizable terms.
\item [{\it ii)}] The HDOs must be organized in the form of a ``convergent'' series. The meaning  of this will be immediately clear.
\end{itemize}
Let us explain our rationale in more detail. If we rewrite eq.\,(\ref{eq:TempHDO1}) as follows
\begin{align}\label{eq:TempHDO2}
\tilde{V}(\phi) = 
\frac{a_4}{(1+\xi \phi^2)^2}\left[\tilde{a}_2\phi^2 + \tilde{a}_3\phi^3 +\phi^4
\left(1 + \sum_{n=5}^{\mathcal{N}}\tilde{a}_n\phi^{n-4}
\right)
\right]\,,
\end{align}
the conditions {\it i)} and 
{\it ii)}  translate into the order relation
\begin{align}\label{eq:OrderRelation}
\dots < \tilde{a}_n\phi^{n-4} < \dots <\tilde{a}_5\phi < 1\,,
\end{align} 
meaning that, at each order $n\geqslant 5$ in the expansion, the coefficient $\tilde{a}_n$ 
has to be small enough (compared to the previous one) to compensate the additional power of $\phi$, which can easily be 
$O(10)$ at CMB scales. 
The  description in terms of the effective operators breaks down at large field values where eq.\,(\ref{eq:OrderRelation}) ceases to be valid. 
Driven by a pure phenomenological approach, one can, for instance, fix the coefficient $\tilde{a}_5$ to a very small number such that 
the condition $\tilde{a}_5\phi < 1$ is satisfied all along the inflationary trajectory while setting the remaining coefficients $\tilde{a}_{n \geqslant 6}$ to zero. 
It is indeed easy to check that, if we take for simplicity the same values for $c_{2,3}$ 
and $\phi_0$ chosen in eq.\,(\ref{eq:Example1}), the presence of a dimension-five HDO with coefficient $\tilde{a}_5 \sim O(10^{-3})$ 
(together with $\tilde{a}_{n \geqslant 6} = 0$) is enough to give acceptable solutions with $n_s \simeq 0.960$ and the correct mass and abundance of PBHs.
However, it is better to follow some organization principle that may help elucidating the physical interpretation of 
 eq.\,(\ref{eq:OrderRelation}). 
Our take on this is the following:
broadly  speaking, the ultraviolet theory that generates the HDOs in eq.\,(\ref{eq:TempHDO2}) will be described, at least, by a mass scale $M$ and a 
dimensionless coupling $g$. 
Let us discuss how these fundamental quantities enter in our construction. We rewrite each HDO as
\begin{align}\label{eq:EffOp}
\mathcal{O}_{n} = a_n\phi^n = \frac{\phi^n}{\Lambda_n^{n-4}}\,,\quad n\geq 5,
\end{align}
where (for each operator) we introduce a suppression scale $\Lambda_n$ (that is not necessarily 
equal to $M_{\rm Pl}$). 
The scale $\Lambda_n$ defines the strength of the effective interaction $\mathcal{O}_{n}$, and it is given by the  ratio  between  a mass scale and  a  certain  power  of  couplings. A simple but pertinent example is that of the electroweak scale $v$. Its inverse squared, the Fermi constant $G_F = 1/v^2$, defines the strength 
of the dimension-six four-fermion operator in the Fermi theory, and can be defined by means of the ratio between the $W^{\pm}$ mass and the weak gauge coupling.
 As anticipated, we make the simplified assumption that the ultraviolet completion that is responsible for the generation of the effective 
operators in eq.\,(\ref{eq:EffOp}) is characterized by a single coupling $g$ and a single mass scale $M$.
In such a case, by means of dimensional analysis\,\cite{Brodsky:2010zk}, we expect in the weak coupling limit the scaling
\begin{align}\label{eq:ExplicitScal}
\frac{1}{\Lambda_n} = \frac{g^{\frac{n-2}{n-4}}}{M}~~~~~~\Longrightarrow~~~~~~\frac{1}{\Lambda_n^{n-4}} = g^2\left(\frac{g}{M}\right)^{n-4}
\equiv \frac{g^2}{\Lambda^{n-4}}\,,
\end{align}
where $\Lambda \equiv M/g$. We refer to appendix\,\ref{app:B} for a detailed derivation of the scaling in eq.\,(\ref{eq:ExplicitScal}).
The mass scale $M$ can be considered as the mass associated to new degrees of freedom populating the ultraviolet theory, 
while $g$ characterizes their self-coupling as well as their coupling with $\phi$. 
Consequently, 
if we compare --in the spirit of eq.\,(\ref{eq:TempHDO2})--  the HDO $\mathcal{O}_{n}$ with the renormalizable term $a_4\phi^4$, we can write 
(keeping only track of powers of $M$ and $g$, and neglecting $O(1)$ proportionality coefficients)
\begin{align}\label{eq:EffOp2}
a_4 \phi^4 + \mathcal{O}_{n} =a_4 \phi^4 \bigg[1+ \underbrace{\bigg(\frac{g^2}{a_4}\bigg)\bigg(\frac{g}{M}\bigg)^{n-4}}_{= \tilde{a}_n}\phi^{n-4}\bigg]\,.
\end{align}
From dimensional analysis, we know that the coefficient $a_4$ has dimension of a coupling squared. This implies that the ratio $c \equiv g^2/a_4$ is a genuine dimensionless number (see appendix \ref{app:B} for details). 
The hierarchy among the coefficients $\tilde{a}_n$ for $n\geq 5$ can
be obtained if $\Lambda > \phi$. 
 More precisely, the conditions in eq.\,(\ref{eq:OrderRelation}) translate into
\begin{align}\label{eq:ConceptualBound}
 \frac{1}{\Lambda} < {\rm min}\left\{
 \frac{a_4}{g^2}\frac{1}{\phi_{\rm in}},\frac{1}{\phi_{\rm in}}
 \right\}\,,
 \end{align}
 where $\phi_{\rm in} \equiv \phi(\Delta N_e^{0.001})$ corresponds to the field value at the largest observable comoving scale (since we need to trust 
  our theoretical description at least up to such field values). Clearly, as we already mentioned, for sufficiently large values of $\phi$ the hierarchy \eq{eq:OrderRelation} will break down. This is just a manifestation of an  old problem of initial conditions in large-field inflation.  We stress that this problem is by no means unique to our model, but completely generic for large field inflation, and we do not aim to solve it in this paper. We simply assume that the slow-roll approximation is valid at $\phi=\phi_{\rm in}$. 
  
  Finally, let us notice that one further condition needs to be satisfied.  As it is clear from the previous discussion, we expect new states associated to the 
ultraviolet completion of our effective  theory  to lie at the mass scale $M$. 
We have to check, therefore, that the energy density during inflation is not high enough to excite these states (which could alter our effective inflationary potential). 
This means that the relation $H < M = g \Lambda$ should be imposed.

Bearing in mind these conceptual limitations, let us investigate some numerical consequences of the potential of eq.\,(\ref{eq:TempHDO2}).
If we set for simplicity $g^2 = a_4$, the HDOs are controlled by one single dimensionful free parameter, the inverse scale $1/\Lambda$,
and we have
\begin{align}\label{eq:TempHDOLambda}
\tilde{V}(\phi) = 
\frac{a_4}{(1+\xi \phi^2)^2}\left[\tilde{a}_2\phi^2 + \tilde{a}_3\phi^3 +\phi^4
\left(1 + \sum_{n=5}^{\mathcal{N}}\frac{c_n\phi^{n-4}}{\Lambda^{n-4}}
\right)
\right]\,.
\end{align}
The coefficients $c_n$ are $O(1)$ dimensionless numbers whose exact values cannot be computed with dimensional analysis alone. 
We consider two benchmark examples. In the first one, we assume $c_n = 1\,\,\forall n$.
\begin{figure}[t]
\begin{center}
$$\includegraphics[width=.45\textwidth]{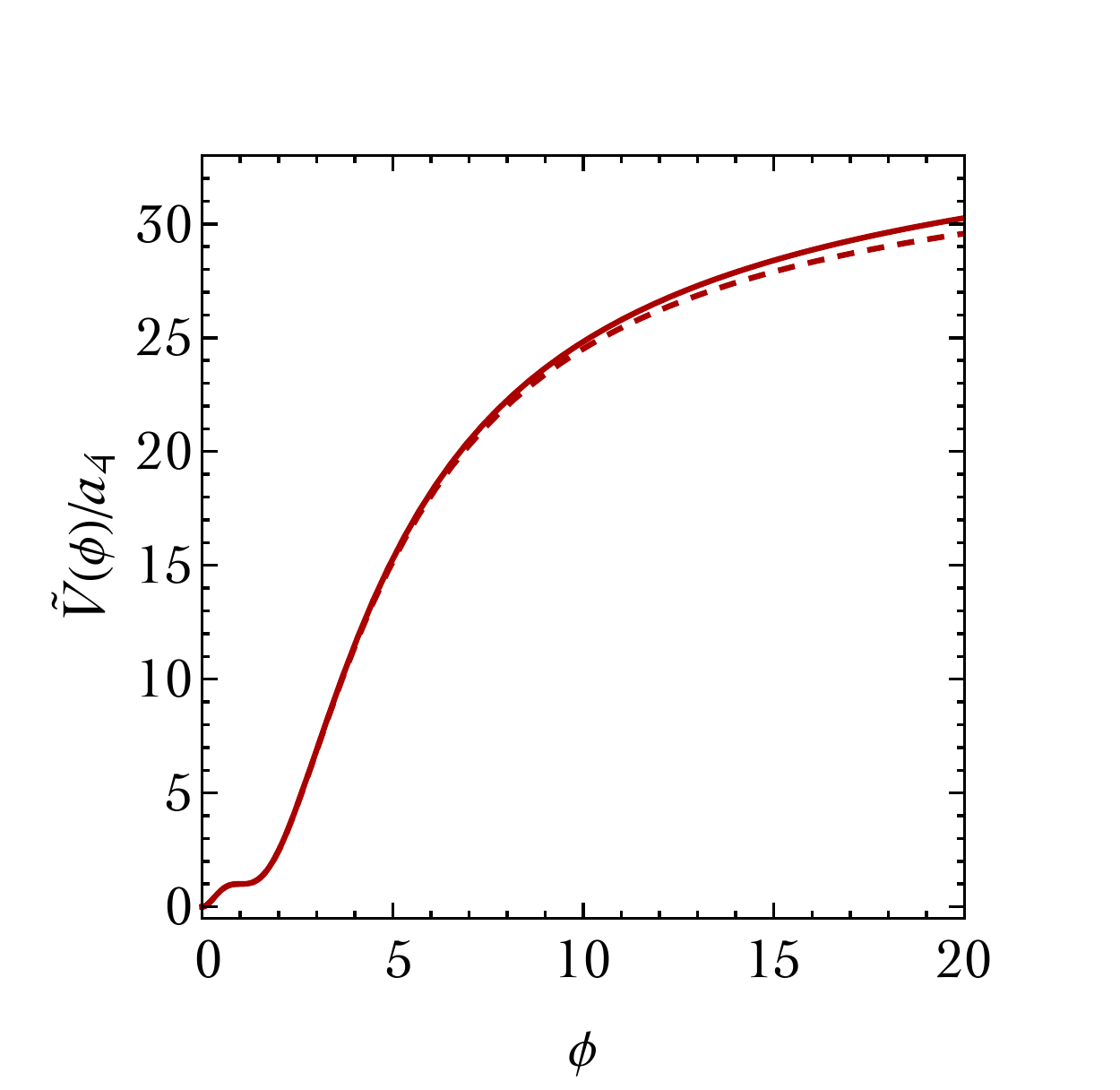}
\qquad\includegraphics[width=.45\textwidth]{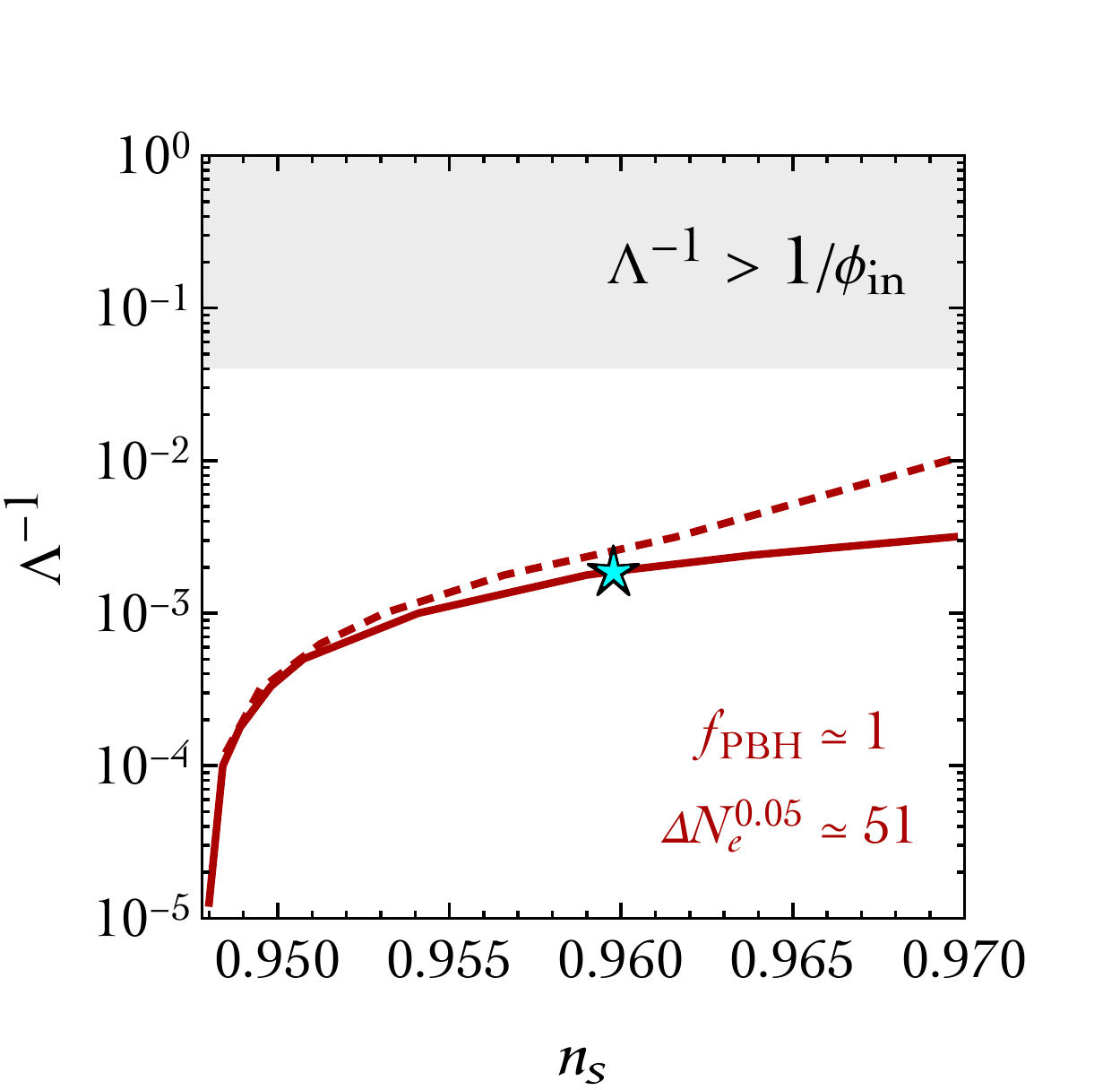}$$ 
\caption{\em \label{fig:HDO} 
{\color{VioletRed4}{
Left panel. 
Corrections to the potential in eq.\,(\ref{eq:TempHDO2}) due to the presence of HDOs.
Right panel. Inflationary solutions that give 100\% of dark matter in the form of PBHs in the plane $(1/\Lambda,n_s)$ where 
$\Lambda = M/g$ (in units of $M_{\rm Pl}$) defines the suppression scale that controls the impact of HDOs
(with large values of $\Lambda$ that correspond to more and more suppressed HDOs).
All remaining cosmological observables respect their corresponding CMB constraints. 
We show solutions with fixed number of $e$-folds $\Delta N_e^{0.05} \simeq 51$. 
We show two representative cases. 
The first case corresponds to the solid red line, and represents the impact of the HDOs in eq.\,(\ref{eq:TempHDOLambda}) 
with $g^2 = a_4$ and $c_n = 1$. 
The second case corresponds to the solid red line, and represents the impact of the HDOs in eq.\,(\ref{eq:TempHDOLambda}) 
with $g^2 = a_4$ and $c_n = (-1)^{n+1}$.
The cyan star marks the solution whose PBH abundance and induced signal of gravitational waves are plotted in fig.\ref{fig:Ka} (cyan regions with
 dashed boundary).
}}
 }
\end{center}
\end{figure} 
Qualitatively, the effect of the HDOs is shown in the left panel of fig.\,\ref{fig:HDO} where we consider for illustration $\Lambda^{-1} = 10^{-3}$ 
(we take $\xi = 0.3$ while the coefficients $\tilde{a}_2$ and $\tilde{a}_3$ are fixed in order to have a stationary inflection point at the field value $\phi_0=1$, see appendix\,\ref{app:A}). 
The important point is that, at small field values, the stationary inflection point is not affected by the presence of the HDOs since it is controlled 
by the quartic and cubic coefficients (which must have opposite signs). At large field values, on the contrary, the presence of the HDOs introduces a small deviation with respect to the 
renormalizable case (solid versus dashed line), and alters the first and second derivatives of the potential, thus changing the slow-roll parameters at the CMB pivot scale $k_* = 0.05$ Mpc$^{-1}$.
In our numerical analysis,  for each value of $\Lambda^{-1}$ we consider inflationary solutions which give rise to $f_{\rm PBH} \simeq 1$ and consistently fit CMB observables at large scales. 
In order to facilitate the comparison with the renormalizable case, in fig.\,\ref{fig:HDO}  we show solutions with fixed number of $e$-folds $\Delta N_e^{0.05} \simeq 51$. 
The values of $c_{2,3}$ and $\phi_0$ are the same used in eq.\,(\ref{eq:Example1}) for the renormalizable case while $\lambda$ and $\xi$ are tuned, for each value of $\Lambda$, in such a way to obtain, respectively, the correct normalization of the power spectrum at CMB scales and the condition $f_{\rm PBH} \simeq 1$ on the abundance of PBHs. Furthermore, it is important to remark that all solutions shown in the right panel of fig.\,\ref{fig:HDO} have, by construction, the position of the peak of the power spectrum, $k_{\rm peak}$, fixed at the value 
$k_{\rm peak} \simeq 1.5\times 10^{14}$ Mpc$^{-1}$. This choice gives an abundance of PBHs peaked at 
$M_{\rm PBH} \simeq 5\times 10^{17}$ g, which is compatible with the possibility of having 100\% of dark matter in the form of PBHs. Moreover, it eliminates all those solutions, like the ones in fig.\,\ref{fig:CMBTension}, in which larger 
values of $n_s$ are obtained at the expense of a larger $k_{\rm peak}$ (and larger $\Delta N_e^{0.05}$).
We consider HDOs up to $\mathcal{N} = 8$, and check that our results remain stable if further  higher-order terms are added. 
In the right panel of fig.\,\ref{fig:HDO} we show, for each one of these solutions, the corresponding value of $n_s$. 
If $\Lambda^{-1}$ is too small, the impact of the HDOs is negligible and it is possible to have 100\% of dark matter in the form of PBHs only for values of the spectral index that 
are 3\,$\sigma$ away from the central value of Planck, as already discussed in section\,\ref{sec:Disc} and shown in fig.\,\ref{fig:CMBTension}.
However, by increasing the value of $\Lambda^{-1}$ without clashing against eq.\,(\ref{eq:OrderRelation}) (region shaded in gray), 
the small correction introduced at large $\phi$ gives values of the spectral index that are in perfect agreement with the current observational bounds.  
This is shown by the red solid line in fig.\,\ref{fig:HDO}.

Let us now consider specifically the solution marked by the cyan star which has $n_s \simeq 0.96$ and $\Lambda^{-1}\simeq 2\times 10^{-3}$. As specified before, we include in our analysis HDOs up to $\mathcal{N} = 8$ but for 
$\Lambda^{-1}\simeq 
2\times 10^{-3}$ it is possible to see that the first two with $\mathcal{N} = 5,6$ dominate over the remaining ones.
 The corresponding population of PBHs is shown in the left panel of fig.\,\ref{fig:Ka} while the induced gravitational wave signal is shown in the right panel of the same figure (cyan regions with dashed boundary in both cases). 
 The value of $a_4$ is fixed by the amplitude of the power spectrum at CMB scales, and we find $a_4 \simeq 10^{-10}$.
 Since we assumed $g^2 = a_4$, we have $g \simeq 10^{-5}$.
 From our discussion, it follows that for the mass scale $M = g\Lambda$ we have $M \simeq 10^{-2}$ in units of $M_{\rm Pl}$. 
 The condition $H< M$ is, therefore, verified since $H \sim \sqrt{a_4}/\xi \simeq 10^{-5}$. 
The same conclusion holds true for all solutions in the right panel of fig.\,\ref{fig:HDO}.
\begin{figure}[t]
\begin{center}
$$\includegraphics[width=.45\textwidth]{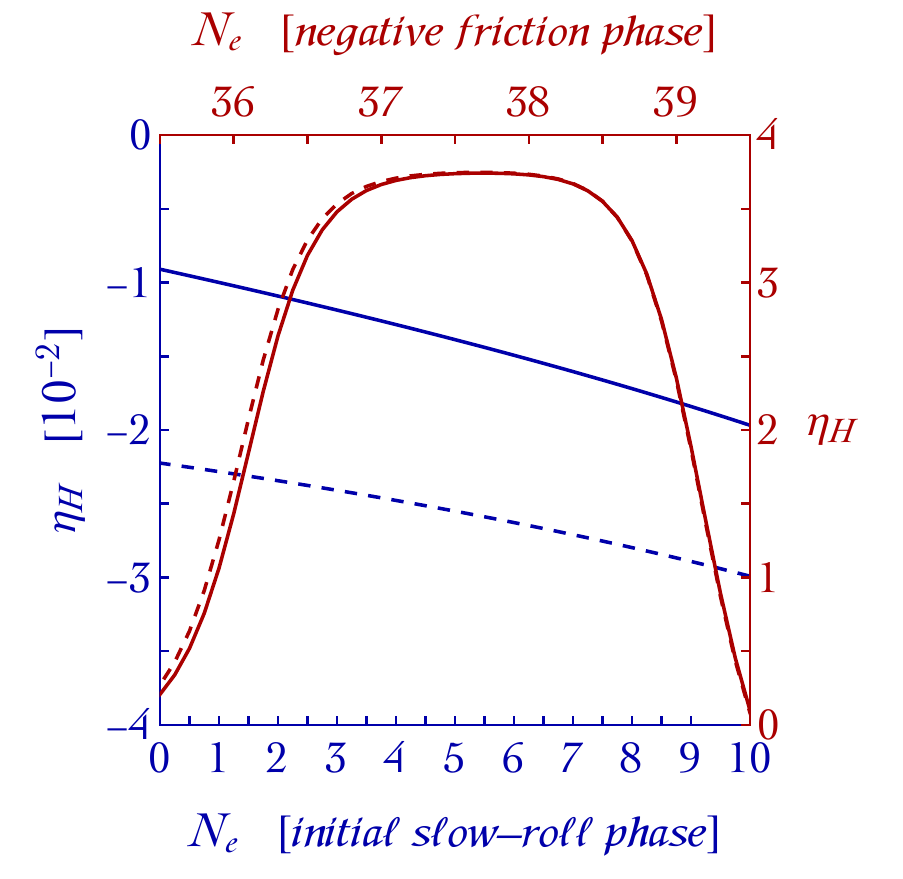}
\qquad\includegraphics[width=.45\textwidth]{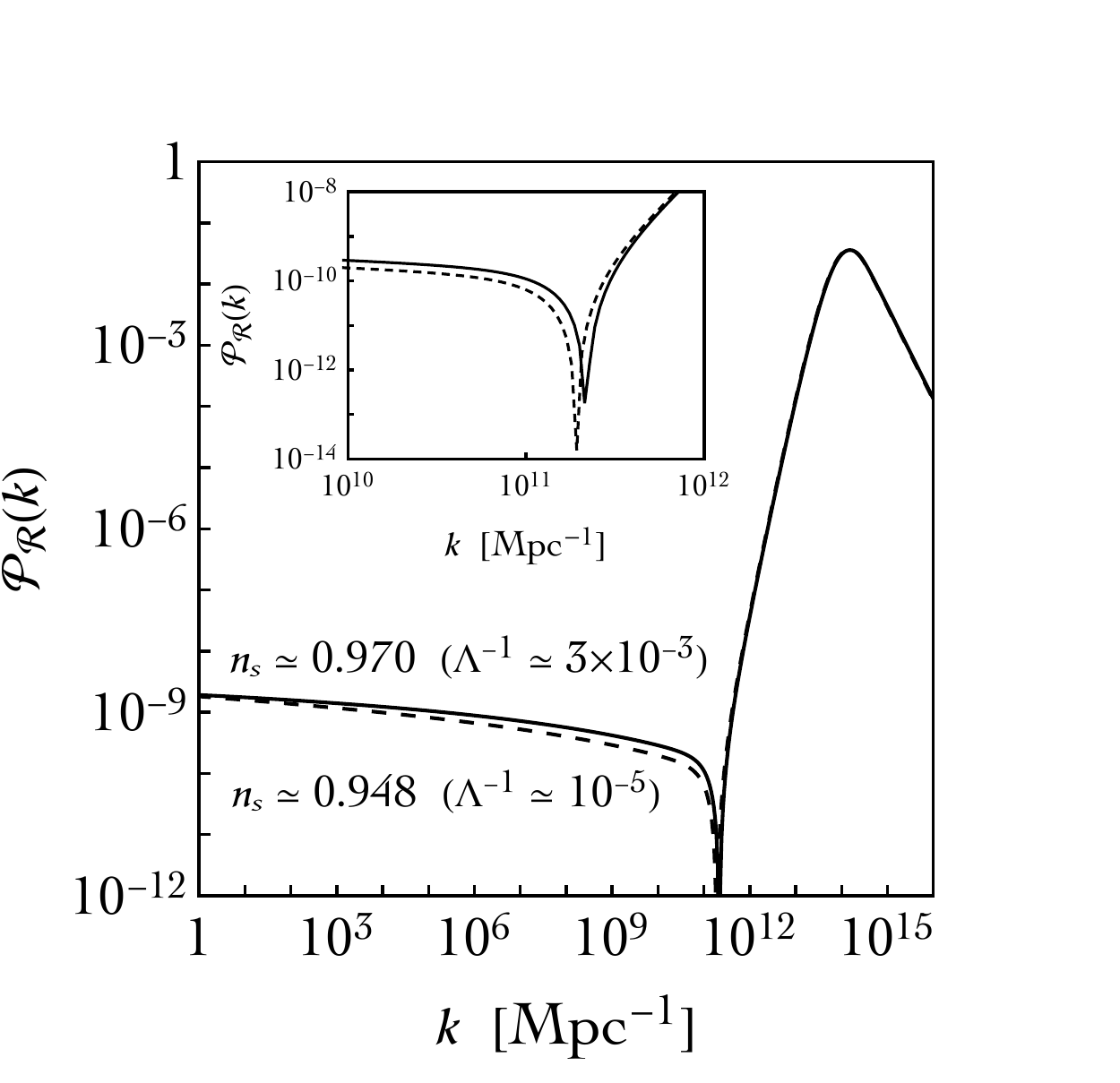}$$ 
\caption{\em \label{fig:HDOdyn} 
{\color{VioletRed4}{
Left panel. Hubble parameter $\eta_H$ as function of the number of $e$-folds. 
We consider two inflationary solutions (among those constructed in the right panel of fig.\,\ref{fig:HDO}) with $n_s \simeq 0.970$ (solid line) and $n_s \simeq 0.948$ (dashed line).
Right panel. Power spectra corresponding to the two solutions shown in the left panel.
}}
 }
\end{center}
\end{figure}

To better understand how the presence of HDOs solves the $n_s$ tension and gives an abundance of PBHs in the right mass window, it is instructive to look at the dynamics of the inflaton field in two extreme cases shown in the right panel of fig.\,\ref{fig:HDO}.
The former, with $n_s \simeq 0.948$, has been obtained with $\Lambda^{-1}\simeq 10^{-5}$ or lower and it is marked with a dashed line in fig.\,\ref{fig:HDOdyn}. The latter, with $n_s \simeq 0.970$ and obtained with $\Lambda^{-1}\simeq 3\times 10^{-3}$, is shown with a solid line in fig.\,\ref{fig:HDOdyn}. 
Four points are worth emphasizing.
\begin{itemize}

\item [{\small{1.}}] We first consider the Hubble parameter $\eta_H$ in the initial slow-roll phase.
This is shown in the left panel of fig.\,\ref{fig:HDOdyn} where on the left-side $y$-axis we plot in blue the values of $\eta_H$ as function of the first 10 $e$-folds of inflation (bottom-side $x$-axis with $N_e = 0$ corresponding to the time at which we fit the CMB observables). 
The modification of the potential induced by HDOs increases the value of $\eta_H$ towards less negative values.
This $O(1)$ change in $\eta_H$ is enough to modify, at the {\it percent} level, the value of $n_s$ from $n_s \simeq 0.948$ to $n_s \simeq 0.970$
 (we remind the reader that in the initial slow-roll phase we have $n_s = 1 -4\epsilon_H + 2\eta_H\simeq 1 + 2\eta_H$).
 
 \item [{\small{2.}}] 
 The power spectrum at large scales considerably flattens, going from $n_s \simeq 0.948$ to $n_s \simeq 0.970$.
 This can be appreciated in the right panel of fig.\,\ref{fig:HDOdyn}. As a consequence, the dip of the power spectrum
 shifts towards larger values of $k$. This is more evident in the inset plot, where we zoom in around the position of the dip of $\mathcal{P}_{\mathcal{R}}(k)$. 
 Keeping in mind the analytical discussion carried out in section\,\ref{sec:Ana},
 this is simply because the beginning of the negative friction phase gets shifted to larger $N_{\rm in}$. Consequently, assuming $\Theta$ fixed, for the same $\epsilon_k$ the value of $N_0$ also increases.
 
\item [{\small{3.}}] The shift in the position of the dip of $\mathcal{P}_{\mathcal{R}}(k)$, for a negative friction phase of the same duration, would also shift the position of the peak of $\mathcal{P}_{\mathcal{R}}(k)$ towards larger $k$, producing PBHs with unacceptably small mass. 
In light of the analytical discussion carried out in section\,\ref{sec:Ana}, 
this happens because for a negative friction phase of the same duration, shifting $N_{\rm in}$ towards larger values will also shift $N_{\rm end}$ which, in turn, controls the position of the peak.
 However, it is possible to compensate the shift in the position of the dip without changing the position of the peak at the prize of a shorter negative friction phase.
 As shown in appendix\,\ref{app:C}, we have the approximate analytical relation 
 $k_{\rm dip}/k_{\rm peak}\propto e^{-5\Delta \Theta_{<}/2}$ where $k_{\rm dip}$ 
 is the position of the dip of the power spectrum and $\Delta\Theta_{<} \equiv N_{\rm end} - N_{\rm in}$ is the duration of the negative friction phase.
 Since all our solutions in fig.\,\ref{fig:HDO} are characterized, by construction, by the same value of $k_{\rm peak}$, 
 the previous relation implies that we need a smaller 
 $\Delta\Theta_{<}$ to compensate the larger value of $k_{\rm dip}$. 
 This is indeed confirmed in the left panel of fig.\,\ref{fig:HDOdyn} where 
 in the right-side $y$-axis
 we plot in red the value of $\eta_H$ at around the negative friction phase (top-side $x$-axis). 
 Numerically, we find $\Delta\Theta_{<}\simeq 3.140$ for $n_s \simeq 0.948$ and $\Delta\Theta_{<} \simeq 3.091$ for 
$n_s \simeq 0.970$ and the relation $k_{\rm dip}/k_{\rm peak}\propto e^{-5\Delta\Theta_{<}/2}$ is verified 
since we have  $k_{\rm dip} \simeq 1.9\times 10^{11}$ Mpc$^{-1}$  for $n_s \simeq 0.948$ and 
$k_{\rm dip} \simeq 2.1\times 10^{11}$ Mpc$^{-1}$  for $n_s \simeq 0.970$. 
Concretely, a shorter phase during which the power spectrum grows is obtained by means 
of a steeper slope of the latter as a function of $k$. This behavior can be seen in the inset plot 
in the right panel of fig.\,\ref{fig:HDOdyn}, especially where $\mathcal{P}_{\mathcal{R}}(k)$ starts to rise just after the dip.

\item [{\small{4.}}] Without including HDOs, we do not have the freedom to change the potential at large field values.
Consequently, the simplest way to get a larger $n_s$ is to fit CMB observables at larger field values.
If we assume the renormalizable potential to be dominated by the quartic term, it is indeed possible to obtain, by means of the slow-roll approximation, the qualitative scaling $n_s = 1 - 16/[(1+6\xi)\phi_{\rm CMB}^2] + O(1/\phi_{\rm CMB}^3)$.
However, increasing $\phi_{\rm CMB}$ increases, in turn, the value of $\Delta N_e^{0.05}$ and shifts the peak of the power spectrum towards smaller scales. This qualitative behavior characterizes all solutions in fig.\,\ref{fig:CMBTension} which produce a sizable abundance of PBHs.
Although we do not have a solid-rock mathematical proof, our numerical scan suggests that this is a generic feature of the renormalizable model.
\end{itemize}
These simple points make clear that the $n_s$ tension can  be  solved with HDOs while obtaining 
PBHs as the totality of dark matter in perfect agreement with observational constraints.

In the second example that we consider we take $c_n = (-1)^{n+1}$. This choice can be justified, for instance, 
in the context of toy ultraviolet completions as the one discussed in appendix\,\ref{app:B} (see in particular eq.\,(\ref{eq:ExplicitOn})). 
We perform the same analysis discussed before, and our result is shown by the red dashed line in fig.\,\ref{fig:HDO}. 
Also in this case the series of HDOs can fix the $n_s$ tension. For large values of $\Lambda$, the impact of HDOs is negligible, as in the case with $c_n = 1$. Larger values of $\Lambda^{-1}$, in comparison to the case $c_n = 1$, are needed in order to obtain the same 
$n_s$. The reason is that in the case $c_n = (-1)^{n+1}$ the alternating signs cause a partial cancellation of the HDOs.

It is worth recalling now that the perturbative unitarity breaking scale for the potential of \eq{eq:MasterPot} is $\Lambda_{U4}=1/\xi$ \cite{Burgess:2009ea,Barbon:2009ya}. New dynamics must arise at a scale lower or equal than $\Lambda_{U4}$ in order to restore unitarity. The values of $\xi\sim O(0.1)$ that we find in our solutions  push $\Lambda_{U4}$ formally above $M_{\rm Pl}$. It is thus tantalizing that $M \sim 10^{-2} \ll\Lambda_{U4}$, as we can speculate with the possibility that the new states arising at the mass scale $M$ (and the corresponding HDOs) may harbinger a UV completion ensuring unitarity beyond $\Lambda_{U4}$.

Although simplistic, the numerical analysis carried out in this section shows that small corrections to the inflaton potential generated by 
HDOs --whose presence, in particular in the context of large-field models of inflation, has no reason to be neglected-- have the capability to fix the 
$n_s$ tension pointed out in section\,\ref{sec:Disc}. 
Before concluding, it is worth mentioning some possible extension of our analysis. 
First, the condition $g^2 = a_4$, that we imposed for simplicity, has no fundamental reason to be true, and 
relaxing it would open an additional direction in the parameter space $(g,\Lambda)$ that would be interesting to explore.
Second, one can in principle add HDOs that include field derivatives; for instance an effective operator of this kind with canonical mass-dimension $d$ would have the 
general form 
$\mathcal{O}_d^{\partial} = (1/\Lambda_{\partial}^{d-4})\partial^{n_{\partial}}\phi^{n_{\phi}}$ with $d= n_{\partial} +n_{\phi}$. 
In light of our discussion, the relevance of these operators could be related to the fact that we expect for the suppression 
scale $\Lambda_{\partial}$ a different scaling 
in terms of powers of $M$ and $g$ according to $1/\Lambda_{\partial} = g^{n_{\phi}-2/d-4}/M$.

%%%%%%%%%%%%%%%%%%%%%%%%%%%%%%%%%%%%%%%%%%%%%%%%%%%%%%%%%%%%%%%%%%%%
\section{Summing up}\label{sec:Sum}
%%%%%%%%%%%%%%%%%%%%%%%%%%%%%%%%%%%%%%%%%%%%%%%%%%%%%%%%%%%%%%%%%%%%

In this paper we have analyzed the possibility to generate PBHs from single-field inflation with a quartic polynomial potential and a non-minimal coupling to gravity. We conclude with a summary list of our findings.

\begin{itemize}

\item [$\circ$] Single-field inflation with the simple potential $V(\phi) = a_2\phi^2 + a_3\phi^3 + a_4\phi^4$ and the general non-minimal coupling to gravity $ \xi\,\sqrt{-g}\, \phi^2R$ (in addition to the usual Einstein-Hilbert term) allows the formation of an  abundant population of PBHs in the window $10^{17} \lesssim M_{\rm PBH}\,[\,{\rm g}\,] \lesssim 10^{21}$, which can comprise the totality of the dark matter. However, the possible values of $n_s$ at $0.05$~Mpc$^{-1}$ are smaller than indicated  by  the latest Planck analysis for the standard  6-parameters Base $\Lambda$CDM model, which  amounts to a $\sim 3\sigma$ tension. This discrepancy is a rather common feature of PBH inflationary models with  a near-stationary inflection point.\footnote{The same applies to the radiative inflation model of \cite{Ballesteros:2017fsr}, but see \cite{Ballesteros:2019hus} for an exception.}

\item [$\circ$] Extensions of the Base $\Lambda$CDM model can alleviate the tension with $n_s$. In particular, values of $N_{\rm eff}$ smaller than the standard $3.046$ can allow $n_s$ to be closer to 0.95.

\item [$\circ$] A well-motivated fix for this $n_s$ tension is obtained by considering the simplest extension of the potential $V(\phi)$, with the inclusion higher-order dimensional operators which are naturally  suppressed. Remarkably, the PBH dark matter predictions are robust under the addition of these operators. Indeed, a small five-dimensional term in the potential is already sufficient to enhance $n_s$, eliminating the tension while producing $f_{\rm PBH}\simeq 1$ in the above window and enough inflation.

\item [$\circ$] The stochastic gravitational wave background induced by the curvature fluctuations in the examples with $f_{\rm PBH}\sim$~1 could be observed in the future by LISA and DECIGO. Moreover, this signal is orders of magnitude larger than the expected background from astrophysical  binary   black hole and neutron star mergers. Examples producing light PBHs which would have evaporated by now, but that are nonetheless in  agreement with BBN  bounds, generate a stochastic background of gravitational  waves that may be observed by Advanced LIGO.

\end{itemize}

%%%%%%%%%%%%%%%%%%%%%%%%
\begin{acknowledgments}
We thank José Ramón Espinosa and Piero Ullio for discussions and Hasti Khoraminezhad for discussions and help with Planck data. The work of G.B.\ and J.R.\ is funded by a {\it Contrato de Atracci\'on de Talento (Modalidad 1) de la Comunidad de Madrid} (Spain), with number 2017-T1/TIC-5520, by {\it MINECO} (Spain) under contract FPA2016-78022-P, {\it MCIU} (Spain) through contract PGC2018-096646-A-I00 and by the IFT UAM-CSIC Centro de Excelencia Severo Ochoa SEV-2016-0597 grant.
The research of A.U. is supported in part by the MIUR under contract 2017\,FMJFMW (``{New Avenues in Strong Dynamics},'' PRIN\,2017) 
and by the INFN grant 
``SESAMO -- 
SinergiE di SApore e Materia Oscura.''
M.T. acknowledges support from the INFN grant ``LINDARK,'' the research grant ``The Dark Universe: A Synergic Multimessenger Approach No. 2017X7X85'' funded by MIUR, and the project ``Theoretical Astroparticle Physics (TAsP)'' funded by the INFN. G.B. thanks KITP for hospitality during the final stage of this work and support from the National Science Foundation under Grant No. NSF-1748958. 
\end{acknowledgments}

\appendix

\section{General polynomial potential with an approximate stationary inflection point}\label{app:A}

Consider the scalar potential
\begin{align}
\tilde{V}(\phi) = \frac{a_4}{(1+\xi \phi^2)^2}\left(\tilde{a}_2\phi^2 + \tilde{a}_3\phi^3 + \phi^4 + \sum_{n=5}^{\mathcal{N}}\tilde{a}_n\phi^n\right)\,,
\end{align} 
with $\tilde{a}_{i} \equiv a_{i}/a_4$.
We impose the condition that $\tilde{V}(\phi)$ must have an exact stationary inflection point at some field value $\phi = \phi_0$, meaning that $\tilde{V}^{\prime}(\phi_0)=\tilde{V}^{\prime\prime}(\phi_0)=0$.
It is possible to recast these two conditions in terms of the coefficients $\tilde{a}_{i=2,3}$.
Some trivial algebra then gives the following potential
\begin{align}\label{eq:MasterPot1}
\tilde{V}(\phi) = & \frac{a_4\phi^4}{(1+\xi \phi^2)^2(3 + \xi^2 \phi_0^4)}\times\nn \\
&\bigg\{
\frac{\phi_0^2}{\phi^2}\bigg[
2(3+\xi\phi_0^2) +  \sum_{n=5}^{\mathcal{N}}\tilde{a}_n \mathcal{F}_n(\phi_0,\xi)
\bigg] + \frac{\phi_0}{\phi}\bigg[
-8 +   \sum_{n=5}^{\mathcal{N}}\tilde{a}_n \mathcal{G}_n(\phi_0,\xi)
\bigg] + \big(3+\xi^2\phi_0^4\big) \bigg[1 +  \sum_{n=5}^{\mathcal{N}}\tilde{a}_n\phi^{n-4}\bigg]
\bigg\}\,,
\end{align}
where
\begin{align}
\mathcal{F}_n(\phi_0,\xi) & \equiv  \frac{3}{2}n(n-3)\phi_0^{n-4} + (6-5n+n^2)\xi\phi_0^{n-2} - \frac{1}{2}(n-4)(n-3)\xi^2\phi_0^n\,, \\
\mathcal{G}_n(\phi_0,\xi) & \equiv (2-n)n\phi_0^{n-4} + (8-6n + n^2)\xi^2\phi_0^n\,.
\end{align}
The coefficients $\tilde{a}_3$ and $\tilde{a}_2$ are given by the following expressions
\begin{align}
\tilde{a}_2 & = \frac{2\phi_0^2(3+\xi \phi_0^2)}{3+\xi^2\phi_0^4} + \frac{1}{3+\xi^2\phi_0^4}\sum_{n=5}^{\mathcal{N}}
\tilde{a}_n\left[
\frac{3}{2}n(n-3)\phi_0^{n-2} + (6-5n+n^2)\xi\phi_0^n + \left(-6 + \frac{7n}{2}- \frac{n^2}{2}\right)\xi^2 \phi_0^{n+2}
\right]\,,\\
\tilde{a}_3 & = \frac{6\phi_0(-1+\xi \phi_0^2)}{3-8\xi \phi_0^2 + \xi^2 \phi_0^5} + 
\frac{\tilde{a}_2(-1+8\xi \phi_0^2 - 3\xi^2 \phi_0^4)}{3-8\xi \phi_0^2 + \xi^2 \phi_0^5} \nn \\ & \quad +
\frac{1}{3-8\xi \phi_0^2 + \xi^2 \phi_0^5}\sum_{n=5}^{\mathcal{N}}\tilde{a}_n\left[
\frac{1}{2}(1-n)n \phi_0^{n-2} + (2+5n -n^2)\xi\phi_0^n 
-\frac{1}{2}(n-5)(n-4)\xi^2\phi_0^{n+2}
\right]\,.
\end{align}
Without loss of generality, we define $\lambda/4! \equiv a_4/(3 + \xi^2 \phi_0^4)$.
A slight deformation of the the exact stationary inflection point in eq.\,(\ref{eq:MasterPot2}) can be described introducing
two small parameters $c_{2,3}$ defined in such a way that our final ansatz for the inflaton potential is
\begin{align}\label{eq:MasterPot2}
&\tilde{V}(\phi) = \frac{\lambda\phi^4}{4!(1+\xi \phi^2)^2}\times\nn \\
&\bigg\{
\frac{\phi_0^2}{\phi^2}(1+c_2)\bigg[
2(3+\xi\phi_0^2) + \sum_{n=5}^{\mathcal{N}}\tilde{a}_n \mathcal{F}_n(\phi_0,\xi)
\bigg] - \frac{\phi_0}{\phi}(1+c_3)\bigg[
8 -   \sum_{n=5}^{\mathcal{N}}\tilde{a}_n \mathcal{G}_n(\phi_0,\xi)
\bigg] + \big(3+\xi^2\phi_0^4\big) \bigg[1 +  \sum_{n=5}^{\mathcal{N}}\tilde{a}_n\phi^{n-4}\bigg]
\bigg\}\,.
\end{align}
If we restrict this  expression to the renormalizable case, we have (see ref.\,\cite{Ballesteros:2017fsr})
\begin{align}
\tilde{V}(\phi) = \frac{\lambda \phi^4}{4!(1+\xi \phi^2)^2}\left[
3 + \xi^2 \phi_0^4 -8(1+c_3)\frac{\phi_0}{\phi} +
2(1+c_2)(3 + \xi\phi_0^2)\frac{\phi_0^2}{\phi^2} 
\right]~~= \resizebox{50mm}{!}{
\parbox{25mm}{
\begin{tikzpicture}[]
\node (label) at (0,0)[draw=white]{ 
       {\fd{2.8cm}{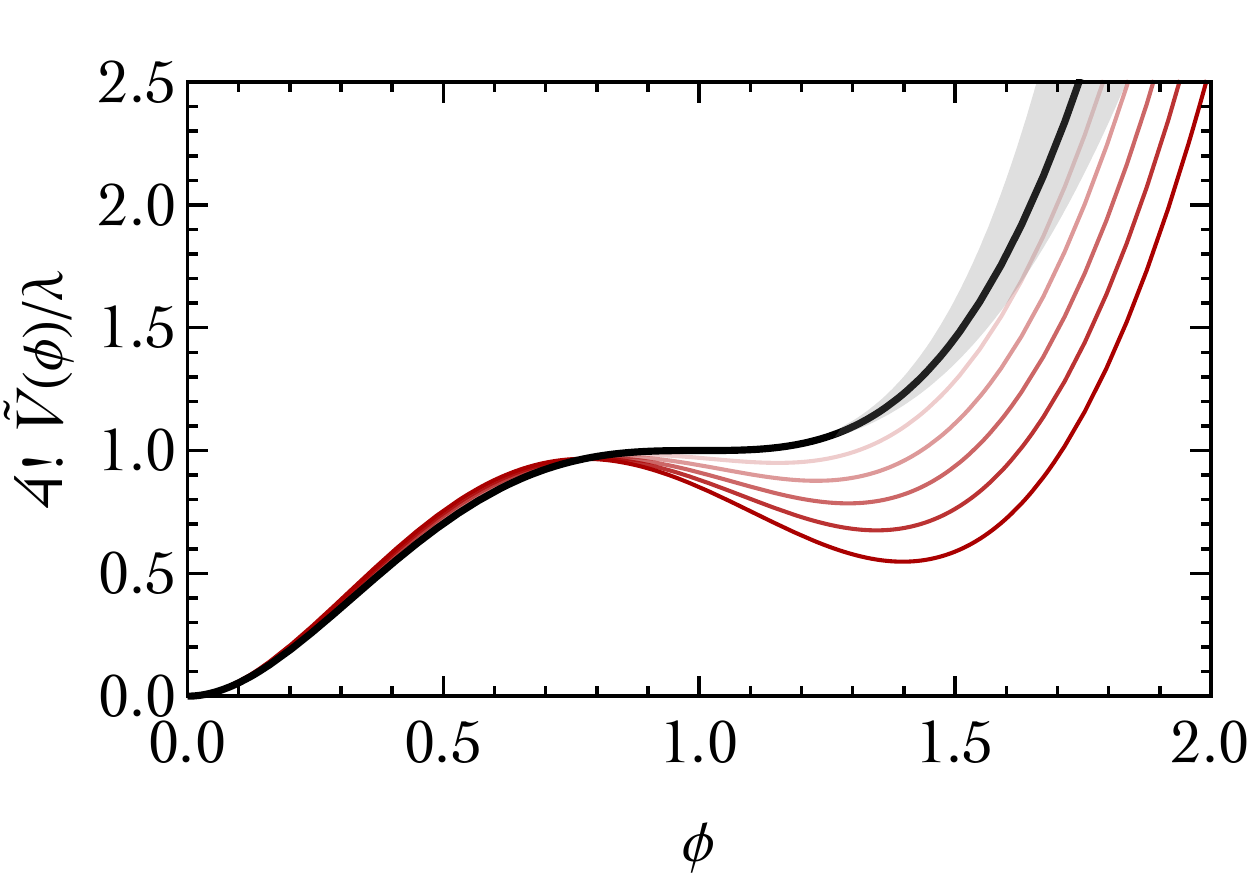}} 
      };
\end{tikzpicture}
}}\nonumber
\end{align}
where in the inset plot we show (for $\phi_0 = 1$ and $\xi = 0.1$) the shape of the potential in the presence of an exact 
stationary inflection point ($c_2 = c_3 =0$, solid black line) and for a slight deformation of it ($0.02 \leqslant c_2 = c_3 \leqslant 0.1$, red lines with 
increasing tonality of red). The region shaded in light gray 
illustrates the effect of changing $\xi$ in  the range {$\xi = 0.1 \pm 0.75$ (while keeping $c_2 = c_3 = 0$), with
larger values of $\xi$ corresponding to milder slopes of the potential. Increasing the value of $\xi$, therefore, has the effect of reducing the classical velocity with which the inflaton field reaches the approximate stationary inflection point. Consequently, being the classical velocity smaller, quantum fluctuations} play {a more prominent role increasing the height of the peak of the primordial curvature power spectrum. However, too large values of $\xi$ eventually trap the inflaton field in the local minimum.}

\section{Power counting and higher-dimensional operators}\label{app:B}

Power counting is a simple but powerful method to organize HDOs\,\cite{Manohar:1983md,Cohen:1997rt,Luty:1997fk,Giudice:2007fh}.
To this end, it is 
convenient to restore the appropriate powers of $\hbar$ (while keeping $c = \epsilon_0 = 1$, with $c$ the speed of light 
 and $\epsilon_0$ the vacuum permittivity). 
 Equivalently, we can introduce  a unit of energy $\mathrm{E}$  and length $\mathrm{L}$, with $[\hbar] = \mathrm{E\,L}$ (natural units correspond to $\mathrm{E} = \mathrm{L}^{-1}$ with $\hbar$ dimensionless). 
The main advantage of this way of counting is that, in contrast to the case of natural units, couplings, 
and not only masses, are dimensionful quantities. It is, therefore, easy to keep track of the correct scaling in terms of couplings that are needed in a given expression just  from dimensional arguments. 
 
 A canonically normalized scalar field has dimension 
 $[\phi] = \mathrm{E}^{1/2}\mathrm{L}^{-1/2}$ (this can be seen 
 using the fact that a generic Lagrangian density 
 has dimension $[\mathcal{L}] = \mathrm{E}\,\mathrm{L}^{-3}$). Similarly, the dimension 
 of a coupling constant (like a gauge coupling) is $[\mathrm{g}] =  \mathrm{E}^{-1/2}\mathrm{L}^{-1/2}$
 while for a scalar quartic (like $a_4$ in our parametrization) we have $[\lambda] =  \mathrm{E}^{-1}\mathrm{L}^{-1}$ 
 (and $[\mathrm{g}^2] = [\lambda]$, thus justifying our comment below eq.\,(\ref{eq:EffOp2}) about the dimensionality of the
 quartic  coefficient $a_4$). 
The Planck scale, as the electroweak scale or the scale $\Lambda$ introduced in eq.\,(\ref{eq:ExplicitScal}), has dimension $[\Lambda] =  \mathrm{E}^{1/2}\mathrm{L}^{-1/2}$. 
This way of counting, as anticipated before, shows that couplings are dimensionful quantities. 
It is, therefore, useful to introduce --instead of $\mathrm{E}$ and $\mathrm{L}$-- units of mass $\mathrm{M} \equiv  \mathrm{L}^{-1}$ and coupling $\mathrm{C} \equiv \mathrm{E}^{-1/2}\mathrm{L}^{-1/2}$\,\cite{Panico:2015jxa}. 
Consequently, we have $[\mathcal{L}] = \mathrm{M}^4\mathrm{C}^{-2}$ and $[\phi] = \mathrm{M}\mathrm{C}^{-1}$. 
Furthermore, notice that a loop diagram is always accompanied by the factor $\kappa \equiv \hbar/(4\pi)^2$ which has dimension of an inverse coupling squared $[\kappa] = \mathrm{C}^{-2}$. 
For the operator in eq.\,(\ref{eq:EffOp}) we have 
\begin{align}
[\mathcal{O}_n] = \left[
\frac{1}{\Lambda_n}
\right]^{n-4}\mathrm{M}^n\mathrm{C}^{-n}\overset{!}{=} \mathrm{M}^4\mathrm{C}^{-2}~~~~\Longrightarrow~~~~
\left[
\frac{1}{\Lambda_n}
\right] = \frac{\mathrm{C}^{\frac{n-2}{n-4}}}{\mathrm{M}}\,.
\end{align}
This scaling can be obtained in two ways (a third one, involving field derivatives, 
was briefly discussed at the end of section\,\ref{sec:HDO})
\begin{align}\label{eq:ScalingOneCouplOneScale}
\mathcal{O}_n = \frac{\mathrm{C}^{n-2}}{\mathrm{M}^{n-4}}\phi^n =
\left\{
\begin{array}{ccc}
 \frac{g^{n-2}}{M^{n-4}}\phi^n &\equiv& \mathcal{O}_n^{\rm tree}  \\
  &     \\
\frac{\kappa g^{n}}{M^{n-4}}\phi^n  &\equiv& \mathcal{O}_n^{\rm loop} 
\end{array}
\right.
\end{align}
where in $\mathcal{O}_n^{\rm tree}$ we genuinely use $g$ to keep track of the powers of coupling $\mathrm{C}$ 
and $M$ to keep track of the powers of mass $\mathrm{M}$ while 
$\mathcal{O}_n^{\rm loop}$ pays the prize of the one-loop suppression $\kappa$ that is 
compensated by the presence of two additional powers of $g$.
We have $\mathcal{O}_n^{\rm loop}/\mathcal{O}_n^{\rm tree} = \kappa g^2$, and we expect 
$\mathcal{O}_n^{\rm loop} \ll \mathcal{O}_n^{\rm tree}$ 
in a weakly coupled theory with $g \ll 4\pi$. We consider only operators with scaling $\mathcal{O}_n^{\rm tree}$ 
in eq.\,(\ref{eq:ExplicitScal}).

An explicit example --that, needless to say it, is just illustrative and by no means intends to give a comprehensive description of the ultraviolet theory-- clarifies the distinction between the two classes of operators. 
Consider the interactions of $\phi$ with a scalar field $\Phi$ whose dynamics is described by the Lagrangian density
 \begin{align}\label{eq:LagrPhi}
 \mathcal{L}[\Phi] = \frac{1}{2}(\partial_{\mu}\Phi)(\partial^{\mu}\Phi) - V(\Phi,\phi)\,,~~~~~~~~
V(\Phi,\phi) =\frac{1}{2}M^2\Phi^2 + \frac{c_1}{2}gM\phi\,\Phi^2 + \frac{c_2}{2} gM\phi^2\Phi\,,
 \end{align}
 where the coefficient of each term is a proper combination 
of $g$ and $M$ fixed --as explained before, and modulo $O(1)$ coefficients $c_{1,2}$-- by dimensional analysis.
This potential produces, upon integrating out $\Phi$, the operators $\mathcal{O}_n^{\rm tree}$ 
 that arise from the geometric series generated by the expansion of the $\Phi$ propagator.
  For instance, 
 for $n = 5$ we have (we include a schematic diagram to better visualize how the power counting of
  $g$, $M$, $\phi$ works)
 \begin{align}
\resizebox{40mm}{!}{
\parbox{35mm}{
\begin{tikzpicture}[]
\node (label) at (0,0)[draw=white]{ 
       {\fd{2.75cm}{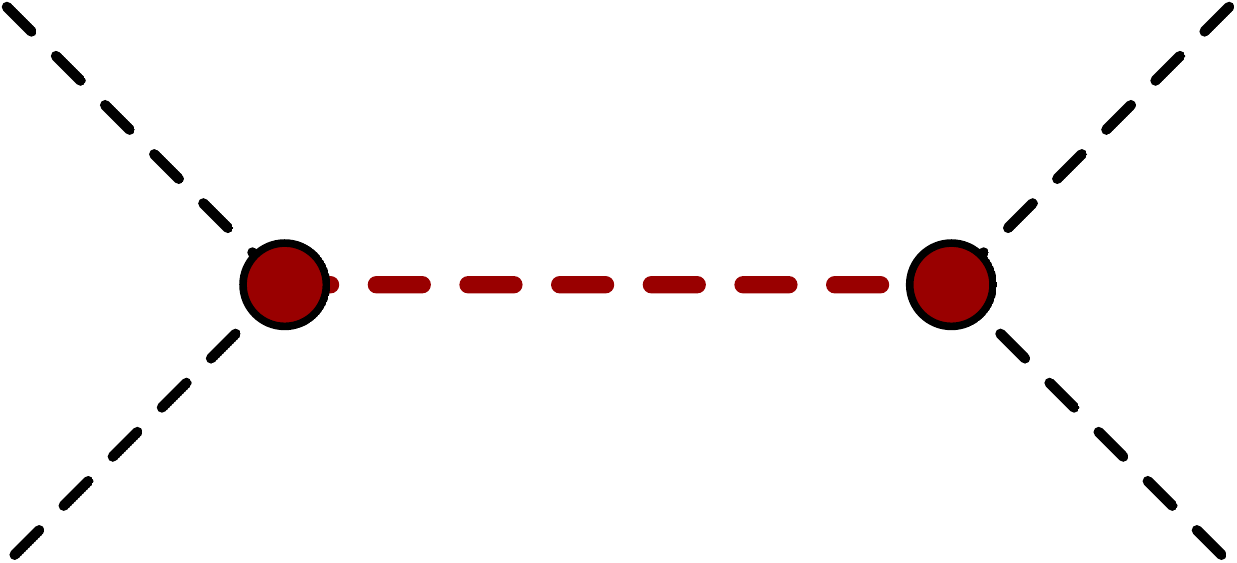}} 
      };
\node[anchor=east] at (-1.3,0.48) {$\phi$};
\node[anchor=east] at (-1.3,-0.48) {$\phi$};
\node[anchor=west] at (1.3,0.48) {$\phi$};
\node[anchor=west] at (1.3,-0.48) {$\phi$};
\node[anchor=south] at (0.,-0.65) {{\color{oucrimsonred}{\scalebox{0.85}{$\frac{1}{M^2 + g\phi M}$}}}};
\node[anchor=south] at (-0.6,0) {{\color{oucrimsonred}{\scalebox{0.65}{$gM$}}}};
\node[anchor=south] at (0.7,0) {{\color{oucrimsonred}{\scalebox{0.65}{$gM$}}}};
\end{tikzpicture}
}} ~~~~~~\overset{M \gg g\phi}{\Longrightarrow}~~~~~~ 
\resizebox{23mm}{!}{
\parbox{35mm}{
\begin{tikzpicture}[]
\node (label) at (0,0)[draw=white]{ 
       {\fd{2.75cm}{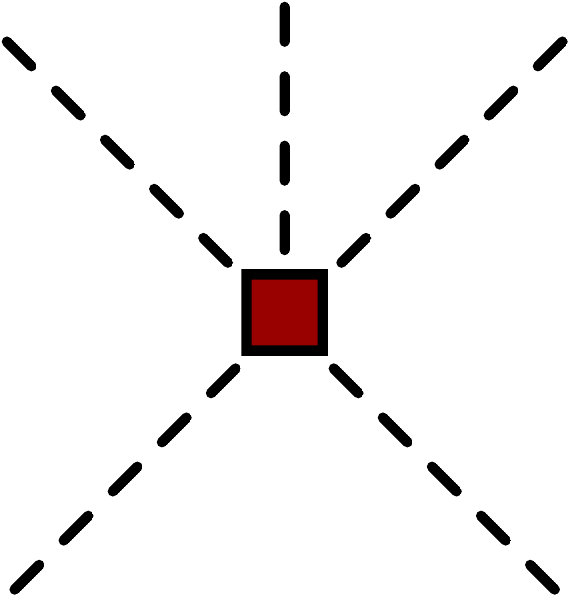}} 
      };
      \node[anchor=south] at (-1.2,-0.3) {{\color{oucrimsonred}{\scalebox{1.4}{$g^3/M$}}}};
      \node[anchor=east] at (-1.3,1) {\scalebox{1.65}{$\phi$}};
      \node[anchor=west] at (1.3,1) {\scalebox{1.65}{$\phi$}};
      \node[anchor=east] at (-1.3,-1) {\scalebox{1.65}{$\phi$}};
      \node[anchor=west] at (1.3,-1) {\scalebox{1.65}{$\phi$}};
      \node[anchor=north] at (0.25,1.5) {\scalebox{1.65}{$\phi$}};
\end{tikzpicture}
}} ~~~ = ~~
 \mathcal{O}_5^{\rm tree}
   = \frac{g^3}{M}\phi^5\,,
\end{align}
where each trilinear elementary vertex, marked by a red dot on the left-side diagram, contributes with a power of $gM$.
In the diagrammatic approach, the HDOs arise from the contraction of heavy propagators inside diagrams of the full theory.
In more detail, the solution of the classical equation of motion for $\Phi$ is 
 \begin{align}
\Phi_{\rm EoM} = (\Box + M^2 + c_1 gM\phi)^{-1}(-c_2 gM \phi^2/2) \simeq \left[
\frac{1}{M^2} - \frac{(\Box + c_1 gM\phi)}{M^4} +
\frac{(\Box + c_1 gM\phi)^2}{M^6}+ \dots
\right](-c_2 gM \phi^2/2)\,.
 \end{align}
 This solution, plugged back into eq.\,(\ref{eq:LagrPhi}), generates the effective Lagrangian
\begin{align}
\mathcal{L}[\Phi_{\rm EoM}(\phi)] & = \frac{c_2^2}{8}g^2M^2\phi^2\left[
\frac{1}{M^2} - \frac{(\Box + c_1 gM\phi)}{M^4} + \frac{(\Box + c_1 gM\phi)^2}{M^6}+ \dots
\right]\phi^2\nn\\
& = \frac{c_2^2}{8}g^2\phi^4 - \frac{c_1 c_2^2}{8}\frac{g^3}{M}\phi^5 + \frac{c_1^2c_2^2}{8}\frac{g^4}{M^2}\phi^6 + \dots 
+ {\rm operators\,with\,derivatives}\nn\\
& = \sum_{n = 4}^{\infty}\frac{(-1)^n c_1^{n-4}c_2^2}{8}\frac{g^{n-2}}{M^{n-4}}\phi^n + {\rm operators\,with\,derivatives} \,,
\label{eq:EffLex}
 \end{align}
 so that we have non-derivative HDOs (with an extra overall -1 since written in terms of the potential for $\phi$)
 \begin{align}\label{eq:ExplicitOn}
 \mathcal{O}^{\rm tree}_{n \geqslant 4} = \frac{(-1)^{n+1} c_1^{n-4}c_2^2}{8}\frac{g^{n-2}}{M^{n-4}}\phi^n\,.
  \end{align}
 Some comments are in order. 
 The effective Lagrangian in eq.\,(\ref{eq:EffLex}) contains the whole tower of operators $\mathcal{O}_n^{\rm tree}$
  starting from the dimension-four operator 
 $\mathcal{O}_{4}^{\rm tree} = g^2\phi^4$.
 This is  not in contradiction with the decoupling theorem. 
 This is because the dimensionful scalar trilinear coupling, as a consequence of our working assumptions, 
 has the structure $g M \phi^2\Phi$, and does not decouple in the limit $M\to \infty $\,\cite{Aoki:1997er}.
The equality $g^2 = a_4$ that we assumed for simplicity in our analysis in section\,\ref{sec:HDO}, therefore, 
is compatible with the possibility that the quartic term $a_4\phi^4$ is generated by the ultraviolet dynamics rather than 
introduced by hand in our renormalizable potential.
Moreover, we see that integrating out $\Phi$ also generates HDOs with derivatives of $\phi$.
These operators, that should be in general included, 
were neglected in our analysis in section\,\ref{sec:HDO}. 
This can be justified based on the fact that in the slow-roll regime (where the effect of the HDOs is more relevant 
for our analysis) terms 
with two derivatives of $\phi$ can be neglected while slow-roll violation occurs at small field values where the effects of HDOs is less relevant since they are more suppressed. Nevertheless, we reiterate that it would be interesting to 
perform a more general analysis that includes also derivative operators.
Finally, the explicit construction shows that the expansion in terms of effective operators 
is valid as long as $M \gg g\phi$. This is precisely the condition that led to eq.\,(\ref{eq:ConceptualBound}) where we 
introduced the scale $\Lambda = M/g$. 
Notice that in our way of counting the product $g\phi$ has dimension of mass, and the ratio $g\phi/M$, therefore, is a genuine dimensionless expansion parameter.

 Together with the tree level operators just discussed, we also have loop-generated operators 
 of the kind
  \begin{align}
\mathcal{O}_5^{\rm loop}
 ~~~ =
\resizebox{40mm}{!}{
\parbox{35mm}{
\begin{tikzpicture}[]
\node (label) at (0,0)[draw=white]{ 
       {\fd{2.75cm}{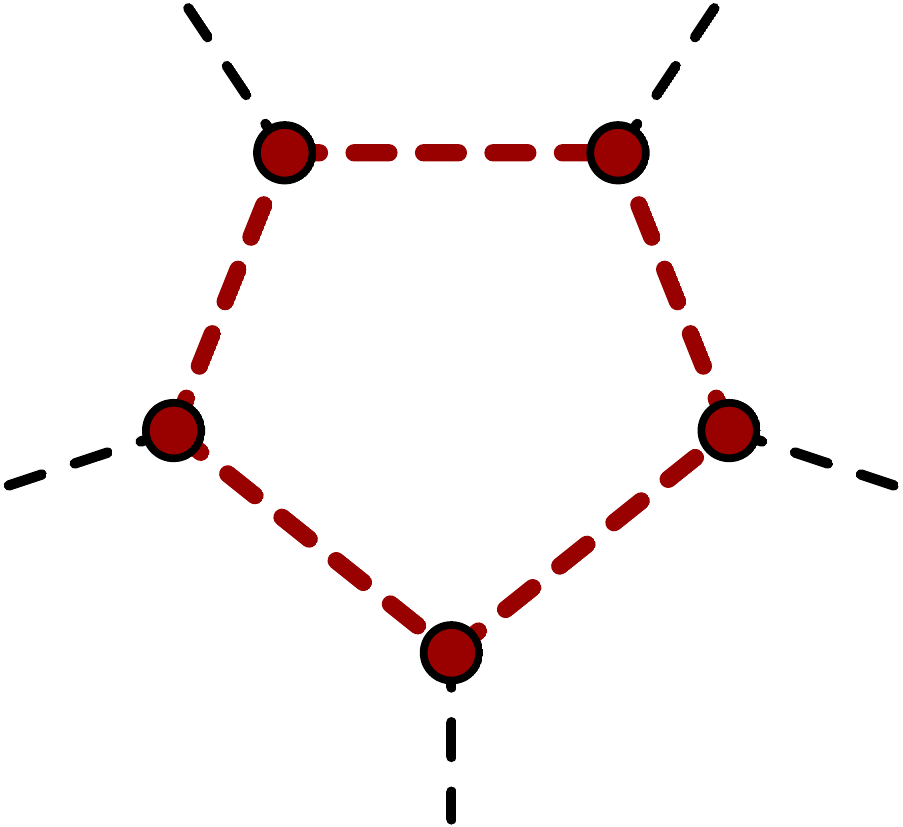}} 
      };
      \node[anchor=east] at (-1.3,0.) {$\phi$};
      \node[anchor=west] at (1.3,0.) {$\phi$};
      \node[anchor=east] at (-0.2,-1.) {$\phi$};
      \node[anchor=east] at (0.55,0.2) {{\color{oucrimsonred}{\scalebox{0.8}{$\kappa/M^6$}}}};
      \node[anchor=east] at (-.8,1.2) {$\phi$};
      \node[anchor=west] at (.8,1.2) {$\phi$};
\node[anchor=east] at (-0.75,0.2) {{\color{oucrimsonred}{\scalebox{0.65}{$gM$}}}}; 
\node[anchor=west] at (0.75,0.2) {{\color{oucrimsonred}{\scalebox{0.65}{$gM$}}}};  
\node[anchor=east] at (-0.55,0.72) {{\color{oucrimsonred}{\scalebox{0.65}{$gM$}}}}; 
\node[anchor=west] at (0.55,0.72) {{\color{oucrimsonred}{\scalebox{0.65}{$gM$}}}};
\node[anchor=south] at (0.,-0.65) {{\color{oucrimsonred}{\scalebox{0.65}{$gM$}}}};      
\end{tikzpicture}
}} = \frac{\kappa g^5}{M}\phi^5\,,
\end{align}
where, on dimensional grounds, the loop integral contributes as $\kappa\int d^4 k/M^{10} = \kappa/M^6$, 
where $k$ is the momentum running in the loop.
In the context of a weakly-coupled ultraviolet completion, we expect $g\ll 4\pi$.
In this case the operator $\mathcal{O}_5^{\rm loop}$ pays, compared to $\mathcal{O}_5^{\rm tree}$, 
the suppression factor $\kappa g^2 \ll 1$.

As an additional comment, notice that in general we expect, instead of the simplified potential in eq.\,(\ref{eq:LagrPhi}), 
the more articulated function
\begin{align}\label{eq:EffPotfullPhi}
 V(\Phi,\phi) = \frac{1}{2}M^2\Phi^2 + \frac{c_{12}}{2}gM\phi\Phi^2 + \frac{c_{21}}{2} gM\phi^2\Phi 
+ \frac{c_{22}}{4}g^2\phi^2\Phi^2 +
 \frac{c_{13}}{3!}g^2\phi\Phi^3+\frac{c_{31}}{3!}g^2\phi^3\Phi + \frac{c_{03}}{3!}gM\Phi^3 + \frac{c_{04}}{4!}g^2\Phi^4\,.
\end{align}
If we integrate out $\Phi$, we find HDOs (neglecting derivative operators and setting for simplicity $c_{ij} = 1$) 
of the form $\mathcal{O}^{\rm tree}_{n \geqslant 4} = b_n (g^{n-2}/M^{n-4})\phi^n$. 
The expected scaling with $g$ and $M$ remains unaltered but non-trivial numerical coefficients $b_n$ are generated.
In the case of the potential in eq.\,(\ref{eq:EffPotfullPhi}), we have  
 $b_{n \geqslant 4} = \{-1/8,1/24,-1/72,-1/144,13/576,\dots\}$.
The presence of these numerical coefficients can be qualitatively understood if we consider
 the diagrammatic origin of the HDOs.
 Consider, for instance, the operators $\mathcal{O}_{n= 4,6,8,\dots}^{\rm tree}$.
  In the limit in which $\Phi$ is heavy, we have additional diagrams of the form (we show only a representative 
  set of them)
 \begin{align}
\mathcal{O}_{n= 4,6,8,\dots}^{\rm tree} = \lim_{M \gg g\phi}
\resizebox{150mm}{!}{
\parbox{35mm}{
\begin{tikzpicture}[]
\node (label) at (0,0)[draw=white]{ 
       {\fd{2.75cm}{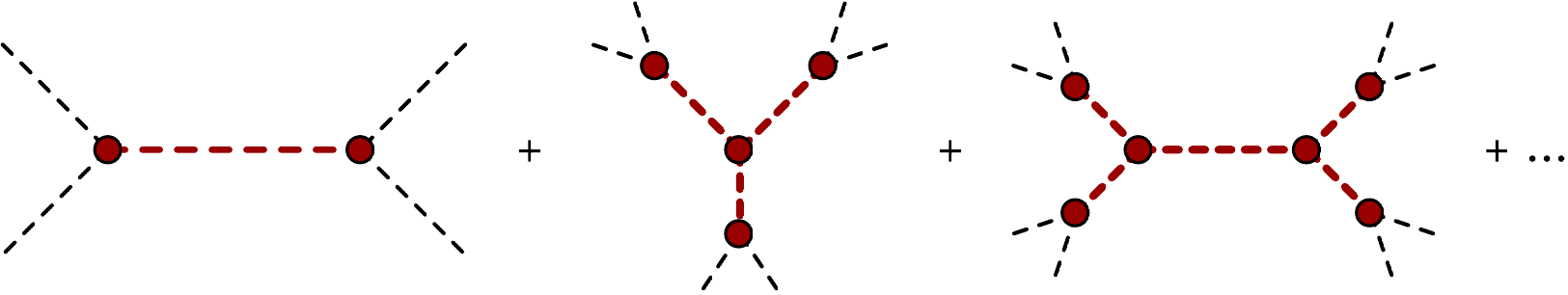}} 
      };
\node[anchor=west] at (-1.55,0.125) {\scalebox{0.25}{$\phi$}}; 
\node[anchor=west] at (-1.55,-0.125) {\scalebox{0.25}{$\phi$}};
\node[anchor=west] at (-.69,0.125) {\scalebox{0.25}{$\phi$}};       
\node[anchor=west] at (-.69,-0.125) {\scalebox{0.25}{$\phi$}};    
\node[anchor=south] at (-0.38,0.04) {\scalebox{0.25}{$\phi$}};    
\node[anchor=south] at (-0.3,0.09) {\scalebox{0.25}{$\phi$}}; 
\node[anchor=south] at (0.23,0.04) {\scalebox{0.25}{$\phi$}};    
\node[anchor=south] at (0.15,0.09) {\scalebox{0.25}{$\phi$}};  
\node[anchor=south] at (-0.18,-0.4) {\scalebox{0.25}{$\phi$}}; 
\node[anchor=south] at (0.025,-0.4) {\scalebox{0.25}{$\phi$}};  
\node[anchor=south] at (1.18,0.01) {\scalebox{0.25}{$\phi$}};    
\node[anchor=south] at (1.10,0.06) {\scalebox{0.25}{$\phi$}};
\node[anchor=south] at (1.18,-0.3) {\scalebox{0.25}{$\phi$}};    
\node[anchor=south] at (1.10,-0.35) {\scalebox{0.25}{$\phi$}};   
\node[anchor=south] at (1.18-0.82,0.01) {\scalebox{0.25}{$\phi$}};    
\node[anchor=south] at (1.10-0.66,0.06) {\scalebox{0.25}{$\phi$}};
\node[anchor=south] at (1.18-0.82,-.3) {\scalebox{0.25}{$\phi$}};    
\node[anchor=south] at (1.10-0.66,-.35) {\scalebox{0.25}{$\phi$}};
\node[anchor=east] at (-0.96,0.06) {{\color{oucrimsonred}{\scalebox{0.18}{$gM$}}}};
\node[anchor=east] at (-1.01,-0.06) {\scalebox{0.18}{$c_{21}$}};
\node[anchor=east] at (-0.6,-0.06) {\scalebox{0.18}{$c_{21}$}};
\node[anchor=east] at (-0.6,0.06) {{\color{oucrimsonred}{\scalebox{0.18}{$gM$}}}};
\node[anchor=east] at (0.,-0.02) {{\color{oucrimsonred}{\scalebox{0.18}{$gM$}}}}; 
\node[anchor=east] at (0.,-0.14) {{\color{oucrimsonred}{\scalebox{0.18}{$gM$}}}};
\node[anchor=east] at (0.03,0.18) {{\color{oucrimsonred}{\scalebox{0.18}{$gM$}}}};
\node[anchor=east] at (.17,0.18) {{\color{oucrimsonred}{\scalebox{0.18}{$gM$}}}}; 
\node[anchor=east] at (1.02,0.045) {{\color{oucrimsonred}{\scalebox{0.18}{$gM$}}}}; 
\node[anchor=east] at (0.89,0.045) {{\color{oucrimsonred}{\scalebox{0.18}{$gM$}}}}; 
\node[anchor=east] at (0.64,0.045) {{\color{oucrimsonred}{\scalebox{0.18}{$gM$}}}};
\node[anchor=east] at (0.64,-0.055) {{\color{oucrimsonred}{\scalebox{0.18}{$gM$}}}}; 
\node[anchor=east] at (0.64+0.6,0.045) {{\color{oucrimsonred}{\scalebox{0.18}{$gM$}}}};
\node[anchor=east] at (0.64+0.6,-0.055) {{\color{oucrimsonred}{\scalebox{0.18}{$gM$}}}}; 
\node[anchor=east] at (.25,.08) {\scalebox{0.18}{$c_{21}$}};
\node[anchor=east] at (-.085,.08) {\scalebox{0.18}{$c_{21}$}};
\node[anchor=east] at (0.16,-0.02) {\scalebox{0.18}{$c_{03}$}};
\node[anchor=east] at (0.17,-0.155) {\scalebox{0.18}{$c_{21}$}};
\node[anchor=east] at (0.81,-0.07) {\scalebox{0.18}{$c_{03}$}};
\node[anchor=east] at (1.05,-0.07) {\scalebox{0.18}{$c_{03}$}};
\node[anchor=east] at (0.73,-0.15) {\scalebox{0.18}{$c_{21}$}};
\node[anchor=east] at (0.73,0.15) {\scalebox{0.18}{$c_{21}$}};
\node[anchor=east] at (1.125,-0.15) {\scalebox{0.18}{$c_{21}$}};
\node[anchor=east] at (1.125,0.15) {\scalebox{0.18}{$c_{21}$}};
\end{tikzpicture}
}}  \nn
\end{align}
where in this case non-trivial combinatorics arises because of the presence of  trilinear self-interactions of $\Phi$.
The coefficients $b_n$ are pure numbers, and power counting alone, without additional assumptions about the full theory, cannot give any clue about their actual value.

Let us close this appendix with a final remark closely related to what we just discussed. 
Although the scaling in eq.\,(\ref{eq:ScalingOneCouplOneScale}) 
can be considered as a generic expectation based on dimensional analysis (given the assumption 
that the sector responsible for the ultraviolet completion is characterized only by one coupling and one scale), important 
structural differences (caused by underlying symmetries) may arise in the series of HDOs as the consequence of specific properties of the ultraviolet completion.
Consider for instance the case in which the heavy field $\Phi$ is even under some $\mathbb{Z}_2$ symmetry.
In this case the potential in eq.\,(\ref{eq:EffPotfullPhi}) reduces to
\begin{align}\label{eq:EffPotfullPhiZ2}
 V(\Phi,\phi) = \frac{1}{2}M^2\Phi^2 + \frac{c_{12}}{2}gM\phi\Phi^2 
+ \frac{c_{22}}{4}g^2\phi^2\Phi^2 +\frac{c_{04}}{4!}g^2\Phi^4\,.
\end{align}
In this case, only loop-suppressed operators $\mathcal{O}_n^{\rm loop} = \kappa g^{n}\phi^n/M^{n-4}$ are generated.
One can try to perform an analysis similar to the one presented in section\,\ref{sec:HDO}. 
The most relevant difference is that the condition $g^2 = a_4$ would be now replaced by 
$\kappa g^4 = a_4$. This implies $g\sim O(10^{-2})$.

\section{Solving the Mukhanov-Sasaki equation analytically}\label{app:C}

The Mukhanov-Sasaki equation must be solved numerically in order to obtain a precise estimate of the abundance of PBHs.
Nevertheless, analytical solutions --although approximated and not best-suited for accurate computations-- can be useful to highlight some of the properties of the 
power spectrum.
We refer to \cite{Biagetti:2018pjj,Ballesteros:2018wlw,Byrnes:2018txb,Ozsoy:2019lyy} for some recent discussions in the context of PBH formation. See also \cite{Leach:2000yw,Leach:2001zf} for earlier numerical an analytical analyses.

\begin{figure}[t]
\begin{center}
$$\includegraphics[width=.45\textwidth]{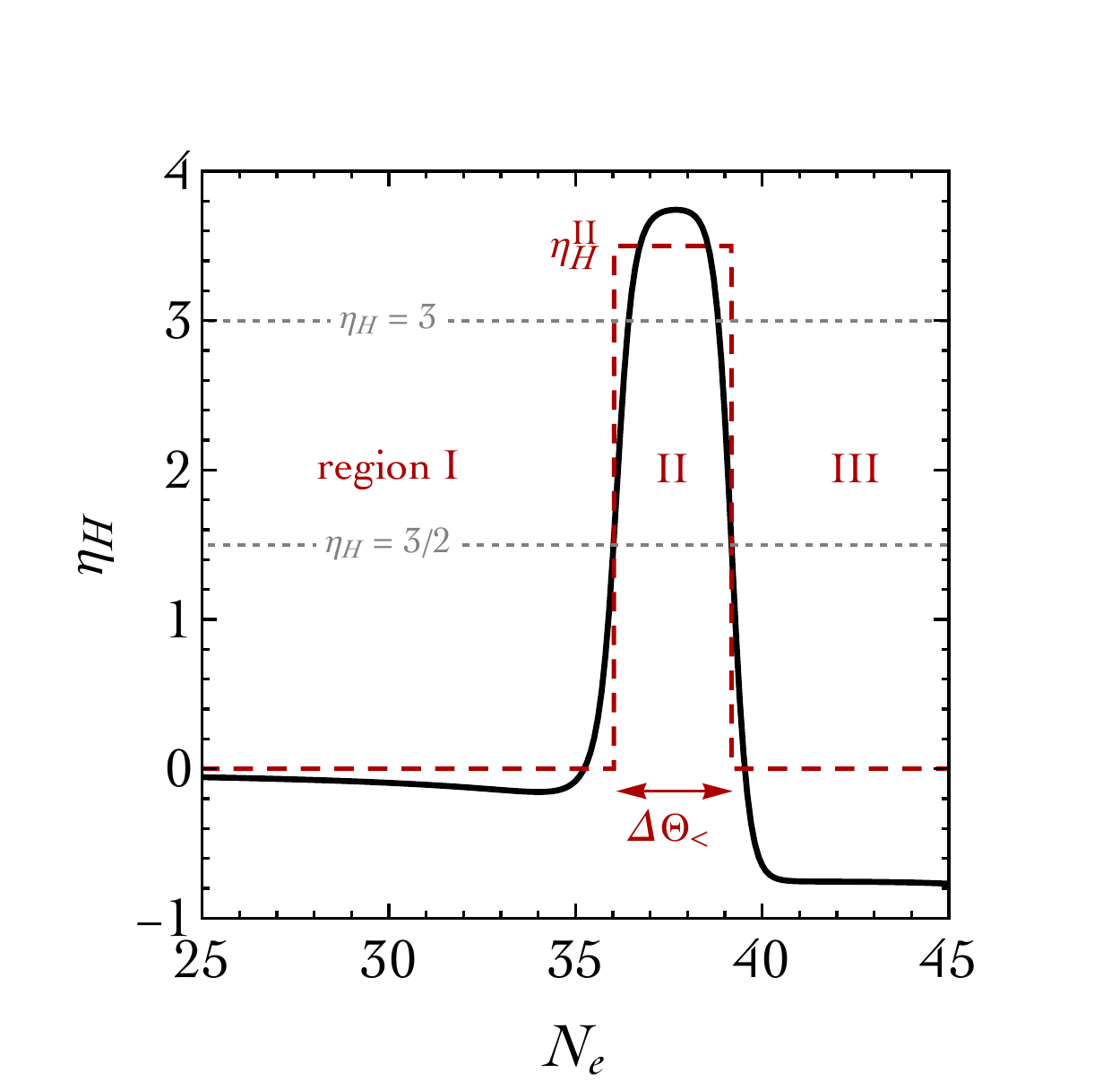}
\qquad\includegraphics[width=.45\textwidth]{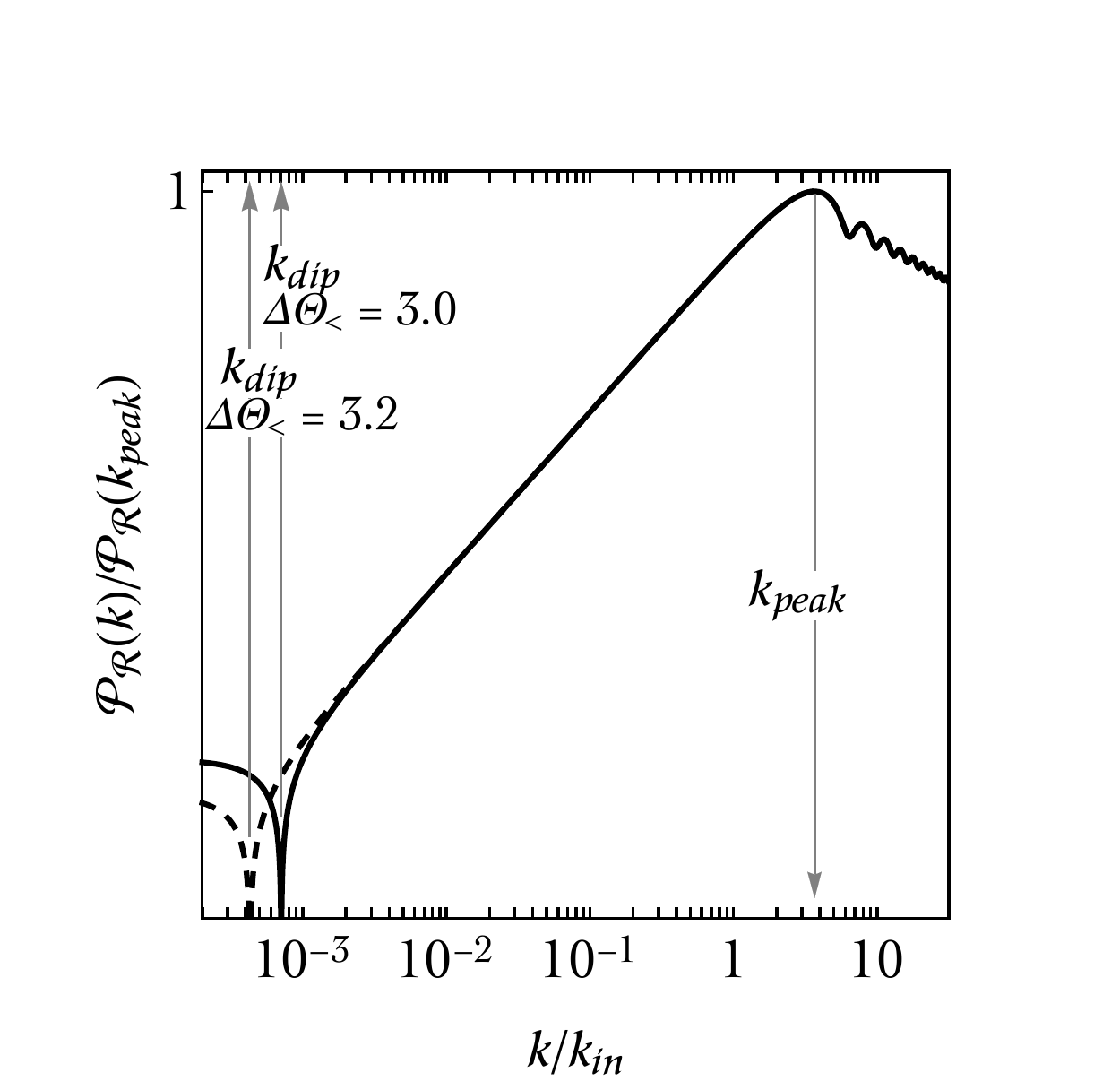}$$ 
\caption{\em \label{fig:analPS} 
{\color{VioletRed4}{
Left panel: Hubble parameter $\eta_H$ as function of the number of $e$-folds. 
We show in solid black the exact numerical result (corresponding to the inflationary solution with $n_s \simeq 0.970$ studied in section\,\ref{sec:HDO}) 
while the dashed red line refers to the piecewise approximation used in appendix\,\ref{app:C}. 
Right panel: Power spectrum obtained by means of the approximation in eq.\,(\ref{eq:analPS}) for $\eta_H^{\rm II} = 4$ and two different values of $\Delta\Theta_<$.
}}
 }
\end{center}
\end{figure}

Using the confomal time $\tau$ as time variable (with $dt/d\tau = a$, $dN_e/d\tau = aH$), the Mukhanov-Sasaki equation in Fourier space is 
\begin{align}\label{eq:MStau}
\frac{d^2 u_{\mathbf{k}}}{d\tau^2} + \left(
k^2 - \frac{1}{z}\frac{d^2 z}{d\tau^2}
\right)u_{\mathbf{k}} = 0\,,~~~~~~~~{\rm with}\,\,\,z = a\frac{dh}{dN_e}   =a\sqrt{2\epsilon_H}\,,
\end{align}
where we have the following exact result
\begin{align}
\frac{1}{z}\frac{d^2 z}{d\tau^2} = a^2 H^2 (1 + \epsilon_H -\eta_H)(2-\eta_H) +  a^2 H^2\frac{d}{dN_e}(\epsilon_H -\eta_H)\,,~~~~~~~~{\rm with}\,\,\,
\eta_H  = \epsilon_H - \frac{1}{2}\frac{d\log \epsilon_H}{dN_e}\,.
\end{align}
If we integrate by parts the expression $d\tau = (1/a^2H)da$, we find for the conformal time
\begin{align}\label{eq:ConformalTime}
\tau = \int \frac{da}{a^2 H} = -\frac{1}{aH} + \int \epsilon_H d\tau\,,
\end{align}
where the first term corresponds to the approximation in which $H$ is constant, the second term comes from
 $aH dH/da = -\epsilon_H$. If we now consider $\epsilon_H$ to be constant, we can integrate eq.\,(\ref{eq:ConformalTime}), and we find the following expression for 
 the evolution of the scale factor:
\begin{align}\label{eq:ConformalTime2}
a(\tau) = -\frac{1}{\tau H(1-\epsilon_H)}\,,~~~~~~ {\rm constant}\,\epsilon_H \neq 0\,,
\end{align}
with $a = -1/\tau H$ in the limit with constant $H$ ($\epsilon_H = 0$). 
If we use this relation in eq.\,(\ref{eq:MStau}) and expand for small $\epsilon_H \ll 1$, we find that
\begin{align}
\frac{d^2 u_{\mathbf{k}}}{d\tau^2} + \left[
k^2 -\frac{1}{\tau^2}\left(
\nu^2 - \frac{1}{4}
\right)
\right]u_{\mathbf{k}} = 0\,,~~~~~~~{\rm with}\,\,\, \nu^2 = \frac{9}{4} - 3\eta_H + \eta_H^2\,
\end{align}
 and we have the canonically normalized solution
\begin{align}\label{eq:HankelGen}
v_{\mathbf{k}}(\tau) = \frac{\sqrt{\pi}}{2}e^{i(\nu + 1/2)\pi/2}\sqrt{-\tau}H_{\nu}^{(1)}(-k\tau)\,,
\end{align}
where $H_{\nu}^{(1)}$ is the Hankel function of the first kind.  The most general solution for $u_{\mathbf{k}}$ has the form
\begin{align}
u_{\mathbf{k}}(\tau) = \alpha_k v_{\mathbf{k}}(\tau) + \beta_k v^*_{\mathbf{k}}(\tau)\,,~~~~~~~~~~|\alpha_k|^2 - |\beta_k|^2 =1\,,
\end{align}
where $\alpha_k$ and $\beta_k$ are complex coefficients subject to the Wronskian condition above.
Eq.\,(\ref{eq:HankelGen}) is normalized in such a way to match the plane-wave solution that we expect in flat Minkowski space-time for $k\gg aH$ (i.e.\ $-k\tau \gg 1$).
For $\eta_H$ constant,  $\nu = \pm (3 - 2 \eta_H)/2$ is just a (real) number independent on $\tau$. 
In the following, we shall exploit the fact that $\eta_H$ can be approximated by a piecewise function, as shown in the left panel of fig.\,\ref{fig:analPS} (dashed red line, compared to the numerical result in solid black).
In each of the three regions indicated in the plot above, $\eta_H$ is approximated with a constant value.
\begin{itemize}
\item [$\circ$] Region\,I. 
This region extends up to the beginning of the phase during which the friction becomes negative (see discussion in section\,\ref{sec:Ana}). 
In this region we approximate $\eta_H^{\,{\rm I}} = 0$ and, correspondingly, we have $\nu_{\rm I} = +3/2$. 
The $\epsilon_H$ parameter takes a constant value that we indicate with $\epsilon_H^{\,{\rm I}}$. 
The comoving curvature perturbation $\mathcal{R}=-u/z$ in region\,I takes the form
\begin{align}\label{eq:RegI}
\mathcal{R}_{\mathbf{k}}^{({\rm I})}(\tau) = \frac{-i}{a\tau\sqrt{4\epsilon_H^{\,{\rm I}}k^3}}e^{-ik\tau}(1+ik\tau) 
\simeq \frac{iH}{\sqrt{4\epsilon_H^{\,{\rm I}}k^3}}e^{-ik\tau}(1+ik\tau)
\,,
\end{align}
where in the second step we used eq.\,(\ref{eq:ConformalTime2}) at the leading order in $\epsilon_H$ (meaning that, in eq.\,(\ref{eq:RegI}), 
$H$ can be considered as constant).
\item [$\circ$] Region\,II. 
The parameter $\eta_H$ takes the constant value $\eta_H^{{\rm II}} \geqslant 3$ (consequently, 
we take $\nu_{\rm II} = (3 - 2 \eta_H^{{\rm II}})/2  \leqslant -3/2$). 
In terms of the number of $e$-folds, let us indicate the transition between region\,I and region\,II --corresponding to the beginning of the negative friction phase-- with $N_{\rm in}$ (and with $\tau_{\rm in}$ the corresponding conformal time). In this region the $\epsilon_H$ parameter evolves (following the equation $d\epsilon_H/dN_e = 2\epsilon_H(\epsilon_H-\eta_H)$) starting from the constant value $\epsilon_H^{\,{\rm I}}$ in region\,I. Solving this equation for constant $\eta_H^{{\rm II}}$ we find the scaling
\begin{align}\label{eq:EspEv}
\epsilon_H^{{\rm II}}(\tau) = \epsilon_H^{\,{\rm I}}\left(\frac{\tau}{\tau_{\rm in}}\right)^{2\eta_H^{{\rm II}}}\,.
\end{align}
The comoving curvature perturbation in region\,II takes the form
\begin{align}
\mathcal{R}_{\mathbf{k}}^{({\rm II})}(\tau) = - H(-\tau)^{3/2}\sqrt{\frac{\pi}{8\epsilon_H^{\,{\rm I}}}}
\left(\frac{\tau_{\rm in}}{\tau}\right)^{\eta_{\rm II}}
\left[
\alpha_k^{\rm II}e^{i\pi(\nu_{\rm II} + 1/2)/2}H_{\nu_{\rm II}}^{(1)}(-k\tau)  + 
\beta_k^{\rm II}e^{-i\pi(\nu_{\rm II} + 1/2)/2}H_{\nu_{\rm II}}^{(2)}(-k\tau)
\right]\,,
\end{align}
where $H_{\nu}^{(2)}$ is the Hankel function of the second kind (for real argument, we have $H_{\nu}^{(1)\,*}= H_{\nu}^{(2)}$).
The complex coefficients $\alpha_k^{\rm II}$ and $\beta_k^{\rm II}$ can be obtained via matching with region\,I by imposing Israel junction conditions.
\item [$\circ$] Region\,III. 
The simplest possibility is to go back to a slow-roll phase with $\eta_{H}^{{\rm III}} = 0$. This is not a perfect approximation but it is enough for our purposes.
Consequently, in region\,III the parameter $\epsilon_H$ is again constant. Its value is given by eq.\,(\ref{eq:EspEv}) evaluated at the end of the negative friction phase. 
Let us indicate with $\tau_{\rm end}$ the end of the negative friction phase in conformal time (and with $N_{\rm end}$ the corresponding number of $e$-folds). We have
\begin{align}
\epsilon_H^{{\rm III}}  = \epsilon_H^{{\rm II}}(\tau_{\rm end}) = \epsilon_H^{\,{\rm I}}\left(\frac{\tau_{\rm end}}{\tau_{\rm in}}\right)^{2\eta_H^{{\rm II}}}\,.
\end{align}
The comoving curvature perturbation in region\,III takes the form
\begin{align}\label{eq:CurvatureRegioIII}
\mathcal{R}_{\mathbf{k}}^{({\rm III})}(\tau) = -\frac{iH}{\sqrt{4\epsilon_H^{\,{\rm I}}k^3}}
\left(\frac{\tau_{\rm in}}{\tau_{\rm end}}\right)^{\eta_H^{{\rm II}}}
\left[
\alpha_k^{\rm III}(1+ik\tau)e^{-ik\tau} - \beta_k^{\rm III}(1-ik\tau)e^{ik\tau}
\right]\,.
\end{align}
The complex coefficients $\alpha_k^{\rm III}$ and $\beta_k^{\rm III}$ can be obtained via matching with region\,II by imposing Israel junction conditions. 
By integrating  $dN_e/d\tau = aH$ we get 
\begin{align}
\frac{\tau_{\rm in}}{\tau_{\rm end}} = e^{N_{\rm end} - N_{\rm in}}
=e^{\Delta\Theta_{<}}\,,~~~~~~\tau_{\rm in} = -\frac{1}{k_{\rm in}}\,,
\end{align}
where $k_{\rm in}$ is the scale that exit the Hubble horizon at time $N_{\rm in}$.
\end{itemize}
On super-horizon scales ($-k\tau \ll 1$), using eq.\,(\ref{eq:CurvatureRegioIII}) with the appropriate
 coefficients $\alpha_k^{\rm III}$ and $\beta_k^{\rm III}$,
the power spectrum is given by the analytical formula
\begin{align}\label{eq:analPS}
  \mathcal{P}_{\mathcal{R}}(k) = \frac{H^2}{8\pi^2\epsilon_H^{\,{\rm I}}}
  \underbrace{
  e^{2\eta_H^{{\rm II}}\Delta\Theta_{<}}\left[
  \alpha_k^{\rm III}\left(\alpha_k^{\rm III}\right)^* + 
  \beta_k^{\rm III}\left(\beta_k^{\rm III}\right)^* - 
  \alpha_k^{\rm III}\left(\beta_k^{\rm III}\right)^* - \left(\alpha_k^{\rm III}\right)^*\beta_k^{\rm III}
  \right]}_{\equiv\,\,\mathcal{F}(\eta_H^{\rm II},\Delta\Theta_{<},k/k_{\rm in})}\,,
\end{align}
where we have $\lim_{x\to 0}\mathcal{F}(\eta_H^{\rm II},\Delta\Theta_{<},x) = 1$. 
The analytical expressions for $\alpha_k^{\rm III}$ and 
$\beta_k^{\rm III}$ for generic $\eta_H^{\rm II}$ are quite lengthy.
An important simplification takes place if we consider $\eta_H^{\rm II} = 4$ since in this case $\nu_{\rm II} = -5/2$ and we can use the explicit expressions
\begin{align}
H_{-5/2}^{(1)}(-k\tau) = \frac{\sqrt{2/\pi}e^{-ik\tau}}{k^2\tau^2\sqrt{-k\tau}}(3+3ik\tau - k^2\tau^2)\,,
~~~~~~~~~\left[H_{-5/2}^{(1)}(-k\tau)\right]^* =  H_{-5/2}^{(2)}(-k\tau)\,.
\end{align}
If we expand for small $x\equiv k/k_{\rm in}$,  we find
\begin{align}\label{eq:AppAnalPS}
\mathcal{F}(\eta_H^{\rm II}=4,\Delta\Theta_{<},x) = 
\left(
1- \frac{64 e^{5\Delta\Theta_{<}}}{105}x^2
\right)^2 + \frac{8}{5}x^2
+\left(\frac{28}{75} 
+\frac{512 e^{3\Delta\Theta_{<}}}{2205} - 
\frac{2048 e^{5\Delta\Theta_{<}}}{4725}
\right)x^4 + O(x^6)\,,
\end{align}
where we neglected all terms suppressed by positive powers of $e^{-\Delta N_{\rm USR}}$.
The position of the dip corresponds to the value of $x$ such that the leading term in eq.\,(\ref{eq:AppAnalPS}) vanishes.
We find
\begin{align}
\left.\frac{k_{\rm dip}}{k_{\rm in}}\right|_{\eta_H^{\rm II} = 4} = \sqrt{\frac{105}{64}}e^{-5\Delta\Theta_{<}/2}\,.
\end{align}
This is the expression used in section\,\ref{sec:HDO} (third point, where 
we used the approximation $k_{\rm peak}\approx k_{\rm in} \gg k_{\rm dip}$).
In the right panel of fig.\,\ref{fig:analPS} 
we show the analytical approximation in eq.\,(\ref{eq:analPS}) with $\eta_H^{\rm II} = 4$ for two different durations 
of the negative friction phase, namely $\Delta\Theta_{<} = 3.2$ (dashed line) and $\Delta\Theta_{<} = 3.0$ (solid line).
Despite the simplicity of the approximations used, we see that, at the qualitative level (that is, as far as the presence of the peak and the dip in the power spectrum and their separation as a function of the duration of the negative friction phase are concerned), the 
numerical result shown in fig.\,\ref{fig:HDOdyn}
is well mimicked
(compared to fig.\,\ref{fig:HDOdyn}, 
here we used two more distant values $\Delta\Theta_{<} = 3.2$ and $\Delta\Theta_{<} = 3.0$  to visualize better the difference in the position of $k_{\rm dip}$). 
In particular, we confirm what we already argued in section\,\ref{sec:HDO}. 
For fixed $k_{\rm peak}$, a shorter negative friction phase moves the position of the dip towards smaller scales thus 
allowing to fit larger values of $n_s$ at large scales. 
Of course, at the quantitative level the crude approximation used in this appendix cannot compete with the exact numerical result. For instance, using 
$\eta_H^{\rm II} = 4$, which greatly simplifies the math, generates a peak that is too high and approximating $\eta_H$ with a piecewise function does not capture the right slope of the power spectrum.

%%%%%%%%%%%%%%%%%%%%%%%%%%%%
%%%%%%%%%%%%%%%%%%%%%%%%%%%%%%%%%%%%%%

%%%%%%%%%%%%%%%%%%%%%%%%%%%%%%%%%%%%%%%%%%%%%%


\begin{thebibliography}{99}
 
 %\cite{Niikura:2017zjd,Katz:2018zrn}
\bibitem{Niikura:2017zjd} 
  H.~Niikura {\it et al.},
  ``{Microlensing constraints on primordial black holes with Subaru/HSC Andromeda observations,}''
  Nat.\ Astron.\  {\bf 3}, no. 6, 524 (2019)
  %doi:10.1038/s41550-019-0723-1
  [\hhref{arXiv:1701.02151} [astro-ph.CO]].
  %%CITATION = doi:10.1038/s41550-019-0723-1;%%
  %144 citations counted in INSPIRE as of 31 Oct 2019

 
 %\cite{Katz:2018zrn}
\bibitem{Katz:2018zrn} 
  A.~Katz, J.~Kopp, S.~Sibiryakov and W.~Xue,
  ``{Femtolensing by Dark Matter Revisited,}''
  JCAP {\bf 1812}, 005 (2018)
  %doi:10.1088/1475-7516/2018/12/005
  [\hhref{arXiv:1807.11495} [astro-ph.CO]].
  %%CITATION = doi:10.1088/1475-7516/2018/12/005;%%
  %54 citations counted in INSPIRE as of 31 Oct 2019
  
  %\cite{Montero-Camacho:2019jte}
\bibitem{Montero-Camacho:2019jte}
  P.~Montero-Camacho, X.~Fang, G.~Vasquez, M.~Silva and C.~M.~Hirata,
  ``Revisiting constraints on asteroid-mass primordial black holes as dark matter candidates,''
  JCAP {\bf 1908} (2019) 031
  % doi:10.1088/1475-7516/2019/08/031
  [\hhref{arXiv:1906.05950} [astro-ph.CO]].
  %%CITATION = doi:10.1088/1475-7516/2019/08/031;%%
  %21 citations counted in INSPIRE as of 17 Dec 2019

  %\cite{Sasaki:2018dmp}
  \bibitem{Sasaki:2018dmp} 
  M.~Sasaki, T.~Suyama, T.~Tanaka and S.~Yokoyama,
  ``{Primordial black holes and perspectives in gravitational wave astronomy,}''
  Class.\ Quant.\ Grav.\  {\bf 35}, no. 6, 063001 (2018)
  %doi:10.1088/1361-6382/aaa7b4
  [\hhref{arXiv:1801.05235} [astro-ph.CO]].
  %%CITATION = doi:10.1088/1361-6382/aaa7b4;%%
  %136 citations counted in INSPIRE as of 31 Oct 2019

  \bibitem{Starobinsky:1992ts}
  A.~A.~Starobinsky,
  ``Spectrum of adiabatic perturbations in the universe when there are singularities in the inflation potential,''
  JETP Lett. {\bf{55}} (1992) 489-494.
    
  %\cite{Ivanov:1994pa}
  \bibitem{Ivanov:1994pa}
  P.~Ivanov, P.~Naselsky and I.~Novikov,
  ``Inflation and primordial black holes as dark matter,''
  Phys.\ Rev.\ D {\bf 50} (1994) 7173.
  % doi:10.1103/PhysRevD.50.7173
  %%CITATION = doi:10.1103/PhysRevD.50.7173;%%
  %161 citations counted in INSPIRE as of 19 Dec 2019
  
  \bibitem{Saito_2008}
  Ryo Saito, Jun’ichi Yokoyama, and Ryo Nagata.
  ``Single-field inflation, anomalous enhancement of superhorizon fluctuations and non-gaussianity in primordial black hole formation.''
  \newblock {\em Journal of Cosmology and Astroparticle Physics}, 2008(06):024, Jun 2008.
  [\hhref{arXiv:0804.3470}].
  
    %\cite{Garcia-Bellido:2017mdw,Gong:2017qlj,Kannike:2017bxn}
\bibitem{Garcia-Bellido:2017mdw} 
  J.~Garcia-Bellido and E.~Ruiz Morales,
  ``{Primordial black holes from single field models of inflation,}''
  Phys.\ Dark Univ.\  {\bf 18}, 47 (2017)
  %doi:10.1016/j.dark.2017.09.007
  [\hhref{arXiv:1702.03901} [astro-ph.CO]].
  %%CITATION = doi:10.1016/j.dark.2017.09.007;%%
  %104 citations counted in INSPIRE as of 07 Nov 2019
  
    \bibitem{mishra2019primordial}
  Swagat~S. Mishra and Varun Sahni.
  ``Primordial black holes from a tiny bump in the inflaton potential,''
  2019.
  [\hhref{arXiv:1911.00057}].
  
 %\cite{Ozsoy:2018flq}
\bibitem{Ozsoy:2018flq} 
  O.~Ozsoy, S.~Parameswaran, G.~Tasinato and I.~Zavala,
  ``{Mechanisms for Primordial Black Hole Production in String Theory,}''
  JCAP {\bf 1807}, 005 (2018)
  %doi:10.1088/1475-7516/2018/07/005
  [\hhref{arXiv:1803.07626} [hep-th]].
  %%CITATION = doi:10.1088/1475-7516/2018/07/005;%%
  %28 citations counted in INSPIRE as of 31 Oct 2019

%\cite{Ballesteros:2019hus}
\bibitem{Ballesteros:2019hus}
  G.~Ballesteros, J.~Rey and F.~Rompineve,
  ``Detuning primordial black hole dark matter with early matter domination and axion monodromy,''
  \hhref{arXiv:1912.01638} [astro-ph.CO].
  %%CITATION = ARXIV:1912.01638;%%
 
  %\cite{Cicoli:2018asa}
\bibitem{Cicoli:2018asa} 
  M.~Cicoli, V.~A.~Diaz and F.~G.~Pedro,
  ``{Primordial Black Holes from String Inflation,}''
  JCAP {\bf 1806}, 034 (2018)
  %doi:10.1088/1475-7516/2018/06/034
  [\hhref{arXiv:1803.02837} [hep-th]].
  %%CITATION = doi:10.1088/1475-7516/2018/06/034;%%
  %28 citations counted in INSPIRE as of 31 Oct 2019

%\cite{Dalianis:2018frf}
\bibitem{Dalianis:2018frf} 
  I.~Dalianis, A.~Kehagias and G.~Tringas,
  ``{Primordial black holes from $\alpha$-attractors,}''
  JCAP {\bf 1901}, 037 (2019)
  %doi:10.1088/1475-7516/2019/01/037
  [\hhref{arXiv:1805.09483} [astro-ph.CO]].
  %%CITATION = doi:10.1088/1475-7516/2019/01/037;%%
  %26 citations counted in INSPIRE as of 11 Dec 2019



%\cite{Ballesteros:2015noa}
\bibitem{Ballesteros:2015noa} 
  G.~Ballesteros and C.~Tamarit,
  ``{Radiative plateau inflation,}''
  JHEP {\bf 1602}, 153 (2016)
  %doi:10.1007/JHEP02(2016)153
  [\hhref{arXiv:1510.05669} [hep-ph]].
  %%CITATION = doi:10.1007/JHEP02(2016)153;%%
  %16 citations counted in INSPIRE as of 31 Oct 2019



%\cite{Ballesteros:2017fsr}
\bibitem{Ballesteros:2017fsr} 
  G.~Ballesteros and M.~Taoso,
  ``{Primordial black hole dark matter from single field inflation,}''
  Phys.\ Rev.\ D {\bf 97}, no. 2, 023501 (2018)
  %doi:10.1103/PhysRevD.97.023501
  [\hhref{arXiv:1709.05565} [hep-ph]].
  %%CITATION = doi:10.1103/PhysRevD.97.023501;%%
  %63 citations counted in INSPIRE as of 31 Oct 2019
  
  %\cite{Spokoiny:1984bd}
\bibitem{Spokoiny:1984bd}
  B.~L.~Spokoiny,
  ``Inflation And Generation Of Perturbations In Broken Symmetric Theory Of Gravity,''
  Phys.\ Lett.\  {\bf 147B} (1984) 39.
 % doi:10.1016/0370-2693(84)90587-2
  %%CITATION = doi:10.1016/0370-2693(84)90587-2;%%
  %219 citations counted in INSPIRE as of 17 Dec 2019
 

%\cite{Carr:2009jm}
\bibitem{Carr:2009jm} 
  B.~J.~Carr, K.~Kohri, Y.~Sendouda and J.~Yokoyama,
  ``{New cosmological constraints on primordial black holes,}''
  Phys.\ Rev.\ D {\bf 81}, 104019 (2010)
  %doi:10.1103/PhysRevD.81.104019
  [\hhref{arXiv:0912.5297} [astro-ph.CO]].
  %%CITATION = doi:10.1103/PhysRevD.81.104019;%%
  %503 citations counted in INSPIRE as of 31 Oct 2019
  
%\cite{Arbey:2019vqx}
\bibitem{Arbey:2019vqx} 
  A.~Arbey, J.~Auffinger and J.~Silk,
  ``{Constraining primordial black hole masses with the isotropic gamma ray background,}''
  \hhref{arXiv:1906.04750} [astro-ph.CO].
  %%CITATION = ARXIV:1906.04750;%%
  %1 citations counted in INSPIRE as of 01 Nov 2019  
  
  %\cite{Ballesteros:2019exr}
\bibitem{Ballesteros:2019exr}
  G.~Ballesteros, J.~Coronado-Bl\'azquez and D.~Gaggero,
  ``X-ray and gamma-ray limits on the primordial black hole abundance from Hawking radiation,''
  \hhref{arXiv:1906.10113} [astro-ph.CO].
  %%CITATION = ARXIV:1906.10113;%%
  %5 citations counted in INSPIRE as of 17 Dec 2019
  

  
 %\cite{Clark:2016nst}
\bibitem{Clark:2016nst}
  S.~Clark, B.~Dutta, Y.~Gao, L.~E.~Strigari and S.~Watson,
  ``Planck Constraint on Relic Primordial Black Holes,''
  Phys.\ Rev.\ D {\bf 95} (2017) no.8,  083006
  % doi:10.1103/PhysRevD.95.083006
  [\hhref{arXiv:1612.07738} [astro-ph.CO]].
  %%CITATION = doi:10.1103/PhysRevD.95.083006;%%
  %24 citations counted in INSPIRE as of 19 Dec 2019 
  
%\cite{Boudaud:2018hqb}
\bibitem{Boudaud:2018hqb}
  M.~Boudaud and M.~Cirelli,
  ``Voyager 1 $e^\pm$ Further Constrain Primordial Black Holes as Dark Matter,''
  Phys.\ Rev.\ Lett.\  {\bf 122} (2019) no.4,  041104
  %doi:10.1103/PhysRevLett.122.041104
  [\hhref{arXiv:1807.03075} [astro-ph.HE]].
  %%CITATION = doi:10.1103/PhysRevLett.122.041104;%%
  %16 citations counted in INSPIRE as of 19 Dec 2019
  
  %\cite{DeRocco:2019fjq}
\bibitem{DeRocco:2019fjq}
  W.~DeRocco and P.~W.~Graham,
  ``Constraining primordial black hole abundance with the Galactic 511 keV line,''
  Phys.\ Rev.\ Lett.\  {\bf 123} (2019) 251102
  % doi:10.1103/PhysRevLett.123.251102
  [\hhref{arXiv:1906.07740} [astro-ph.CO]].
  %%CITATION = doi:10.1103/PhysRevLett.123.251102;%%
  %8 citations counted in INSPIRE as of 19 Dec 2019
  
  %\cite{Laha:2019ssq}
\bibitem{Laha:2019ssq}
  R.~Laha,
  ``Primordial black holes as a dark matter candidate are severely constrained by the Galactic Center 511 keV gamma-ray line,''
  Phys.\ Rev.\ Lett.\  {\bf 123} (2019) 251101
 % doi:10.1103/PhysRevLett.123.251101
  [\hhref{arXiv:1906.09994} [astro-ph.HE]].
  %%CITATION = doi:10.1103/PhysRevLett.123.251101;%%
  %6 citations counted in INSPIRE as of 19 Dec 2019
  
  %\cite{Dasgupta:2019cae}
\bibitem{Dasgupta:2019cae}
  B.~Dasgupta, R.~Laha and A.~Ray,
  ``Neutrino and positron constraints on spinning primordial black hole dark matter,''
  \hhref{arXiv:1912.01014} [hep-ph].
  %%CITATION = ARXIV:1912.01014;%%
  
%\cite{Poulter:2019ooo}
\bibitem{Poulter:2019ooo}
  H.~Poulter, Y.~Ali-Haïmoud, J.~Hamann, M.~White and A.~G.~Williams,
  ``CMB constraints on ultra-light primordial black holes with extended mass distributions,''
  \hhref{arXiv:1907.06485} [astro-ph.CO].
  %%CITATION = ARXIV:1907.06485;%%
  %4 citations counted in INSPIRE as of 21 Dec 2019


  %\cite{Carr:2017jsz}
\bibitem{Carr:2017jsz} 
  B.~Carr, M.~Raidal, T.~Tenkanen, V.~Vaskonen and H.~Veerm\"ae,
  ``{Primordial black hole constraints for extended mass functions,}''
  Phys.\ Rev.\ D {\bf 96}, no. 2, 023514 (2017)
  %doi:10.1103/PhysRevD.96.023514
  [\hhref{arXiv:1705.05567} [astro-ph.CO]].
  %%CITATION = doi:10.1103/PhysRevD.96.023514;%%
  %156 citations counted in INSPIRE as of 01 Nov 2019
  
   %\cite{Aghanim:2018eyx}
\bibitem{Aghanim:2018eyx} 
  N.~Aghanim {\it et al.} [Planck Collaboration],
  ``{Planck 2018 results. VI. Cosmological parameters,}''
  \hhref{arXiv:1807.06209} [astro-ph.CO].
  %%CITATION = ARXIV:1807.06209;%%
  %1725 citations counted in INSPIRE as of 06 Nov 2019 
  
    %\cite{Acquaviva:2002ud,Ananda:2006af,Baumann:2007zm}
\bibitem{Acquaviva:2002ud} 
  V.~Acquaviva, N.~Bartolo, S.~Matarrese and A.~Riotto,
  ``{Second order cosmological perturbations from inflation,}''
  Nucl.\ Phys.\ B {\bf 667}, 119 (2003)
  %doi:10.1016/S0550-3213(03)00550-9
  [\hhref{astro-ph/0209156}].
  %%CITATION = doi:10.1016/S0550-3213(03)00550-9;%%
  %569 citations counted in INSPIRE as of 23 Oct 2019
  
 %\cite{Ananda:2006af,Baumann:2007zm}
\bibitem{Ananda:2006af} 
  K.~N.~Ananda, C.~Clarkson and D.~Wands,
  ``{The Cosmological gravitational wave background from primordial density perturbations,}''
  Phys.\ Rev.\ D {\bf 75}, 123518 (2007)
  %doi:10.1103/PhysRevD.75.123518
  [\hhref{gr-qc/0612013}].
  %%CITATION = doi:10.1103/PhysRevD.75.123518;%%
  %165 citations counted in INSPIRE as of 23 Oct 2019 
  
  %\cite{Baumann:2007zm}
\bibitem{Baumann:2007zm} 
  D.~Baumann, P.~J.~Steinhardt, K.~Takahashi and K.~Ichiki,
  ``{Gravitational Wave Spectrum Induced by Primordial Scalar Perturbations,}''
  Phys.\ Rev.\ D {\bf 76}, 084019 (2007)
  %doi:10.1103/PhysRevD.76.084019
  [\hhref{hep-th/0703290}].
  %%CITATION = doi:10.1103/PhysRevD.76.084019;%%
  %183 citations counted in INSPIRE as of 23 Oct 2019
 
  \bibitem{saito2008gravitational}
  Ryo Saito and Jun'ichi Yokoyama.
  ``Gravitational wave background as a probe of the primordial black hole abundance,'' 2008.
  [\hhref{arXiv:0812.4339}].

  \bibitem{Saito_2010}
  R.~Saito and J.~Yokoyama.
  ``Gravitational-wave constraints on the abundance of primordial black holes.''
  \newblock {\em Progress of Theoretical Physics}, 123(5):867–886, May 2010.
  [\hhref{arXiv:0912.5317}].
  
    \bibitem{Kohri_2018}
  Kazunori Kohri and Takahiro Terada.
  ``Semianalytic calculation of gravitational wave spectrum nonlinearly induced from primordial curvature perturbations.''
  \newblock {\em Physical Review D}, 97(12), Jun 2018.
  [\hhref{arXiv:1804.08577}].
  
 %\cite{Caprini:2015zlo}
\bibitem{Caprini:2015zlo} 
  C.~Caprini {\it et al.},
  ``{Science with the space-based interferometer eLISA. II: Gravitational waves from cosmological phase transitions,}''
  JCAP {\bf 1604}, 001 (2016)
  %doi:10.1088/1475-7516/2016/04/001
  [\hhref{arXiv:1512.06239} [astro-ph.CO]].
  %%CITATION = doi:10.1088/1475-7516/2016/04/001;%%
  %279 citations counted in INSPIRE as of 05 Nov 2019 
  
 %\cite{Yagi:2011wg}
\bibitem{Yagi:2011wg} 
  K.~Yagi and N.~Seto,
  ``{Detector configuration of DECIGO/BBO and identification of cosmological neutron-star binaries,}''
  Phys.\ Rev.\ D {\bf 83}, 044011 (2011)
  Erratum: [Phys.\ Rev.\ D {\bf 95}, no. 10, 109901 (2017)]
  %doi:10.1103/PhysRevD.95.109901, 10.1103/PhysRevD.83.044011
  [\hhref{arXiv:1101.3940} [astro-ph.CO]].
  %%CITATION = doi:10.1103/PhysRevD.95.109901, 10.1103/PhysRevD.83.044011;%%
  %88 citations counted in INSPIRE as of 05 Nov 2019
  
 %\cite{Coleman:2018ozp}
\bibitem{Coleman:2018ozp} 
  J.~Coleman [MAGIS-100 Collaboration],
  ``{Matter-wave Atomic Gradiometer InterferometricSensor (MAGIS-100) at Fermilab,}''
  PoS ICHEP {\bf 2018}, 021 (2019)
  %doi:10.22323/1.340.0021
  [\hhref{arXiv:1812.00482} [physics.ins-det]].
  %%CITATION = doi:10.22323/1.340.0021;%%
  %6 citations counted in INSPIRE as of 05 Nov 2019 

%\cite{Chen:2018rzo}
\bibitem{Chen:2018rzo}
  Z.~C.~Chen, F.~Huang and Q.~G.~Huang,
  ``{Stochastic Gravitational-wave Background from Binary Black Holes and Binary Neutron Stars and Implications for LISA,}''
  Astrophys.\ J.\  {\bf 871} (2019) no.1,  97
  %doi:10.3847/1538-4357/aaf581
  [\hhref{arXiv:1809.10360} [gr-qc]].
  %%CITATION = doi:10.3847/1538-4357/aaf581;%%
  %10 citations counted in INSPIRE as of 24 Dec 2019  
    
  %\cite{LIGOScientific:2019vic}
\bibitem{LIGOScientific:2019vic} 
  B.~P.~Abbott {\it et al.} [LIGO Scientific and Virgo Collaborations],
  ``{Search for the isotropic stochastic background using data from Advanced LIGO's second observing run,}''
  Phys.\ Rev.\ D {\bf 100}, no. 6, 061101 (2019)
  %doi:10.1103/PhysRevD.100.061101
  [\hhref{arXiv:1903.02886} [gr-qc]].
  %%CITATION = doi:10.1103/PhysRevD.100.061101;%%
  %31 citations counted in INSPIRE as of 01 Nov 2019
  
    \bibitem{aLIGOdesign}
  L.~Barsotti, P.~Fritschel, M.~Evans and S.~Gras, 
  \href{https://dcc.ligo.org/T1800044-v5/public}{Updated Advanced LIGO sensitivity design curve}.
  
    %\cite{Abbott:2017xzg}
\bibitem{Abbott:2017xzg} 
  B.~P.~Abbott {\it et al.} [LIGO Scientific and Virgo Collaborations],
  ``{GW170817: Implications for the Stochastic Gravitational-Wave Background from Compact Binary Coalescences,}''
  Phys.\ Rev.\ Lett.\  {\bf 120}, no. 9, 091101 (2018)
  %doi:10.1103/PhysRevLett.120.091101
  [\hhref{arXiv:1710.05837} [gr-qc]].
  %%CITATION = doi:10.1103/PhysRevLett.120.091101;%%
  %68 citations counted in INSPIRE as of 06 Nov 2019
  
  %\cite{Ballesteros:2014yva}
\bibitem{Ballesteros:2014yva}
  G.~Ballesteros and J.~A.~Casas,
  ``Large tensor-to-scalar ratio and running of the scalar spectral index with Instep Inflation,''
  Phys.\ Rev.\ D {\bf 91} (2015) 043502
  % doi:10.1103/PhysRevD.91.043502
  [\hhref{arXiv:1406.3342} [astro-ph.CO]].
  %%CITATION = doi:10.1103/PhysRevD.91.043502;%%
  %14 citations counted in INSPIRE as of 17 Dec 2019
  
  %\cite{Kinney:2005vj}
\bibitem{Kinney:2005vj}
  W.~H.~Kinney,
  ``Horizon crossing and inflation with large eta,''
  Phys.\ Rev.\ D {\bf 72} (2005) 023515
 % doi:10.1103/PhysRevD.72.023515
  [\hhref{gr-qc/0503017}].
  %%CITATION = doi:10.1103/PhysRevD.72.023515;%%
  %167 citations counted in INSPIRE as of 18 Dec 2019
  
%\cite{Chongchitnan:2006wx,Motohashi:2017kbs,Germani:2017bcs,Ballesteros:2017fsr}
\bibitem{Chongchitnan:2006wx} 
  S.~Chongchitnan and G.~Efstathiou,
  ``{Accuracy of slow-roll formulae for inflationary perturbations: implications for primordial black hole formation,}''
  JCAP {\bf 0701}, 011 (2007)
  %doi:10.1088/1475-7516/2007/01/011
  [\hhref{astro-ph/0611818}].
  %%CITATION = doi:10.1088/1475-7516/2007/01/011;%%
  %30 citations counted in INSPIRE as of 16 Sep 2019


%\cite{Motohashi:2017kbs,Germani:2017bcs,Ballesteros:2017fsr}
\bibitem{Motohashi:2017kbs} 
  H.~Motohashi and W.~Hu,
  ``{Primordial Black Holes and Slow-Roll Violation,}''
  Phys.\ Rev.\ D {\bf 96}, no. 6, 063503 (2017)
  %doi:10.1103/PhysRevD.96.063503
  [\hhref{arXiv:1706.06784} [astro-ph.CO]].
  %%CITATION = doi:10.1103/PhysRevD.96.063503;%%
  %58 citations counted in INSPIRE as of 16 Sep 2019
  
  %\cite{Germani:2017bcs,Ballesteros:2017fsr}
\bibitem{Germani:2017bcs} 
  C.~Germani and T.~Prokopec,
  ``{On primordial black holes from an inflection point,}''
  Phys.\ Dark Univ.\  {\bf 18}, 6 (2017)
  %doi:10.1016/j.dark.2017.09.001
  [\hhref{arXiv:1706.04226} [astro-ph.CO]].
  %%CITATION = doi:10.1016/j.dark.2017.09.001;%%
  %70 citations counted in INSPIRE as of 16 Sep 2019

%%\cite{Carr:1975qj}
%\bibitem{Carr:1975qj} 
%  B.~J.~Carr,
%  ``{The Primordial black hole mass spectrum,}''
%  Astrophys.\ J.\  {\bf 201}, 1 (1975).
%  %doi:10.1086/153853
%  %%CITATION = doi:10.1086/153853;%%
%  %573 citations counted in INSPIRE as of 22 Oct 2019
  
  %\cite{Ballesteros:2018wlw}
\bibitem{Ballesteros:2018wlw}
  G.~Ballesteros, J.~Beltran Jimenez and M.~Pieroni,
  %``Black hole formation from a general quadratic action for inflationary primordial fluctuations,''
  JCAP {\bf 1906} (2019) 016
  doi:10.1088/1475-7516/2019/06/016
  [arXiv:1811.03065 [astro-ph.CO]].
  %%CITATION = doi:10.1088/1475-7516/2019/06/016;%%


%\cite{Green:2004wb,Musco:2004ak}
\bibitem{Green:2004wb} 
  A.~M.~Green, A.~R.~Liddle, K.~A.~Malik and M.~Sasaki,
  ``{A New calculation of the mass fraction of primordial black holes,}''
  Phys.\ Rev.\ D {\bf 70}, 041502 (2004)
  %doi:10.1103/PhysRevD.70.041502
  [\hhref{astro-ph/0403181}].
  %%CITATION = doi:10.1103/PhysRevD.70.041502;%%
  %73 citations counted in INSPIRE as of 23 Oct 2019
  
  %\cite{Musco:2004ak}
\bibitem{Musco:2004ak} 
  I.~Musco, J.~C.~Miller and L.~Rezzolla,
  ``{Computations of primordial black hole formation,}''
  Class.\ Quant.\ Grav.\  {\bf 22}, 1405 (2005)
  %doi:10.1088/0264-9381/22/7/013
  [\hhref{gr-qc/0412063}].
  %%CITATION = doi:10.1088/0264-9381/22/7/013;%%
  %118 citations counted in INSPIRE as of 23 Oct 2019
  
  %\cite{Musco:2018rwt}
\bibitem{Musco:2018rwt}
I.~Musco,
%``Threshold for primordial black holes: Dependence on the shape of the cosmological perturbations,''
Phys. Rev. D \textbf{100} (2019) no.12, 123524
doi:10.1103/PhysRevD.100.123524
[arXiv:1809.02127 [gr-qc]].
%40 citations counted in INSPIRE as of 05 May 2020

\bibitem{Escriva:2019phb}
A.~Escrivà, C.~Germani and R.~K.~Sheth,
%``Universal threshold for primordial black hole formation,''
Phys. Rev. D \textbf{101} (2020) no.4, 044022
doi:10.1103/PhysRevD.101.044022
[arXiv:1907.13311 [gr-qc]].

%\cite{Atal:2019erb}
\bibitem{Atal:2019erb}
V.~Atal, J.~Cid, A.~Escrivà and J.~Garriga,
``{\it PBH in single field inflation: the effect of shape dispersion and non-Gaussianities,}''
[\hhref{arXiv:1908.11357} [astro-ph.CO]].
%10 citations counted in INSPIRE as of 20 Apr 2020

%\cite{Franciolini:2018vbk}
\bibitem{Franciolini:2018vbk}
G.~Franciolini, A.~Kehagias, S.~Matarrese and A.~Riotto,
``{\it Primordial Black Holes from Inflation and non-Gaussianity,}''
JCAP \textbf{03} (2018), 016
%doi:10.1088/1475-7516/2018/03/016
[\hhref{arXiv:1801.09415} [astro-ph.CO]].
%61 citations counted in INSPIRE as of 20 Apr 2020

%\cite{Akrami:2018odb}
\bibitem{Akrami:2018odb} 
  Y.~Akrami {\it et al.} [Planck Collaboration],
  ``{Planck 2018 results. X. Constraints on inflation,}''
  \hhref{arXiv:1807.06211} [astro-ph.CO].
  %%CITATION = ARXIV:1807.06211;%%
  %545 citations counted in INSPIRE as of 04 Nov 2019

%  %\cite{Gong:2017qlj,Drees:2019xpp}
%\bibitem{Gong:2017qlj} 
%  H.~Di and Y.~Gong,
%  ``{Primordial black holes and second order gravitational waves from ultra-slow-roll inflation,}''
%  JCAP {\bf 1807}, 007 (2018)
%  %doi:10.1088/1475-7516/2018/07/007
%  [\hhref{arXiv:1707.09578} [astro-ph.CO]].
%  %%CITATION = doi:10.1088/1475-7516/2018/07/007;%%
%  %34 citations counted in INSPIRE as of 06 Nov 2019
  
  %\cite{Drees:2019xpp}
\bibitem{Drees:2019xpp} 
  M.~Drees and Y.~Xu,
  ``{Overshooting, Critical Higgs Inflation and Second Order Gravitational Wave Signatures,}''
  \hhref{arXiv:1905.13581} [hep-ph].
  %%CITATION = ARXIV:1905.13581;%%
  %4 citations counted in INSPIRE as of 06 Nov 2019

%\cite{Espinosa:2018eve}
\bibitem{Espinosa:2018eve} 
  J.~R.~Espinosa, D.~Racco and A.~Riotto,
  ``{A Cosmological Signature of the SM Higgs Instability: Gravitational Waves,}''
  JCAP {\bf 1809}, 012 (2018)
  %doi:10.1088/1475-7516/2018/09/012
  [\hhref{arXiv:1804.07732} [hep-ph]].
  %%CITATION = doi:10.1088/1475-7516/2018/09/012;%%
  %30 citations counted in INSPIRE as of 04 Nov 2019






%\cite{Inomata:2018epa}
\bibitem{Inomata:2018epa} 
  K.~Inomata and T.~Nakama,
  ``{Gravitational waves induced by scalar perturbations as probes of the small-scale primordial spectrum,}''
  Phys.\ Rev.\ D {\bf 99}, no. 4, 043511 (2019)
  %doi:10.1103/PhysRevD.99.043511
  [\hhref{arXiv:1812.00674} [astro-ph.CO]].
  %%CITATION = doi:10.1103/PhysRevD.99.043511;%%
  %27 citations counted in INSPIRE as of 04 Nov 2019
  
  
  
  %\cite{Biagetti:2018pjj}
\bibitem{Biagetti:2018pjj}
  M.~Biagetti, G.~Franciolini, A.~Kehagias and A.~Riotto,
  ``{Primordial Black Holes from Inflation and Quantum Diffusion,}''
  JCAP {\bf 1807} (2018) 032
  %doi:10.1088/1475-7516/2018/07/032
  [\hhref{arXiv:1804.07124} [astro-ph.CO]].
  %%CITATION = doi:10.1088/1475-7516/2018/07/032;%%
  %33 citations counted in INSPIRE as of 16 Dec 2019
 

\bibitem{Cruces:2018cvq}
  D.~Cruces, C.~Germani and T.~Prokopec,
  ``{Failure of the stochastic approach to inflation beyond slow-roll,}''
  JCAP {\bf 1903} (2019) 048
  %doi:10.1088/1475-7516/2019/03/048
  [\hhref{arXiv:1807.09057} [gr-qc]].
  %%CITATION = doi:10.1088/1475-7516/2019/03/048;%%
  %18 citations counted in INSPIRE as of 16 Dec 2019


%\cite{Ezquiaga:2018gbw}
\bibitem{Ezquiaga:2018gbw}
  J.~M.~Ezquiaga and J.~Garc\'ia-Bellido,
  ``{Quantum diffusion beyond slow-roll: implications for primordial black-hole production,}''
  JCAP {\bf 1808} (2018) 018
  %doi:10.1088/1475-7516/2018/08/018
  [\hhref{arXiv:1805.06731} [astro-ph.CO]].
  %%CITATION = doi:10.1088/1475-7516/2018/08/018;%%
  %37 citations counted in INSPIRE as of 16 Dec 2019

%\cite{Ezquiaga:2019ftu}
\bibitem{Ezquiaga:2019ftu}
  J.~M.~Ezquiaga, J.~Garc\'ia-Bellido and V.~Vennin,
  ``{The exponential tail of inflationary fluctuations: consequences for primordial black holes,}''
  [\hhref{arXiv:1912.05399} [astro-ph.CO]].
  %%CITATION = ARXIV:1912.05399;%%

  
  
 
  %\cite{Gerbino:2016sgw}
\bibitem{Gerbino:2016sgw} 
  M.~Gerbino, K.~Freese, S.~Vagnozzi, M.~Lattanzi, O.~Mena, E.~Giusarma and S.~Ho,
  ``{Impact of neutrino properties on the estimation of inflationary parameters from current and future observations,}''
  Phys.\ Rev.\ D {\bf 95}, no. 4, 043512 (2017)
  %doi:10.1103/PhysRevD.95.043512
  [\hhref{arXiv:1610.08830} [astro-ph.CO]].
  %%CITATION = doi:10.1103/PhysRevD.95.043512;%%
  %38 citations counted in INSPIRE as of 06 Nov 2019
 

%\cite{Mangano:2005cc}
\bibitem{Mangano:2005cc} 
  G.~Mangano, G.~Miele, S.~Pastor, T.~Pinto, O.~Pisanti and P.~D.~Serpico,
  ``{Relic neutrino decoupling including flavor oscillations,}''
  Nucl.\ Phys.\ B {\bf 729}, 221 (2005)
  %doi:10.1016/j.nuclphysb.2005.09.041
  [\hhref{hep-ph/0506164}].
  %%CITATION = doi:10.1016/j.nuclphysb.2005.09.041;%%
  %488 citations counted in INSPIRE as of 06 Nov 2019
  
    \bibitem{Kawasaki_1999}
  M.~Kawasaki, K.~Kohri, and Naoshi Sugiyama.
  ``Cosmological constraints on late-time entropy production.''
  \newblock {\em Physical Review Letters}, 82(21):4168–4171, May 1999.
  [\hhref{astro-ph/9811437}].
  
  \bibitem{Hasegawa_2019}
  Takuya Hasegawa, Nagisa Hiroshima, Kazunori Kohri, Rasmus~S.L. Hansen, Thomas
  Tram, and Steen Hannestad.
  ``Mev-scale reheating temperature and thermalization of oscillating neutrinos by radiative and hadronic decays of massive particles,''
  \newblock {\em Journal of Cosmology and Astroparticle Physics},
  2019(12):012–012, Dec 2019.
  [\hhref{arXiv:1908.10189}].


  %\cite{deSalas:2015glj}
\bibitem{deSalas:2015glj} 
  P.~F.~de Salas, M.~Lattanzi, G.~Mangano, G.~Miele, S.~Pastor and O.~Pisanti,
  ``{Bounds on very low reheating scenarios after Planck,}''
  Phys.\ Rev.\ D {\bf 92}, no. 12, 123534 (2015)
  %doi:10.1103/PhysRevD.92.123534
  [\hhref{arXiv:1511.00672} [astro-ph.CO]].
  %%CITATION = doi:10.1103/PhysRevD.92.123534;%%
  %72 citations counted in INSPIRE as of 06 Nov 2019
  

%\cite{Wong:2019kwg}
\bibitem{Wong:2019kwg} 
  K.~C.~Wong {\it et al.},
  ``{H0LiCOW XIII. A 2.4\% measurement of $H_{0}$ from lensed quasars: $5.3\sigma$ tension between early and late-Universe probes,}''
  \hhref{arXiv:1907.04869} [astro-ph.CO].
  %%CITATION = ARXIV:1907.04869;%%
  %60 citations counted in INSPIRE as of 06 Nov 2019

%\cite{Riess:2019cxk}
\bibitem{Riess:2019cxk} 
  A.~G.~Riess, S.~Casertano, W.~Yuan, L.~M.~Macri and D.~Scolnic,
  ``{Large Magellanic Cloud Cepheid Standards Provide a 1\% Foundation for the Determination of the Hubble Constant and Stronger Evidence for Physics beyond $\Lambda$CDM,}''
  Astrophys.\ J.\  {\bf 876}, no. 1, 85 (2019)
  %doi:10.3847/1538-4357/ab1422
  [\hhref{arXiv:1903.07603} [astro-ph.CO]].
  %%CITATION = doi:10.3847/1538-4357/ab1422;%%
  %195 citations counted in INSPIRE as of 06 Nov 2019


%\cite{Verde:2019ivm}
\bibitem{Verde:2019ivm} 
  L.~Verde, T.~Treu and A.~G.~Riess,
  ``{Tensions between the Early and the Late Universe,}''
  Nature Astronomy 2019
  %doi:10.1038/s41550-019-0902-0
  [\hhref{arXiv:1907.10625} [astro-ph.CO]].
  %%CITATION = doi:10.1038/s41550-019-0902-0;%%
  %29 citations counted in INSPIRE as of 06 Nov 2019

%\cite{DiValentino:2019dzu}
\bibitem{DiValentino:2019dzu} 
  E.~Di Valentino, A.~Melchiorri and J.~Silk,
  ``{Cosmological constraints in extended parameter space from the Planck 2018 Legacy release,}''
  \hhref{arXiv:1908.01391} [astro-ph.CO].
  %%CITATION = ARXIV:1908.01391;%%
  %4 citations counted in INSPIRE as of 06 Nov 2019
  
  %\cite{Germani:2018jgr}
\bibitem{Germani:2018jgr}
C.~Germani and I.~Musco,
%``Abundance of Primordial Black Holes Depends on the Shape of the Inflationary Power Spectrum,''
Phys. Rev. Lett. \textbf{122} (2019) no.14, 141302
doi:10.1103/PhysRevLett.122.141302
[arXiv:1805.04087 [astro-ph.CO]].
 
 %\cite{Brodsky:2010zk}
\bibitem{Brodsky:2010zk} 
  S.~J.~Brodsky and P.~Hoyer,
  ``{The $\hbar$ Expansion in Quantum Field Theory,}''
  Phys.\ Rev.\ D {\bf 83}, 045026 (2011)
  %doi:10.1103/PhysRevD.83.045026
  [\hhref{arXiv:1009.2313}].
  %%CITATION = doi:10.1103/PhysRevD.83.045026;%%
  %35 citations counted in INSPIRE as of 20 Dec 2019
  
  %\cite{Burgess:2009ea}
\bibitem{Burgess:2009ea}
  C.~P.~Burgess, H.~M.~Lee and M.~Trott,
  ``Power-counting and the Validity of the Classical Approximation During Inflation,''
  JHEP {\bf 0909} (2009) 103
  % doi:10.1088/1126-6708/2009/09/103
  [\hhref{arXiv:0902.4465} [hep-ph]].
  %%CITATION = doi:10.1088/1126-6708/2009/09/103;%%
  %275 citations counted in INSPIRE as of 21 Dec 2019
  
  %\cite{Barbon:2009ya}
\bibitem{Barbon:2009ya}
  J.~L.~F.~Barbon and J.~R.~Espinosa,
  ``On the Naturalness of Higgs Inflation,''
  Phys.\ Rev.\ D {\bf 79} (2009) 081302
 % doi:10.1103/PhysRevD.79.081302
  [\hhref{arXiv:0903.0355} [hep-ph]].
  %%CITATION = doi:10.1103/PhysRevD.79.081302;%%
  %322 citations counted in INSPIRE as of 21 Dec 2019
  
  %\cite{Tisserand:2006zx}
\bibitem{Tisserand:2006zx}
  P.~Tisserand {\it et al.} [EROS-2 Collaboration],
  ``Limits on the Macho Content of the Galactic Halo from the EROS-2 Survey of the Magellanic Clouds,''
  Astron.\ Astrophys.\  {\bf 469} (2007) 387
  % doi:10.1051/0004-6361:20066017
  [\hhref{astro-ph/0607207}].
  %%CITATION = doi:10.1051/0004-6361:20066017;%%
  %511 citations counted in INSPIRE as of 19 Dec 2019
  
 

%\cite{Kannike:2017bxn}
\bibitem{Kannike:2017bxn} 
  K.~Kannike, L.~Marzola, M.~Raidal and H.~Veerm\"ae,
  ``{Single Field Double Inflation and Primordial Black Holes,}''
  JCAP {\bf 1709}, 020 (2017)
  %doi:10.1088/1475-7516/2017/09/020
  [\hhref{arXiv:1705.06225} [astro-ph.CO]].
  %%CITATION = doi:10.1088/1475-7516/2017/09/020;%%
  %80 citations counted in INSPIRE as of 07 Nov 2019
  
%\cite{Manohar:1983md,Cohen:1997rt,Luty:1997fk,Giudice:2007fh}
\bibitem{Manohar:1983md} 
  A.~Manohar and H.~Georgi,
  ``Chiral Quarks and the Nonrelativistic Quark Model,''
  Nucl.\ Phys.\ B {\bf 234}, 189 (1984).
  %doi:10.1016/0550-3213(84)90231-1
  %%CITATION = doi:10.1016/0550-3213(84)90231-1;%%
  %2038 citations counted in INSPIRE as of 27 Dec 2019  
  
%\cite{Cohen:1997rt,Luty:1997fk,Giudice:2007fh}
\bibitem{Cohen:1997rt} 
  A.~G.~Cohen, D.~B.~Kaplan and A.~E.~Nelson,
  ``Counting 4$\pi$'s in strongly coupled supersymmetry,''
  Phys.\ Lett.\ B {\bf 412}, 301 (1997)
  %doi:10.1016/S0370-2693(97)00995-7
  [\hhref{hep-ph/9706275}].
  %%CITATION = doi:10.1016/S0370-2693(97)00995-7;%%
  %184 citations counted in INSPIRE as of 27 Dec 2019
  
  

%\cite{Luty:1997fk,Giudice:2007fh}
\bibitem{Luty:1997fk} 
  M.~A.~Luty,
  ``Naive dimensional analysis and supersymmetry,''
  Phys.\ Rev.\ D {\bf 57}, 1531 (1998)
  %doi:10.1103/PhysRevD.57.1531
  [\hhref{hep-ph/9706235}].
  %%CITATION = doi:10.1103/PhysRevD.57.1531;%%
  %171 citations counted in INSPIRE as of 27 Dec 2019

%\cite{Giudice:2007fh}
\bibitem{Giudice:2007fh} 
  G.~F.~Giudice, C.~Grojean, A.~Pomarol and R.~Rattazzi,
  ``The Strongly-Interacting Light Higgs,''
  JHEP {\bf 0706}, 045 (2007)
  %doi:10.1088/1126-6708/2007/06/045
  [\hhref{hep-ph/0703164}].
  %%CITATION = doi:10.1088/1126-6708/2007/06/045;%%
  %860 citations counted in INSPIRE as of 27 Dec 2019

%\cite{Panico:2015jxa}
\bibitem{Panico:2015jxa} 
  G.~Panico and A.~Wulzer,
  ``The Composite Nambu-Goldstone Higgs,''
  Lect.\ Notes Phys.\  {\bf 913}, pp.1 (2016)
  %doi:10.1007/978-3-319-22617-0
  [\hhref{arXiv:1506.01961} [hep-ph]].
  %%CITATION = doi:10.1007/978-3-319-22617-0;%%
  %299 citations counted in INSPIRE as of 27 Dec 2019
  
  %\cite{Aoki:1997er}
\bibitem{Aoki:1997er} 
  K.~Aoki,
  ``Nondecoupling effects due to a dimensionful coupling,''
  Phys.\ Lett.\ B {\bf 418}, 125 (1998)
  %doi:10.1016/S0370-2693(97)01495-0
  [\hhref{hep-ph/9709309}].
  %%CITATION = doi:10.1016/S0370-2693(97)01495-0;%%
  %4 citations counted in INSPIRE as of 31 Dec 2019
  
 %\cite{Byrnes:2018txb,Ozsoy:2019lyy}
\bibitem{Byrnes:2018txb} 
  C.~T.~Byrnes, P.~S.~Cole and S.~P.~Patil,
  ``Steepest growth of the power spectrum and primordial black holes,''
  JCAP {\bf 1906}, 028 (2019)
  %doi:10.1088/1475-7516/2019/06/028
  [\hhref{arXiv:1811.11158} [astro-ph.CO]].
  %%CITATION = doi:10.1088/1475-7516/2019/06/028;%%
  %42 citations counted in INSPIRE as of 04 Jan 2020

%\cite{Ozsoy:2019lyy}
\bibitem{Ozsoy:2019lyy} 
  O.~Özsoy and G.~Tasinato,
  ``On the slope of curvature power spectrum in non-attractor inflation,''
  \hhref{arXiv:1912.01061} [astro-ph.CO].
  %%CITATION = ARXIV:1912.01061;%% 
  
  \bibitem{Leach:2000yw}
  S.~M.~Leach and A.~R.~Liddle,
  %``Inflationary perturbations near horizon crossing,''
  Phys.\ Rev.\ D {\bf 63} (2001) 043508
  %doi:10.1103/PhysRevD.63.043508
  [\hhref{astro-ph/0010082}].
  %%CITATION = doi:10.1103/PhysRevD.63.043508;%%
  %100 citations counted in INSPIRE as of 22 Jan 2020
  
  \bibitem{Leach:2001zf}
  S.~M.~Leach, M.~Sasaki, D.~Wands and A.~R.~Liddle,
  ``Enhancement of superhorizon scale inflationary curvature perturbations,''
  Phys.\ Rev.\ D {\bf 64} (2001) 023512
  %doi:10.1103/PhysRevD.64.023512
  [\hhref{astro-ph/0101406}].
  %%CITATION = doi:10.1103/PhysRevD.64.023512;%%
  %139 citations counted in INSPIRE as of 22 Jan 2020
 
%  \bibitem{Novikov1979A}
%  I.~D.~ Novikov, A.~G.~ Polnarev, A.~A.~ Starobinsky and Ia.~B.~ Zeldovich, ``Primordial black holes,''
%   Astronomy and Astrophysics, {\bf{80}} (1979) p. 104-109.

\end{thebibliography}
\end{document}